\documentclass[preprint]{elsarticle}

\usepackage[utf8]{inputenc}  
\usepackage[T1]{fontenc}     
\usepackage{xcolor} 
\usepackage{lmodern}         
\usepackage{microtype}       
\usepackage{soul}
\usepackage[scaled]{helvet} 
\usepackage[english]{babel}  
\usepackage{graphicx}        
\usepackage[bookmarks=false]{hyperref}            
\usepackage{comment}         
\biboptions{sort&compress}   

\usepackage{amsmath,amsfonts,amssymb}

\usepackage{braket}
\usepackage{bm}
\usepackage{gensymb}
\usepackage{float}
\usepackage{booktabs}
\usepackage{subcaption}
\usepackage{booktabs}
\usepackage{siunitx}
\usepackage[capitalise,nameinlink]{cleveref}
\usepackage[toc,page]{appendix}

\usepackage{xcolor}
\usepackage{todonotes}

\graphicspath{{fig/}}

\begin{document}

\begin{frontmatter}

\title{Surrogate Quantum Circuit Design for the Lattice Boltzmann Collision Operator}

\author[inst1]{Monica L\u{a}c\u{a}tu\c{s}\corref{cor1}}
\ead{m.i.l.lacatus@tudelft.nl}

\author[inst1]{Matthias M\"{o}ller}
\ead{m.moller@tudelft.nl}

\cortext[cor1]{Corresponding author.}

\address[inst1]{Delft Institute of Applied Mathematics, Delft University of Technology, Mekelweg 4, 2628 CD Delft, The Netherlands}

\begin{abstract}

This study introduces a framework for learning a low-depth surrogate quantum circuit (SQC) that approximates the nonlinear, dissipative, and hence non-unitary Bhatnagar-Gross-Krook (BGK) collision operator in the lattice Boltzmann method (LBM) for the D$_2$Q$_9$ lattice. By appropriately selecting the quantum state encoding, circuit architecture, and measurement protocol, non-unitary dynamics emerge naturally within the physical population space. This approach removes the need for probabilistic algorithms relying on any ancilla qubits and post-selection to reproduce dissipation, or for multiple state copies to capture nonlinearity. The SQC is designed to preserve key physical properties of the BGK operator, including mass conservation, scale equivariance, and D$_8$ equivariance, while momentum conservation is encouraged through penalization in the training loss. When compiled to the IBM Heron quantum processor's native gate set, assuming all-to-all qubit connectivity, the circuit requires only 724 native gates and operates locally on the velocity register, making it independent of the lattice size. The learned SQC is validated on two benchmark cases, the Taylor-Green vortex decay and the lid-driven cavity, showing accurate reproduction of vortex decay and flow recirculation. While integration of the SQC into a quantum LBM framework presently requires measurement and re-initialization at each timestep, the necessary steps towards a measurement-free formulation are outlined.

\vspace{1mm}
\emph{Keywords:} Lattice Boltzmann Method, Quantum Computing, Quantum Circuit Learning, Quantum Machine Learning, Quantum CFD, Quantum LBM.


\end{abstract}
\end{frontmatter}


\sethlcolor{yellow}

\section{Introduction}

Despite decades of study, turbulence remains a central challenge in fluid mechanics, both numerically and theoretically. Turbulent fluctuations span a large range of length and time scales, differing by several orders of magnitude. The grid resolution and time-step refinement required for accurate simulations increase steeply with the flow Reynolds number (Re). In practice, the total number of degrees of freedom in a direct numerical simulation (DNS) scale roughly as Re$^3$, meaning that fully resolving the flow field has, to date, been feasible only at low-Re numbers and for simple, canonical geometries \cite{Li2009,Yao2023}. This limitation arises not only from the sheer number of arithmetic operations but also from the prohibitive memory demands of storing high-fidelity fields throughout the simulation. Moreover, as the historic pace of Moore’s Law is predicted to slow down \cite{Theis2017End}, the gap between available classical hardware performance and the requirements of high-Re DNS continues to widen. This has prompted growing interest in alternative computing paradigms such as quantum computing.

Quantum computers exploit superposition and entanglement to perform certain tasks more efficiently than classical computers. For example, Shor’s algorithm computes integer factorization in polynomial time \cite{Shor1994}, and Grover’s algorithm searches unsorted databases with a quadratic speedup \cite{Grover1996}. Both algorithms thus demonstrate proven theoretical advantages over the best-known classical methods. Nevertheless, present quantum devices remain in the noisy intermediate-scale quantum (NISQ) era, with limited qubit counts and shallow circuit depths that preclude their use in real engineering applications. Nevertheless, ongoing developments in hardware design and error-mitigation techniques are steadily improving their reliability \cite{Kandala2019}. Once quantum hardware reaches a mature, fault-tolerant stage, its primary value will be in solving problems that classical computers struggle with. This potential has fueled sustained interest in developing quantum algorithms for computational fluid dynamics (QCFD) aimed at overcoming current limitations in resolving turbulence at high-Re numbers. Broadly speaking, these algorithms fall into three categories: methods that seek to solve the Navier-Stokes equations directly \cite{Gaitan2020NSQuantum,Budinski2022,Liu2023VQLSStokes,Song2024HybridNS,Kyriienko2020DQCsNS}, quantum lattice-gas cellular automata (LGCA) simulations \cite{Yepez2002Burgers,BastidaZamora2024EfficientQLGA,Fonio2025Quantum,SchalkersMoller2024DataEncoding,Love2019QuantumExtensions} and quantum versions of the lattice Boltzmann method (LBM). In this work, we focus exclusively on the latter.

Classically, LBM operates at the mesoscopic level, recovering the Navier-Stokes equations in the weakly compressible limit by streaming and colliding particle distribution functions, often referred to as populations, on a discrete lattice. Unlike the Navier-Stokes formulation, in which nonlinearity and non-locality are coupled by the convection term, LBM decouples these effects. The linear streaming step performs non-local transport, while the collision step handles local, nonlinear relaxation of the populations toward their local equilibrium distribution. This separation makes LBM inherently parallelizable, and GPU‐accelerated implementations have demonstrated speedups of one-to-two orders of magnitude over traditional Navier-Stokes solvers on benchmark cases \cite{Yu2014}. The same parallelism suggests a natural fit for quantum computing, since encoding the populations into the amplitudes of superposed quantum states allows a quantum implementation of the LBM to update an exponentially large configuration space in the same number of steps required classically for a single configuration.

However, the development of an end-to-end quantum LBM (QLBM) algorithm has thus far encountered a major bottleneck. The collision step in LBM is a nonlinear and dissipative relaxation process and is therefore inherently non-unitary. This inherent non-unitarity is fundamentally incompatible with the unitarity of quantum operations. Consequently, early QLBM research has focused exclusively on the streaming step. Todorova and Steijl \cite{TodorovaSteijl2020} implemented this transport as a quantum walk using multiple controlled-NOT (CNOT) gates. Schalkers and Möller \cite{SchalkersMoller2024} later reduced the total number of CNOTs and introduced an object-encoding scheme with robust boundary conditions in their Quantum Transport Method (QTM). In subsequent work \cite{SchalkersMoller2024b}, they developed an efficient technique for computing observables, such as the force acting on an object, by using a quantum version of the momentum-exchange method from classical LBM simulations. 

In recent years, various algorithms have been proposed to implement the collision step in a QLBM simulation. Many of these algorithms truncate the equilibrium distribution at first order in the macroscopic velocity, thereby discarding the quadratic nonlinear terms in the velocity \cite{KocherBryngelson2024, WawrzyniakEtAl2024, Budinski2021}. The result is a fully linear collision operator such that the full Navier-Stokes equations can no longer be recovered. This confines QLBM to low-Re number regimes. Furthermore, despite its linearity, the collision operator remains non-unitary due to its dissipative nature. The common solution to deal with the non-unitarity is to use the linear combination of unitaries (LCU) technique \cite{childs2012hamiltonian}. With LCU, a non-unitary matrix can be implemented as a weighted sum of unitaries. This, however, comes at the cost of an extra ancilla qubit and the need to perform mid-circuit measurement and post-selection on the ancilla qubit. Since post-selection succeeds only probabilistically, each QLBM time step may require re-initializing and repeating the circuit several times until the desired outcome is obtained. This severely undermines any quantum advantage of a QLBM simulation over a classical LBM simulation, especially for long simulation times where the success probability reduces with each additional time step.

Kumar and Frankel \cite{KumarFrankel2025} follow a slightly different approach and express the QLBM algorithm as a matrix-vector product. The vector represents the populations, and the matrix represents the linear collision and streaming operators. As this matrix is non-unitary, they perform classical singular-value decomposition to break it down into a sum of unitary matrices that can be simulated using LCU. The main advantage of their algorithm is that, compared to earlier implementations of LCU, the number of unitaries needed is fixed (eight per time step) and does not depend on the problem size. Nevertheless, when these unitaries are decomposed into elementary gates, it results in a large circuit depth of $O(10^5)$ per time step.

Xu et al. \cite{Xu2025} propose an ancilla-free algorithm for the D$_1$Q$_3$ and D$_2$Q$_5$ lattices that uses only two registers, $d$ and $q$, and implements the linear collision as a series of local unitary operations. This greatly reduces resource requirements compared to the LCU approach. However, the algorithm remains probabilistic and is driven by measurements on register $q$. A successful outcome on $q$ collapses $d$ into the correctly normalized state for the next iteration, while a failed outcome simply re-runs the same constant-depth circuit on $(q,d)$. Since each re-run does not require re-encoding $d$, the algorithm is significantly less costly than LCU-based methods. 

Alternative algorithms use Carleman linearization to handle the nonlinear collision by introducing a new variable for each monomial in the equilibrium distribution, resulting in a linear but infinite-dimensional system. Itani and Succi \cite{ItaniSucci2024} map each monomial into a bosonic Fock space and truncate the expansion at a chosen total occupation number, discarding all higher order terms. This creates a trade-off between accuracy and qubit count. They show that both collision and streaming can then be implemented unitarily, with qubit requirements scaling only logarithmically in the Re number. However, a complete circuit design has yet to be demonstrated. Sanavio and Succi \cite{SanavioSucci2024} instead assemble the truncated collision into a global relaxation matrix and simulate it via an LCU-based approach. This results in a very deep circuit scaling like $(NQ)^4$ in the number of two-qubit gates, where $N$ is the number of lattice points and $Q$ the number of discrete lattice velocities. They also show results for a single time-step collision circuit that is independent of the lattice size but that still requires on the order of 30,000 two-qubit gates. To reduce circuit depth, Sanavio et al. \cite{SanavioSucci2025} apply block-encoding techniques for sparse operators to embed both the collision and streaming operations into a single unitary. While this lowers the circuit depth, their algorithm still relies on ancilla post-selection, with the overall success probability falling off as the inverse of the fourth power of the number of discrete velocities.

A hybrid approach introduced by Wang et al. \cite{WangMengZhaoYang2025} replaces the LBM collision with the linear collision operator used in LGCA simulations. They demonstrate accurate simulations of vortex pair merging at Re = 350 and 2D decaying homogeneous isotropic turbulence at Re = 51. However, their method requires a corrective step at each iteration to maintain near-equilibrium, which demands significant qubit resources and therefore negates any quantum advantage. Therefore, developing a unitary collision operator that preserves the classical collision’s nonlinear, dissipative relaxation dynamics, eliminates probabilistic post-selection, and requires only a low-depth circuit implementation remains an open challenge.

The motivation behind our current study is to approach the modeling of the collision operator from a new perspective. As discussed above, previous works have mainly focused on reproducing the operator’s dissipative behavior through probabilistic algorithms that require ancilla qubits and post-selection, and are therefore inherently limited by their success probability. Moreover, these works have not directly attempted to model the operator’s nonlinearity. In contrast, our aim is to explore whether both dissipation and nonlinearity can be captured, at least partially, within a framework that avoids probabilistic algorithms as well as techniques such as introducing copies of the quantum state to reproduce the nonlinearity, which would incur additional costs in the number of copies required to perform multiple time steps. In the present formulation, nonlinearity and dissipation are not imposed directly at the level of the quantum state. Instead, they emerge in the classical population space from the combined effects of data encoding, unitary evolution, and the measurement protocol. The proposed formulation provides a low-depth circuit that is independent of lattice size and capable of approximating more than a linear relaxation toward equilibrium, using a circuit depth several orders of magnitude shallower than previous formulations in literature. It is important to note that this formulation still requires measurement and re-initialization at each time step. However, we outline the steps needed to extend it toward a measurement-free framework, where measurement is performed only at the end of the simulation. Although this approach does not yet offer a complete solution to the collision modeling problem, it introduces a new way to conceptualize how non-unitary operators can be realized within the inherently unitary framework of quantum computation.

Building on this motivation, we design and classically train a surrogate quantum circuit (SQC) that approximates the collision operator. Specifically, we focus on the Bhatnagar-Gross-Krook (BGK) collision operator on the D$_2$Q$_9$ lattice, although the method extends naturally to other lattices and collision models. To ensure physical fidelity, we embed mass conservation, scale equivariance, and symmetry preservation, formulated here as D$_8$ equivariance, directly into the circuit design and momentum conservation in the learning objective. This approach encourages the SQC to approximate the classical BGK collision operator faithfully. During the development of this work, a related approach was independently presented in the doctoral thesis of Itani \cite{Itanithesis}, who also investigated learning a quantum circuit for the BGK collision operator. In their formulation, D$_8$ equivariance is introduced through additional ancilla qubits or by averaging outputs over D$_8$-transformed copies of the data. Furthermore, their proposed circuit does not explicitly ensure mass conservation. In contrast, our framework incorporates these physical invariances into the circuit design and achieves D$_8$ equivariance without requiring ancilla qubits or data augmentation, resulting in a more compact and physically consistent approximation of the collision process. Moreover, we demonstrate that the proposed framework can partially reproduce the nonlinear and dissipative behavior of the BGK collision operator.

Our approach to designing and learning the SQC builds on recent work on data-driven modeling of LBM collision operators. Bedrunka et al. \cite{Bedrunka2024} introduced a neural formulation for the multi-relaxation-time (MRT) collision model, where a neural network learns to map local velocity moments to the non-hydrodynamic relaxation times, improving stability and accuracy compared to the classical MRT operator. Horstmann et al. \cite{Horstmann2024} adopted a similar framework focused on the relaxation times associated with bulk viscosity, showing that a locally adaptive bulk viscosity can stabilize weakly compressible flows and reduce Galilean invariance errors. Corbetta et al. \cite{corbetta2023learning} then established a general learning framework for LBM collision operators, focusing on the BGK operator on the D$_2$Q$_9$ lattice. Their work demonstrated that embedding mass and momentum conservation, scale equivariance, and D$_8$ equivariance directly in the network architecture and training procedure is essential for obtaining physically consistent approximations of the true collision dynamics. Their D$_8$ equivariance enforcement through group averaging proved effective in two dimensions but scales poorly in higher dimensions. Ortali et al. \cite{Ortali2024LENNs} addressed this shortcoming by introducing lattice equivariant neural networks with built-in symmetry constraints, enabling efficient and scalable equivariant learning in three dimensions.

For the construction and learning of the SQC, the framework of Corbetta et al. \cite{corbetta2023learning} serves as the foundation, with mass conservation and scale equivariance enforced directly at the circuit level, and momentum conservation promoted through the training objective. Following Ortali et al. \cite{Ortali2024LENNs}, symmetry preservation is embedded in the architecture by designing each layer to be fully equivariant, thereby achieving D$_8$ equivariance without relying on group-averaging strategies. Training is performed within a modified version of the quantum circuit learning (QCL) framework introduced by Mitarai et al. \cite{Mitarai2018Quantum}, which enables the approximation of complex functions with minimal circuit depth.

The remainder of this paper is structured as follows. Section \ref{chap:backThe} introduces the LBM framework, outlines the main properties of the BGK collision operator, and reviews the QCL approach. Section \ref{chap:SQCDesgin} describes the design of the SQC, followed in Section \ref{chap:trainingResults} by the training setup and results. Section \ref{sec:nonlinearity} discusses how nonlinearity and dissipation emerge within our framework, while Section \ref{sec:SQCevaluation} evaluates the performance of the SQC on the Taylor-Green vortex decay and lid-driven cavity benchmark cases. Finally, Section \ref{conclusion} summarizes the main findings and discusses possible directions for future work.

\section{Background and Theory}
\label{chap:backThe}

In this section we cover the theoretical background that our study is based on. We start with a short overview of the LBM and focus primarily on its collision process. Readers seeking a more comprehensive overview of LBM may refer to \cite{krugerLBM}. We then introduce the fundamentals of QCL and explain how we will use QCL to learn an SQC for the BGK collision operator. We assume that readers already have a working knowledge of quantum computing principles and standard operations. For a more detailed introduction to these topics please refer to \cite{Chuang}.

\subsection{Lattice Boltzmann Method}
\label{sub-sec:LBM}

LBM is a mesoscopic fluid-simulation approach in which discrete particle populations evolve on a Cartesian lattice in space and time. Each population $f_i(\mathbf{x},t)$ represents the probability density of particles moving with a discrete velocity $\mathbf{e}_i$ at position $\mathbf{x}$ and time $t$. In two dimensions, the most common lattice is based on the D$_2$Q$_9$ stencil. This stencil consists of nine populations with nine corresponding discrete velocities: a resting state $\mathbf{e}_0=(0,0)$, four axis-aligned velocities $\mathbf{e}_{1,3}=(\pm1,0)$ and $\mathbf{e}_{2,4}=(0,\pm1)$, and four diagonal velocities $\mathbf{e}_{5,6,7,8}=(\pm1,\pm1)$. At each time step $\Delta t$, the method proceeds in two stages:

\begin{enumerate}

 \item \textbf{Collision:} Populations at each lattice node relax toward their local equilibrium distribution, according to a collision operator $\Omega$:
    
    \begin{equation}
    \label{eq:collision}
      f^*_i(\mathbf{x},t) =  f_i(\mathbf{x},t) + \Omega_i\left(\mathbf{x},t\right), \quad i = 0,\ldots,8.
    \end{equation}

  \item \textbf{Streaming:} Each post-collision population $f^*_i(\mathbf{x},t)$ propagates from its current lattice node $\mathbf{x}$ to the neighboring node at $\mathbf{x} + \mathbf{e}_i\,\Delta t$ :
    \begin{equation}
     \label{eq:streaming}
      f_i\bigl(\mathbf{x} + \mathbf{e}_i\,\Delta t,\;t + \Delta t\bigr) =  f^*_i(\mathbf{x},t), \quad i = 0,\ldots,8.
    \end{equation}
    
\end{enumerate}

\noindent 
The simplest collision operator is the BGK operator:

\begin{equation}
\label{eq:bgk}
  \Omega_{i,\mathrm{BGK}}(\mathbf{x},t)
  = -\frac{1}{\tau}
  \left(f_i(\mathbf{x},t) - f_i^{\mathrm{eq}}(\mathbf{x},t)\right), \quad i = 0,\ldots,8,
\end{equation}

\noindent
where $\tau$ is the relaxation time that governs how quickly the populations approach the local equilibrium distribution $f_i^{\mathrm{eq}}(\mathbf{x},t)$. The kinematic viscosity $\nu$ of the fluid is directly related to $\tau$ by

\begin{equation}
  \label{eq:viscosity}
  \nu = c_s^2 \left(\tau - \frac{1}{2}\right),
\end{equation}

\noindent
with $c_s$ denoting the lattice speed of sound (for D$_2$Q$_9$, $c_s=\frac{1}{\sqrt{3}}$). The equilibrium distribution follows from a second-order Hermite expansion of the Maxwell-Boltzmann distribution:
\begin{equation}
  f_{i}^{\mathrm{eq}}(\mathbf{x},t)
  = 
  w_{i}\,\rho
  \left[
  1
  + \frac{\mathbf{e}_{i}\cdot \mathbf{u}}{c_{s}^{2}}
  + \frac{(\mathbf{e}_{i}\cdot \mathbf{u})^{2} - c_{s}^{2}\,(\mathbf{u}\cdot \mathbf{u})}
         {2\,c_{s}^{4}}
  \right], \quad i = 0,\ldots,8.
  \label{eq:equilibrium}
\end{equation}

\noindent
Here, $w_i$ are the lattice weights, which for the D$_2$Q$_9$ lattice are given as follows:

\begin{equation}
\label{eq:latticeweight}
    w_i =
\begin{cases}
\displaystyle \frac{4}{9}, & i = 0,\\[8pt]
\displaystyle \frac{1}{9}, & i = 1,\dots,4,\\[8pt]
\displaystyle \frac{1}{36}, & i = 5,\dots,8.
\end{cases}
\end{equation}

\noindent
The macroscopic density $\rho$ and momentum $\rho \mathbf{u}$ are obtained from the zeroth and first moments of the populations:

\begin{equation}
  \label{eq:moments}
  \rho = \sum_{i=0}^8f_{i}(\mathbf{x},t),
  \qquad
  \rho\,\mathbf{u} = \sum_{i=0}^8 f_{i}(\mathbf{x},t)\,\mathbf{e}_{i}.
\end{equation}

The BGK operator depends on the equilibrium distribution $f_i^{\text{eq}}(\mathbf{x},t)$, which in turn depends on the flow velocity $\mathbf{u}$ both linearly and quadratically, as shown in Eq. \ref{eq:equilibrium}. This quadratic dependence makes the BGK operator inherently nonlinear. The operator also introduces dissipation through the relaxation timescale $\tau$, which is related to the kinematic viscosity $\nu$ by Eq. \ref{eq:viscosity}. The combined effects of nonlinearity and dissipation render the BGK operator non-unitary. The relaxation timescale $\tau$ further determines how strongly the system approaches equilibrium. For $0<\tau<1$, the pre-collision state contracts toward equilibrium, while for $\tau>1$ it overshoots it. When $\tau=1$ the operator becomes a full projection onto the equilibrium state.

In this study, we focus on the case $\tau =1$, such that the post-collision populations satisfy $f^*_i(\mathbf{x},t)=f_i^{\text{eq}}(\mathbf{x},t)$. This choice simplifies the design of the SQC, as the post-collision state is fully determined by the equilibrium distribution. In contrast, for $\tau \neq 1$ the algorithm would need to represent varying degrees of contraction and over-relaxation while maintaining information about the pre-collision state and coherently combining it with the equilibrium state. This is considerably more challenging than directly mapping the system onto equilibrium. The implications of fixing $\tau=1$ for QLBM simulations are discussed in Section \ref{subsubsec:trainingdataset}. Finally, to ensure that the learned SQC accurately approximates the BGK operator, we require it to preserve the same symmetries and conservation properties as the original operator, which are discussed in the following section.

\subsection{Properties of the BGK Operator}
\label{subsec:properties}

\subsubsection{D$_8$ Equivariance}
\label{subsub-sec:D8equiv}

The ability of LBM to recover the weakly compressible Navier-Stokes equations relies on choosing a lattice whose discrete velocities and weights enforce full rotational invariance of both second- and fourth-order moments. This invariance ensures that the pressure tensor and the viscous stress tensor are isotropic. In two dimensions, the D$_2$Q$_9$ stencil achieves this by being invariant under the dihedral group D$_8$, the full symmetry group of a square. D$_8$ is generated by a 90$^{\circ}$ rotation $r$ about the origin and a reflection $s$ across one of the square’s symmetry axes (vertical, horizontal, or diagonal). Every symmetry operation in D$_8$ can be written as $r^k$ or $r^ks$ for $k=0,1,2,3$, giving the eight group elements:

\begin{equation}
\label{eq:D8group}
    D_8 = [I, r, r^2, r^3, s, rs, r^2s,r^3s].
\end{equation}

\begin{figure}
    \centering
    \includegraphics[width=\linewidth]{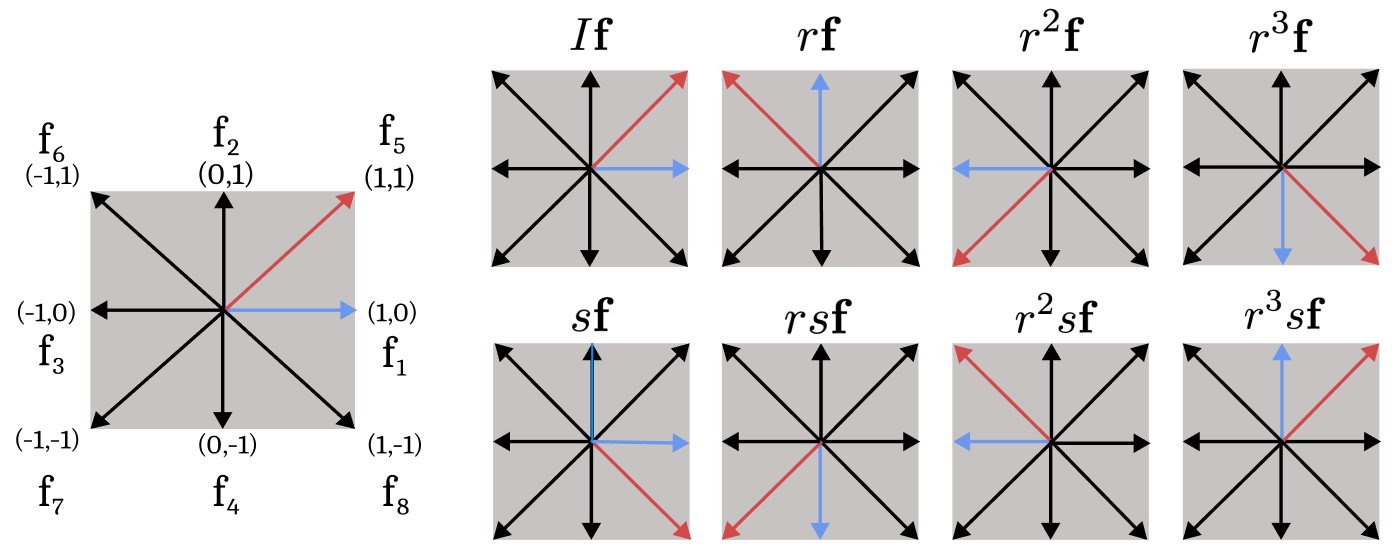}
    \caption{Action of the dihedral group D$_8$ on the D$_2$Q$_9$ populations, highlighting the transformations of the axial $f_1$ (blue) and diagonal $f_5$ (red) under each group element. Here $r$ is a 90$\degree$ anti-clockwise rotation about the origin, and $s$ is a reflection across the horizontal axis.}
    \label{fig:D2Q9_symmetry}
\end{figure}

Figure \ref{fig:D2Q9_symmetry} demonstrates how each of these symmetry operations rotates or reflects the D$_2$Q$_9$ discrete velocity vectors and their corresponding populations. Both the streaming and collision steps in LBM commute with these operations. This property is known as D$_8$ equivariance and can be written as follows for the BGK operator:

\begin{equation}
\label{eq:D8equiv}
  \Omega_{\rm BGK}\bigl(\sigma\cdot \mathbf{f} \bigr)
  \;=\;\sigma\cdot\Omega_{\rm BGK}(\mathbf{f}),
  \quad \forall\,\sigma\in D_8 
\end{equation}

\noindent
where $\sigma \cdot \mathbf{f}$ denotes the action of the symmetry operation  $\sigma$ on the local population vector $\mathbf{f}=[f_0,f_1,...,f_8]^T$.

\subsubsection{Scale Equivariance}
\label{subsub-sec:ScaleEquiv}

The BGK operator is homogeneous of degree one. This means that if all pre-collision populations are scaled by the same positive factor $\lambda>0$, the post‐collision populations scale by the same factor:

\begin{equation}
    \label{eq:scaleIn}
   \Omega_{\rm BGK}(\lambda \mathbf{f}) = \lambda  \Omega_{\rm BGK}(\mathbf{f}), \quad \forall\, \  \lambda>0
\end{equation}

This property ensures that a uniform change of scale, such as a change of units (lattice units, SI units, etc.), in the magnitudes of the pre-collision populations does not affect the relative magnitudes of the post-collision populations.

\subsubsection{Mass and Momentum Invariance}
\label{subsub-sec:rhouInv}

The BGK operator exactly conserves mass and momentum during the collision process: 

\begin{equation}
    \label{eq:rhouInv}
    \sum_{i=0}^8 (f_i^{\text{eq}}(\mathbf{x},t) - f_i(\mathbf{x},t))=0, \quad \sum_{i=0}^8(f_i^{\text{eq}}(\mathbf{x},t) - f_i(\mathbf{x},t))\mathbf {e}_i=\mathbf{0}.
\end{equation}

\subsection{Quantum Circuit Learning}
\label{sub-sec:QCL}

QCL is a hybrid classical-quantum framework first proposed by Mitarai et al. \cite{Mitarai2018Quantum}. In QCL, each input ${x_i}$ is encoded into a low-depth parametrized quantum circuit $U(x_i,\theta)$ with parameters $\theta$. After applying the circuit, the expectation value of a chosen observable $O$ is measured. A classical function $F\bigl(\{\langle O(x_i,\theta)\rangle\}\bigr)$ then produces the output $y_i=y(x_i,\theta)$. A classical optimizer adjusts the parameters $\theta$ to minimize a loss function $L(h(x_i),y(x_i,\theta))$, where $h(x_i)$ is the target function.

A typical parameterized circuit alternates layers of single-qubit rotation gates with layers of CNOT gates. A rotation on a single qubit about axis $x$, $y$ or $z$ is defined by:

\begin{equation}
    U_n(\theta) = \exp\left(-i\frac{\theta}{2}\sigma_n\right),
\end{equation}

\noindent
where $\sigma_n$ is the corresponding Pauli operator ($\sigma_x$, $\sigma_y$, or $\sigma_z$). Applying such rotations in parallel to all $N$ qubits introduces $N$ continuous free parameters per layer. Each qubit is independently mapped to a chosen point on its Bloch sphere, determined by its individual rotation angle and axis. After this layer, the global state becomes a tensor product of $N$ single-qubit states. A subsequent layer of CNOT gates then entangles the qubits by flipping each target qubit whenever its control qubit is in state $\ket{1}$. This operation introduces correlations that cannot be expressed as a simple product state. By repeating these rotation and entangling layers one obtains a universal gate set, meaning that any multi-qubit unitary operation can be approximated to arbitrary precision using only these two types of layers.

In practice, however, the precision of such approximations is constrained by the trainability of the circuits. Increasing circuit depth or the number of qubits often leads to regions known as barren plateaus, characterized by gradients that vanish exponentially with system size or depth, as demonstrated by McClean et al. \cite{McClean2018Barren}. To avoid these plateaus, one must use few qubits and use a shallow but expressive circuit that incorporates the symmetries of the problem \cite{Nguyen2024Theory}. Under these conditions, QCL offers a promising direction for learning an SQC for the BGK operator. We refer to this approximation as a \emph{surrogate quantum circuit} to emphasize that it does not reproduce the exact BGK collision but instead serves as a low-depth, quantum circuit approximation of it.

\section{Surrogate Quantum Circuit Design}
\label{chap:SQCDesgin}

In this section, we present the construction and training framework of the SQC. The circuit is designed and trained to map the pre-collision populations $f_i(\mathbf{x},t)$ to the post-collision populations $\hat f_i^{\text{eq}}(\mathbf{x},t)$, such that $\hat f_i^{\text{eq}}(\mathbf{x},t)$ closely approximate the true post-collision populations $f_i^{\text{eq}}(\mathbf{x},t)$ computed using the BGK operator. The trainable circuit architecture is constructed to satisfy the properties of the BGK operator outlined in Section \ref{subsec:properties}. This section details the quantum state encoding procedure, the design of the SQC architecture, the measurement protocol and the training routine.

\subsection{Quantum State Encoding}
\label{subsec:encoding}

To encode the pre-collision populations into the quantum state, we use a modified version of the rooted-density encoding introduced by Schalkers and Möller in \cite{SchalkersMoller2024b}. This encoding maps the square root of each population to a corresponding quantum amplitude and uses a position register and a velocity register to represent the lattice node positions and the discrete velocity directions, respectively. The quantum state is thus prepared as:

\begin{equation}
\label{eq:rooted-encoding-wholegrid}
  \ket{\psi}
  =
  \frac{1}{\sqrt{M}}
  \sum\limits_{\mathbf{x},\,i} \sqrt{f_i(\mathbf{x},t)} \,
  \ket{\mathbf{e}_i} \otimes \ket{\mathbf{x}}, \quad M = \sum\limits_{\mathbf{x},\,i} f_i(\mathbf{x},t),
\end{equation}

\noindent
where the normalization factor $\sqrt{M}$ equals the square root of the total mass of the system, obtained by summing all population $f_i(\mathbf{x},t)$ over every lattice site $\mathbf{x}$ and discrete velocity direction $\mathbf{e}_i$. Since unitary operations preserve vector norms, the quantum evolution conserves this total mass during both the streaming and collision steps, ensuring mass conservation throughout the QLBM simulation. The streaming step is a nonlocal operation that can be implemented as a unitary transformation permuting the position basis states in a manner consistent with the velocity discretization, without altering the quantum state amplitudes \cite{TodorovaSteijl2020, SchalkersMoller2024}. 

In contrast, the collision step is purely local, meaning that the BGK operator acts independently and identically at each lattice site. To mirror this structure, the SQC architecture is designed to act only on the velocity register, enforcing the same locality principle. The global collision unitary can therefore be expressed as a direct sum of identical SQC unitaries, each operating on the velocity subspace of a single lattice site. The SQC is trained on the locally normalized quantum state of a single site (see Eq. \ref{eq:rooted-encoding}), ensuring that it preserves local mass by maintaining unit norm within that subspace. When these SQC blocks are applied within the globally normalized state (Eq. \ref{eq:rooted-encoding-wholegrid}), each acts independently and identically on its corresponding lattice site. The relative amplitudes within each local velocity subspace remain unaffected by the global normalization, so the post-collision distributions coincide with those produced by the locally trained SQC. Consequently, training on locally normalized states guarantees consistent predictions under global normalization and ensures both local and global mass conservation through unitarity.

\begin{equation}
\label{eq:rooted-encoding}
\ket{\psi} = \frac{1}{\sqrt{\rho}}\sum_i \sqrt{f_i(\mathbf{x},t)} \ket{\textbf{e}_i}, \quad \rho = \sum_i f_i(\mathbf{x},t).
\end{equation}

To encode the nine discrete velocity vectors $\mathbf{e}_i$, we use four qubits spanning a 16-dimensional Hilbert space. Table \ref{tab:velocity_embedding} shows the mapping from the discrete velocities $\mathbf{e}_i$ to the corresponding four-qubit basis states stored in the velocity register. Nine of these basis states represent the discrete velocities $\mathbf{e}_i$ of the D$_2$Q$_9$ lattice, while the remaining seven are left unoccupied (initialized with zero amplitude). The SQC acts on the entire Hilbert space and can couple the occupied and unoccupied basis states, leading to a small transfer of amplitudes into the latter. To reduce this transfer, the circuit is trained using a loss function that penalizes it, as described in Section \ref{subsec:training}. However, this procedure does not guarantee zero amplitude transfer, and discarding the amplitudes in the unused states would lead to mass loss at each simulation time step. To address this issue, after measurement the probability associated with the additional seven basis states is treated as contributing directly to the $f_0(\mathbf{x},t)$ population when computing macroscopic quantities such as density and momentum. Concretely, we assign them the same discrete velocity as the rest population, $\mathbf{e}_i=(0,0)$, so their contribution is effectively absorbed into $f_0$ in the calculations. During a QLBM simulation, any amplitude that transfers into these states is kept at each time step and not re-initialized to zero, ensuring strict mass conservation. Although other ways of treating these additional states are possible, this approach performs well in our study.

\begin{table}[H]
\centering
\caption{Mapping of the nine discrete lattice velocities $\mathbf{e}_i$ to four-qubit basis states $\ket{\mathbf{e}_i}$.}
\label{tab:velocity_embedding}
\begin{tabular}{ccc}
\toprule
Population  & Discrete Velocity & Velocity Basis State \\[-0.2ex]
$f_i$                  & $\mathbf{e}_i$        & $\ket{\mathbf{e}_i}$ \\ 
\midrule
$f_0$ & (0, 0)   & $\ket{0000}$ \\
$f_1$ & (1, 0)   & $\ket{0001}$ \\
$f_2$ & (0, 1)   & $\ket{0010}$ \\
$f_3$ & (-1, 0)  & $\ket{0100}$ \\
$f_4$ & (0, -1)  & $\ket{1000}$ \\
$f_5$ & (1, 1)  & $\ket{0011}$ \\
$f_6$ & (-1, 1)  & $\ket{0110}$ \\
$f_7$ & (-1, -1) & $\ket{1100}$ \\
$f_8$ & (1, -1)  & $\ket{1001}$ \\
\bottomrule
\end{tabular}
\end{table}

\begin{figure}
    \centering
    \includegraphics[width=\linewidth]{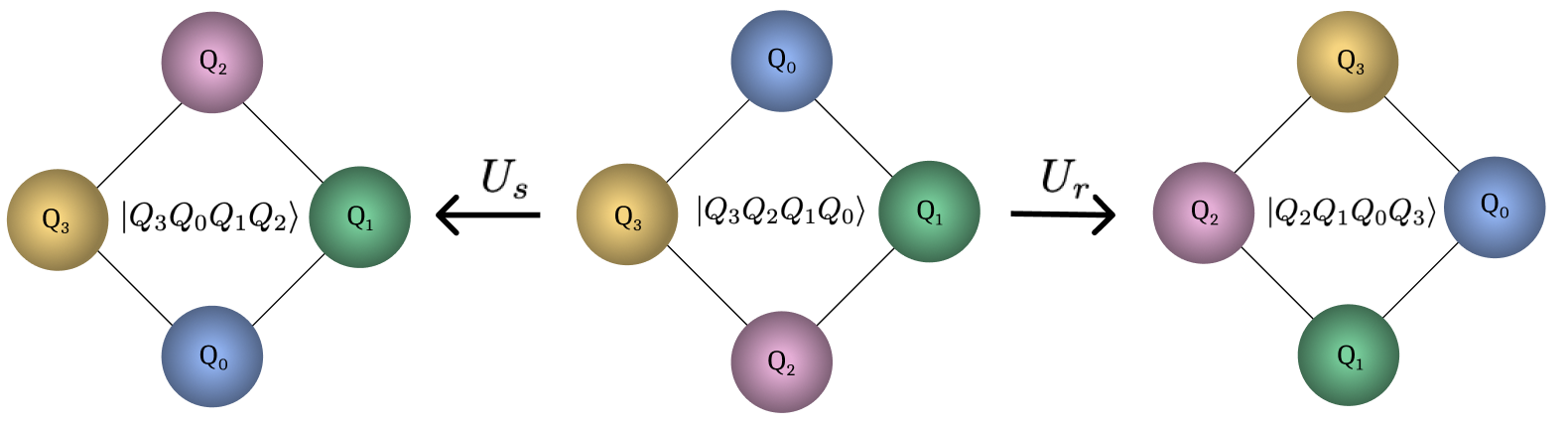}
    \caption{Mapping of lattice symmetries to permutations of qubits in the velocity register: (center) the original 4 qubit basis state $\ket{Q_3Q_2Q_1Q_0}$, (left) the reflection induced by $U_s$ across the horizontal axis permutes qubits $3 \leftrightarrow 1$ (leaving $2$ and $0$ fixed) giving $\ket{Q_1Q_2Q_3Q_0}$, (right) the 90$\degree$ anti-clockwise rotation induced by $U_r$ cyclically permutes the qubits $0 \rightarrow 1 \rightarrow 2 \rightarrow 3 \rightarrow 0$, giving $\ket{Q_2Q_1Q_0Q_3}$.}
    \label{fig:velocity-qubits}
\end{figure}

As already discussed in Section \ref{subsub-sec:D8equiv}, the D$_2$Q$_9$ lattice is invariant under the dihedral group D$_8$. The eight rotations and reflections that constitute D$_8$ (see Eq. \ref{eq:D8group}) act by permuting the discrete lattice velocities. Because each velocity is now represented in Hilbert space by a four qubit basis state, every symmetry element $\sigma\in\mathrm{D}_8$ is realized on the basis states by a unitary $U_{\sigma}$ that permutes those qubits exactly as $\sigma$ permutes the velocities. For example, a $90^{\circ}$ physical rotation is implemented as the cyclic permutation of the four qubits by the unitary $U_r$. A physical reflection across a chosen axis is implemented as a swap of one specific pair of qubits that leaves the other two unchanged by the unitary $U_s$. Figure \ref{fig:velocity-qubits} illustrates this arrangement, where the symmetries of the problem are visualized by placing the four qubits at the vertices of a square. The square symmetry of the D$_2$Q$_9$ lattice is thus replicated in Hilbert space through the square arrangement of the qubits.

To give a concrete example, consider the discrete velocity $(1,1)$, which we embed as the basis state $\ket{0011}$. Under a $90^{\circ}$ anti-clockwise rotation, the physical velocity becomes $(-1,1)$, and embedding it gives $\ket{0110}$. Equivalently, if we start from $\ket{0011}$ and then cyclically permute the four qubits through the unitary $U_{r}$ that applies the $90^{\circ}$ anti-clockwise rotation in the Hilbert space, we arrive at $\ket{0110}$ directly. In other words, embedding first and then applying the unitary representation $U_{r}$ of the rotation, gives the same result as performing a physical rotation first and then embedding. A similar analysis can be made for a reflection operation. This perfect one-to-one correspondence proves that the encoding is equivariant with respect to the D$_8$ group.

\subsection{SQC Architecture Construction}
\label{subsubsec:circuitansatz}

To design an SQC architecture expressive enough to approximate the BGK collision operator, we embed the operator’s equivariances and invariances directly into the circuit design. As discussed in the previous section, mass conservation follows naturally from our encoding and from the unitarity of the quantum operations. Furthermore, since unitary transformations are linear and homogeneous of degree one, the scale equivariance of the collision operator is inherently preserved. Momentum conservation is more challenging to impose and is instead encouraged through a penalization term in the circuit’s loss function, as described in Section~\ref{subsubsec:trainingconditions}. Finally, full D$_8$ equivariance, equivalent to invariance under conjugation by the group representation, is enforced by requiring that each layer of the SQC satisfy the following commutation condition:

\begin{equation}
\label{eq:SQCequiv}
U_{\sigma} U_{\mathrm{SQC}}^{(l)}(\theta) U_{\sigma}^{\dagger} = U_{\mathrm{SQC}}^{(l)}(\theta), \quad \forall \ \sigma \in \mathrm{D}_8.
\end{equation}

\noindent
Here, $U_{\mathrm{SQC}}^{(l)}$ denotes a generic parametrized layer of the circuit, and $U_{\sigma}$ is the unitary that implements the action of a D$_8$ symmetry operation in Hilbert space. This relation follows standard formulations of group-equivariant quantum circuits and quantum neural networks (see, for example, \cite{Meyer2023Symmetry, Nguyen2024Theory}) and forms the basis for enforcing D$_8$ equivariance in our SQC architecture. To satisfy the condition in Eq. \ref{eq:SQCequiv}, the circuit is constructed using single-qubit rotation layers and two-qubit entangling layers. In principle, more complex multi-qubit entangling operations could also satisfy this requirement, but we exclude them here, as the chosen combination of single- and two-qubit layers is sufficient to enforce the symmetry constraints while maintaining good learning performance. Exploring such multi-qubit entangling schemes remains an interesting direction for future work.

\subsubsection{Single-Qubit Rotation Layers}

The single-qubit rotation layers implement uniform single-qubit rotations that act identically on all four qubits, preserving the D$_8$ symmetry by construction. Each layer applies a rotation by angle $\theta$ about a fixed axis $n \in \{x, z\}$:

\begin{equation}
\label{eq:singlequbitrotations}
    U_n(\theta)= \left[\exp \left (-i\frac{\theta}{2}\sigma_n \right)\right]^{\otimes 4}
\end{equation}

\noindent 
where $\sigma_n = \sigma_x$ or $\sigma_z$ denotes the Pauli operator acting on a single qubit.\footnote{Restricting the rotations to the $x$ and $z$ axes is sufficient for expressivity, as any $y$-axis rotation can be decomposed into a sequence of $x$ and $z$ rotations.} Because the same unitary acts on every qubit, the unitaries $U_x(\theta)$ and $U_z(\theta)$ commute with all D$_8$ symmetry operations. Consequently, each rotation layer satisfies the equivariance condition in Eq. \ref{eq:SQCequiv}:

\begin{equation}
    U_\sigma U_n(\theta) U_\sigma^{\dagger}=U_n(\theta), \quad \forall \ \sigma \in \text{D}_8.
\end{equation}

\noindent
For notational simplicity, we denote $U_x(\theta)$ and $U_z(\theta)$ simply by $X$ and $Z$ for the remainder of this paper.

\subsubsection{Two-Qubit Entangling Layers}

To ensure D$_8$ equivariance in the two-qubit entangling layers, we must identify which entangling gates and qubit couplings among the four qubits of the SQC preserve the equivariance condition in Eq. \ref{eq:SQCequiv}. This requires understanding how both the gates and the coupling structure transform under the symmetry operations of the D$_8$ group. An entangling layer is D$_8$-equivariant if and only if it commutes with all unitaries representing the group’s symmetry operations:

    \begin{equation}
        \label{eq:equivUE}
        U_\sigma W_EU_\sigma^{\dagger} = W_E = \prod_{\{i,j\}\in E}W_{ij} , \quad \forall \ \sigma \in \text{D}_8.
    \end{equation}
    
\noindent
Here $W_E$ denotes the layer obtained by applying a two-qubit gate $W_{ij}$ to each coupled qubit pair $\{i,j\}$ in the set $E$. Each unitary $U_\sigma$ permutes or swaps the qubit indices in the velocity register according to the corresponding symmetry operation $\sigma$, as discussed in Section \ref{subsec:encoding}. Conjugating $W_E$ by $U_{\sigma}$ thus relabels the qubit indices in every two-qubit gate:

     \begin{equation}
        U_\sigma W_EU_\sigma^{\dagger} = \prod_{\{i,j\}\in E}W_{\sigma(i)\sigma(j)}, \quad \forall \ \sigma \in \text{D}_8.
    \end{equation}

\noindent
Therefore, the gate originally acting on the qubit pair $\{i,j\}$ is mapped to one acting on $\{\sigma(i), \sigma(j)\}$. For the equivariance condition in Eq. \ref{eq:equivUE} to hold, the conjugated layer must reproduce the original layer exactly:

\begin{equation}
\prod_{\{i,j\}\in E} W_{\sigma(i)\sigma(j)} = \prod_{\{i,j\}\in E} W_{ij}, \quad \forall \ \sigma \in \text{D}_8.
\end{equation}

\noindent
This imposes two conditions on the design of the layers:

\begin{enumerate}
    \item \textbf{Connectivity invariance}: 
    The coupling set $E$ must be invariant under all symmetry operations of the D$_8$ group:
    
    \begin{equation}
        \label{eq:connectviityinv}
        \{i,j\} \in E \Rightarrow \{\sigma(i),\sigma(j)\} \in E, \quad \forall \ \sigma \in \text{D}_8.
    \end{equation}

    This condition means that if a given pair of qubits $\{i,j\}$ is coupled through a two-qubit gate, then every other pair of qubits obtained from it by applying any D$_8$ symmetry operation must also be coupled. In other words, all qubit pairs related by a D$_8$ operation must appear together in $E$.

    \item \textbf{Gate uniformity}: 
    All two-qubit gates acting on such symmetry related pairs, as defined in the condition above, must be identical. This means $W_{ij}=W$ for all $\{i,j\} \in E$. Otherwise a D$_8$ operation could exchange nonidentical gates and violate Eq. \ref{eq:equivUE}.
\end{enumerate}

To satisfy the first condition, consider once again the four-qubit system arranged on the vertices of a square with vertex set $V=\{0,1,2,3\}$. The D$_8$ group acts on the set of unordered qubit pairs:

\begin{equation}
    \label{eq:qubitcoupling}
    E = \{\{0,1\}, \{1,2\}, \{2,3\}, \{3,0\}, \{0,2\}, \{1,3\}\}
\end{equation}

\noindent
by rotating or reflecting the square and thereby permuting or swapping the vertex labels. This geometric action corresponds to the same D$_8$ symmetry operations discussed earlier in the context of the rooted-density encoding, illustrated in Figure \ref{fig:velocity-qubits}. The D$_8$ action partitions the set of pairs into orbits, where each orbit collects all pairs that can be transformed into one another by some D$_8$ operation. Formally, the orbit of a pair $\{i,j\}$ under the D$_8$ symmetry operations is defined as:

\begin{equation}
    O(\{i,j\}) = \{ \{\sigma(i),\sigma(j)\} \ | \ \forall \ \sigma \in \text{D}_8 \}.
\end{equation}

\noindent
Two pairs belong to the same orbit if a D$_8$ operation maps one pair onto the other. Under this action, the six possible qubit couplings in Eq. \ref{eq:qubitcoupling} fall into exactly two distinct orbits:

\begin{enumerate}
    \item The \textbf{axial orbit} consisting of the four nearest-neighbor pairs: 

        \begin{equation}
            O_{\text{axial}} = \{\{0,1\}, \{1,2\}, \{2,3\}, \{3,0\}\}.
        \end{equation}

    \item The \textbf{diagonal orbit} consisting of the two opposite-corner pairs.
        \begin{equation}
            O_{\text{diag}} = \{ \{0,2\}, \{1,3\}\}.
        \end{equation}

\end{enumerate}

\begin{figure}
    \centering
    \includegraphics[width=0.8\linewidth]{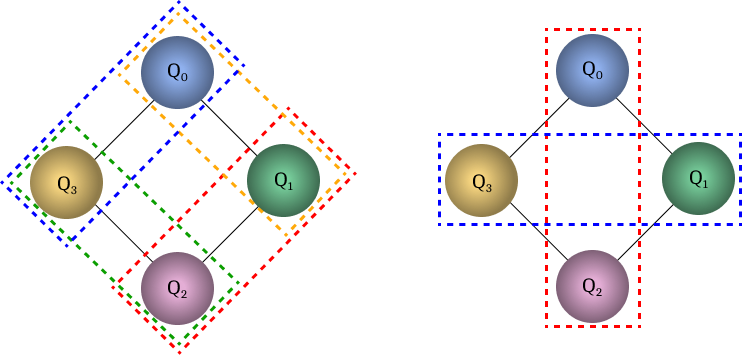}
    \caption{D$_8$-invariant coupling sets on the square qubit configuration: (left) axial coupling pattern $O_{\text{axial}} = \{\{0,1\}, \{1,2\}, \{2,3\}, \{3,0\}\}$, corresponding to the nearest-neighbor pairs along the edge of the square; (right) diagonal coupling pattern $O_{\text{diag}} = \{ \{0,2\}, \{1,3\}\}$, connecting opposite corners.}
    \label{fig:squarequbits}
\end{figure}

To satisfy the connectivity invariance condition in Eq. \ref{eq:connectviityinv}, the coupling set $E$ must remain unchanged under all D$_8$ symmetry operations. This means that if one pair $\{i,j\}$ is included in $E$, then all pairs related to it by any D$_8$ operation must also be included. Because each orbit already collects all such symmetry related pairs, any D$_8$ invariant coupling set must be composed of entire orbits (never partial ones). Consequently, there are only two distinct D$_8$ invariant coupling sets: one formed from the axial orbit $O_{\text{axial}}$, and one from the diagonal orbit $O_{\text{diag}}$. These are visualized in Figure \ref{fig:squarequbits}. In our circuit construction, each layer corresponds to one of these invariant coupling sets, an axial layer or a diagonal layer, ensuring that the connectivity pattern respects the square’s D$_8$ symmetry. These layers therefore take the following form:

\begin{equation}
    W_{O_\text{axial}} = \prod_{\{i,j\}\in O_{\text{axial}}}W_{ij}  \quad  \text{and} \quad  W_{O_\text{diag}} = \prod_{\{i,j\}\in O_{\text{diag}}}W_{ij}.
\end{equation}

Next, we analyze which two-qubit gates satisfy the gate-uniformity condition and therefore preserve D$_8$ equivariance when applied across one of the invariant coupling sets. The commonly used CNOT gate provides a useful counterexample. For a qubit pair $\{i,j\}$ the gate takes the following form:

\begin{equation}
    W_{ij} = U_{\text{CNOT}}^{(i\to j)} = \ket{0}\bra{0}_i \otimes I_j +\ket{1}\bra{1}_i\otimes \mathrm{NOT}_j
\end{equation}

\noindent
which acts asymmetrically on its two tensor factors, using qubit $i$ as the control and $j$ as the target. Applying the unitary $U_s$ that implements the reflection symmetry operation of the D$_8$ group in Hilbert space, by swapping the indices of the two qubits in the velocity basis states, results in:

\begin{equation}
    U_{s} U_{\text{CNOT}}^{(i\to j)} U_s^{\dagger} = U_{\text{CNOT}}^{(j\to i)} \neq U_{\text{CNOT}}^{(i\to j)}.
\end{equation}

\noindent
Therefore, even if identical CNOT gates are placed on every edge of the square to form an axial CNOT layer corresponding to the invariant coupling set $O_{\text{axial}}$

\begin{equation}
    W_{O_\text{axial}} = \prod_{\{i,j\}\in O_{\text{axial}}} U_{\text{CNOT}}^{(i\to j)},
\end{equation}

\noindent
this layer fails to commute with the D$_8$ symmetry operations:

\begin{equation}
    U_\sigma W_{O_\text{axial}} U_\sigma^{\dagger} = \prod_{\{i,j\}\in O_{\text{axial}}} U_{\text{CNOT}}^{(\sigma(i)\to \sigma(j))} \neq W_{O_\text{axial}}, \quad \forall \ \sigma \in \text{D}_8.
\end{equation}

\noindent
The asymmetry between control and target qubits breaks the exchange symmetry of each pair, violating the gate-uniformity condition and hence Eq. \ref{eq:equivUE}. An axial layer of CNOT gates therefore does not form a D$_8$ equivariant entangling layer.

By contrast, the two-qubit Ising gate, denoted as:

\begin{equation}
    W_{ij} = U_{ij}^{n}(\theta) = \exp\left[ -i\frac{\theta}{2} \sigma_{n,i}\sigma_{n,j}\right]
\end{equation}

\noindent
where $\sigma_{n,i}$ denotes a Pauli operator ($\sigma_x$ or $\sigma_z$) acting on qubit $i$, is symmetric under exchange of its two qubits, since $U_{ij}^{(n)}(\theta) = U_{ji}^{(n)}(\theta)$. Because the Ising gate depends only on the product of identical single-qubit Pauli operators, it remains invariant under all D$_8$ symmetry operations. A full Ising layer composed of such two-qubit gates is written as:

\begin{equation}
    W_{\text{Ising}} = \prod_{\{i,j\}\in E} U_{ij}^{(n)}(\theta).
\end{equation}

\noindent
Under the symmetry operation implemented by $U_{\sigma}$, the layer transforms as: 

\begin{equation}
    U_\sigma W_{\text{Ising}} U_\sigma^{\dagger} = \prod_{\{i,j\}\in E} U_{\sigma(i)\sigma(j)}^{(n)}(\theta) = \prod_{\{i,j\}\in E} U_{ij}^{(n)}(\theta) = W_{\text{Ising}}, \quad \forall \ \sigma \in \text{D}_8.
\end{equation}

\noindent
which shows that the layer is fully D$_8$ equivariant and satisfies the commutation relation in Eq. \ref{eq:equivUE}. For convenience, we introduce shorthand notation for Ising layers defined on the two invariant coupling sets:

\begin{equation}
    XX^{A} = \prod_{\{i,j\}\in O_{\text{axial}}} U_{ij}^{(x)}(\theta), \quad ZZ^{A}=\prod_{\{i,j\}\in O_{\text{axial}}} U_{ij}^{(z)}(\theta)
\end{equation}

\noindent
for the axial layers, and

\begin{equation}
    XX^{D} = \prod_{\{i,j\}\in O_{\text{diag}}} U_{ij}^{(x)}(\theta), \quad ZZ^{D}=\prod_{\{i,j\}\in O_{\text{diag}}} U_{ij}^{(z)}(\theta).
\end{equation}

\noindent
for the diagonal layers. The complete SQC architecture is obtained by alternating the single-qubit rotation gate layers and two-qubit Ising gate layers described above. The exact ordering and number of layers that make up the final SQC architecture are analyzed in Section \ref{chap:trainingResults}.

\subsection{Quantum State Measurement}
\label{subsec:measurement}

The SQC is trained to evolve the pre-collision quantum state into its corresponding post-collision state. For the pre-collision quantum state described by Eq. \ref{eq:rooted-encoding}, the amplitudes take the form $a_i = \sqrt{f_i(\mathbf{x},t)/\rho}$. Let the SQC circuit be represented by the unitary operator $U^{\text{SQC}}$, defined by its sequence of single-qubit and entangling layers. The post-collision amplitudes are then given by

\begin{equation}
    \label{eq:aj-finalstate-evolution}
    a_j = \sum_{i=0}^{15} U^{\text{SQC}}_{ji} a_i 
        = \sum_{i=0}^{15} U^{\text{SQC}}_{ji} \sqrt{\frac{f_i(\mathbf{x},t)}{\rho}}
         \quad \ j = 0,\dots, 15.
\end{equation}

\noindent
Here, $U^{\text{SQC}}_{ji}$ denotes the $(j,i)$-th matrix element of the unitary $U^{\text{SQC}}$, representing the contribution of the $i$-th pre-collision amplitude to the $j$-th post-collision amplitude. The indices $i$ and $j$ span the full basis of the velocity register, including the nine discrete velocities and the seven additional basis states introduced by the quantum encoding.

When we train the SQC, we interpret it as evolving the pre-collision quantum state into a post-collision quantum state, whose amplitudes encode the predicted post-collision populations $\hat f_j^{\text{eq}}(\mathbf{x},t)$. These amplitudes can therefore be written equivalently as:

\begin{equation}
    \label{eq:aj-finalstate}
    a_j =  \sqrt{\frac{\hat f_j^{\text{eq}}(\mathbf{x},t)}{\rho}} e^{i\varphi_j}, 
     \quad \ j = 0,\dots, 15.
\end{equation}

\noindent 
Each amplitudes $a_j$ is expressed in polar form, since any complex number $c = a + ib \in \mathbb{C}$ can be written as $c = |c| e^{i\varphi}$, where $|c| = \sqrt{a^2 + b^2}$ is the magnitude and $\varphi = \arctan(b/a)$ the phase. In this representation, the magnitude $|a_j|$ encodes the square-root-normalized predicted post-collision population $\hat f_j^{\text{eq}}(\mathbf{x},t)$, while the phase $\varphi_j$ arises from the single-qubit rotation and Ising gates within the SQC layers and reflects interference effects introduced by the quantum circuit. The significance of these phases for the expressivity of the SQC and for the overall QLBM algorithm will be further discussed in Section~\ref{sec:nonlinearity}. Furthermore, we introduce seven fictitious populations $\hat f_{9-15}^{\text{eq}}(\mathbf{x},t)$ that correspond to the additional seven basis states. As explained in Section \ref{subsec:encoding}, these fictitious populations essentially contribute to the $\hat f_{0}^{\text{eq}}(\mathbf{x},t)$ rest population when computing macroscopic quantities such as mass and momentum.

To recover $\hat f_j^{\text{eq}}(\mathbf{x},t)$ from the quantum state, we perform a projective measurement in the computational basis:

\begin{equation}
    \label{eq:probability}
    p_j = |a_j|^2 = \frac{\hat f_j^{\text{eq}}(\mathbf{x},t)}{\rho} 
     \quad \ j = 0,\dots, 15,
\end{equation}

\noindent
where $p_j$ is the probability of measuring the basis state $\ket{\textbf{e}_j}$. The post-collision populations are then obtained by de-normalizing these probabilities:

\begin{equation}
    \label{eq:probability2}
    \hat f_j^{\text{eq}}(\mathbf{x},t) = \rho p_j  
      \quad \ j = 0,\dots, 15.
\end{equation}

\noindent
This establishes a direct correspondence between the populations recovered in physical space through measurement and the normalized populations encoded in the quantum amplitudes.

\subsection{SQC Training Procedure}
\label{subsec:training}

\subsubsection{Training Dataset}
\label{subsubsec:trainingdataset}

To ensure that the SQC learns a general representation of the BGK operator rather than biasing toward any single flow case, we generate a fully synthetic dataset following the procedure introduced in the study of Corbetta et. al \cite{corbetta2023learning}.\footnote{All variables in this study are given in lattice units (l.u.).} First, we uniformly sample density-velocity pairs $(\rho,\mathbf{u})$ from $\rho\in[0.95,1.05]$ and $|\mathbf{u}|\in[0,0.01]$. To ensure an isotropic sampling of velocity directions, for each sampled velocity magnitude $|\mathbf{u}|$, we draw a random angle $\theta \in [0,2\pi)$ and define the velocity components as:

\begin{equation}
u_x = |\mathbf{u}|\cos\theta, \quad u_y = |\mathbf{u}|\sin\theta.
\end{equation}

\noindent
For each training sample, we first compute the equilibrium populations $f_i^{\mathrm{eq}}$. To prevent the pre-collision populations from equaling their equilibrium values and thus making the BGK collision step trivial, we introduce a non-equilibrium contribution $f_i^{\mathrm{neq}}$. The pre-collision populations are then obtained from:

\begin{equation}
  f_i = f_i^{\mathrm{eq}} + f_i^{\mathrm{neq}}, \quad i = 0, \ldots, 8,
\end{equation}

\noindent
where each $f_i^{\mathrm{neq}}$ is obtained from zero-mean Gaussian samples with standard deviation $\sigma_{\mathrm{neq}}\in[0,5\times10^{-4}]$, followed by a moment projection step that enforces zero density and momentum contribution. We then apply the BGK collision operator (Eq. \ref{eq:bgk}) with a relaxation timescale of $\tau = 1$, to obtain the ground-truth post-collision populations $f_i^{\text{eq}}$. Repeating this procedure $N=10^6$ times yields a training dataset $\{(f_i,f_i^{\text{eq}})_{i=0}^8\}_{n=0}^{N-1}$. Seven additional fictitious populations are appended to each sample to match the 16-dimensional structure of the quantum state encoding (see Section \ref{subsec:encoding}). These extra pre-collision and post-collision values are set to zero, ensuring that the SQC learns to avoid transferring amplitude into the corresponding basis states during training. The final dataset thus takes the form $\{(f_i,f_i^{\text{eq}})_{i=0}^{15}\}_{n=0}^{N-1}$. A complete summary of the parameter settings used for the data-generation, together with a visualization of the resulting dataset, is provided in Appendix \ref{App:DataGen}.

The parameter choices used for dataset generation warrant further explanation. The density range was set to $\rho \in [0.95, 1.05]$ in accordance with the weakly compressible assumption of LBM. In standard LBM formulations, a reference density $\rho_0 = 1$ is used, and only small perturbations $\rho = \rho_0 + \delta\rho$ with $|\delta\rho| \ll 1$ are allowed. Such small variations ensure that the pressure $p = c_s^2 \rho$ remains nearly linear in $\rho$ and that compressibility effects are negligible. Limiting density fluctuations to within $\pm5\%$ corresponds to low-Mach number flows ($\text{Ma} = U/c_s \lesssim 0.1$), for which LBM reduces, to leading order, to the incompressible Navier-Stokes equations.

Next, we discuss the choice made for the velocity range. In LBM, the effective nonlinearity of the collision process is governed by the Mach number rather than the Reynolds number \cite{SucciSanavioLove2025}. This decoupling is one of the main advantages of LBM and is particularly relevant for quantum implementations, where reproducing nonlinear dynamics remains a major challenge. In the Navier-Stokes equations, the nonlinear advection term scales with the Reynolds number, so high-Re flows become intrinsically more nonlinear and therefore more difficult to simulate. In contrast, in LBM the degree of nonlinearity increases with the Mach number. While higher lattice velocities strengthen nonlinear effects, the low-Mach number constraint keeps these velocities small. It is worth emphasizing that these lattice velocities correspond to much larger dimensional velocities once rescaled, such that the low-Mach number requirement does not restrict the physical flow regime that can be represented.

Training the SQC over the full velocity range typically used in LBM would require sampling $|\mathbf{u}|\in[0,0.1]$. Near the upper end of this interval, however, the equilibrium populations become strongly nonlinear, making the BGK relaxation considerably more difficult for the SQC to reproduce accurately. This is because, while the SQC can partially capture the nonlinearity of the relaxation, it cannot fully reproduce it as explained further in Section \ref{sec:nonlinearity}. To avoid this strongly nonlinear regime, we restrict the training data to the smaller velocity range $|\mathbf{u}|\in[0,0.01]$. Within this interval, the flow remains only weakly nonlinear and the circuit can learn the relaxation dynamics more effectively.

This also motivates the choice made for the standard deviation range $\sigma_{\mathrm{neq}}\in[0,5\times10^{-4}]$ used when generating the non-equilibrium component $f_i^{\text{neq}}$. The range is chosen to keep the system close to equilibrium and avoid strongly out-of-equilibrium states, which are difficult for the SQC to learn because the associated relaxation becomes strongly nonlinear and requires much higher effective dissipation. Empirically, these parameter settings provided the most stable and accurate training conditions.

Since we fix $\tau=1$, which in turn fixes the lattice viscosity, and the velocity range is constrained by the low-Mach number condition, the only remaining way to increase the Reynolds number is by refining the spatial discretization. This refinement increases the total number of lattice sites. For LBM simulations executed on classical hardware, the total computational cost scales linearly with both the number of lattice sites $N$ and the number of discrete velocity directions $Q$, resulting in a computational complexity of $O(NQ)$. In contrast, the rooted-density encoding described in Eq. \ref{eq:rooted-encoding-wholegrid} provides an exponential compression of the state representation. The position and velocity registers jointly encode the spatial and velocity information of all lattice populations within a single quantum state. Since the position and velocity registers require only $\lceil \log_2 N \rceil$ and $\lceil \log_2 Q \rceil$ qubits respectively, the total number of qubits scales as $O(\log_2 N + \log_2 Q)$. This encoding thus achieves exponential compression in memory requirements, embedding the entire lattice configuration within a single quantum superposition. The resulting superposition allows unitaries implementing the streaming and collision steps to act coherently across all sites and directions, offering a new intrinsic type of parallelism not present in classical implementations. Therefore, while fixing $\tau=1$ imposes a major constraint on classical LBM simulations, a quantum implementation can exploit the exponential compression of the lattice state to overcome this limitation.

\subsubsection{Training Conditions}
\label{subsubsec:trainingconditions}

The SQC is trained on classical hardware using standard mini batch gradient descent. We emphasize that every aspect of training, including all gradient evaluations, is performed entirely on classical processors and thus avoids the extra cost of running the optimization loop on a quantum computer. Training is performed in batches of size $\mathrm{B}=5$. This batch size allows training to proceed in mini-batches while keeping the compilation cost of training the circuit within practical limits.

The loss function used for training is a mean-squared error (MSE) loss between the SQC’s predicted post-collision populations $\hat f^{\text{eq}}_i$ and the reference populations $f^{\text{eq}}_i$ from the BGK operator, including the seven zero-amplitude populations. By incorporating these additional populations in the loss, the SQC is explicitly penalized for transferring amplitude into basis states that should remain unoccupied. Formally, the MSE is defined as

\begin{equation}
\label{eq:MSE}
\mathrm{MSE}
=
\frac{1}{(Q+7)B}
\sum_{b=0}^{B-1}
\sum_{i=0}^{Q+6}
\bigl(f^{\text{eq},(b)}_{i} - \hat f^{\text{eq},(b)}_{i}\bigr)^{2}
\end{equation}

\noindent
where $Q=9$ is the number of discrete velocities in the D$_2$Q$_9$ lattice.

For each training run, we compute an accuracy metric that measures the fraction of post-collision population predictions falling within a prescribed tolerance. Let $N$ be the number of samples in the test (or validation) set, and for each sample $n$ and population index $i$ let $f_i^{\text{eq},(n)}$ and $\hat f_i^{\text{eq},(n)}$ denote the true and predicted values, respectively. We declare a prediction accurate whenever
\[
\bigl\lvert f_i^{\text{eq},(n)} - \hat f_i^{\text{eq},(n)}\bigr\rvert < \varepsilon,
\]

\noindent
where the tolerance is chosen as $\varepsilon = 10^{-5}$. Using the indicator function

\[
\mathbf{1}\{\lvert f_i^{\text{eq},(n)} - \hat f_i^{\text{eq},(n)}\rvert < \varepsilon\}
=
\begin{cases}
1, & \text{if the prediction error is below } \varepsilon,\\
0, & \text{otherwise}.
\end{cases}
\]

\noindent
the average accuracy for population $i$ is simply the fraction of samples whose prediction is accurate within the given tolerance:
\[
\bigl\langle \mathrm{Accuracy}(f_i)\bigr\rangle
= \frac{1}{N}
\sum_{n=0}^{N-1}
\mathbf{1}\bigl\{\lvert f_i^{\text{eq},(n)} - \hat f_i^{\text{eq},(n)}\rvert < \varepsilon\bigr\},
\quad \ i = 0, \dots, 8.
\]

Since momentum conservation cannot be enforced exactly via the quantum state encoding or the circuit architecture, we add a penalty term to the MSE loss function that quantifies deviations from the exact conservation. For each batch sample $b$ we compute the true momenta 

\begin{equation}
p_{x}^{(b)}
=
\sum_{i=0}^{15} f_{i}^{\text{eq},(b)} \, e_{i,x}, \quad p_{y}^{(b)}
=
\sum_{i=0}^{15} f_{i}^{\text{eq},(b)} \, e_{i,y},
\end{equation}

\noindent
and the predicted momenta

\begin{equation}
\hat p_{x}^{(b)}
=
\sum_{i=0}^{15} \hat f_{i}^{\text{eq},(b)} \, e_{i,x}, \quad \hat p_{y}^{(b)}
=
\sum_{i=0}^{15} \hat f_{i}^{\text{eq},(b)} \, e_{i,y}.
\end{equation}

\noindent
The momentum penalty loss is then defined as:

\begin{equation}
L_{m}
=
\frac{1}{B}
\sum_{b=0}^{B-1}
\Bigl[\bigl(p_{x}^{(b)} - \hat p_{x}^{(b)}\bigr)^{2}
      + \bigl(p_{y}^{(b)} - \hat p_{y}^{(b)}\bigr)^{2}\Bigr].
\end{equation}

\begin{figure}
    \centering
    \includegraphics[width=\linewidth]{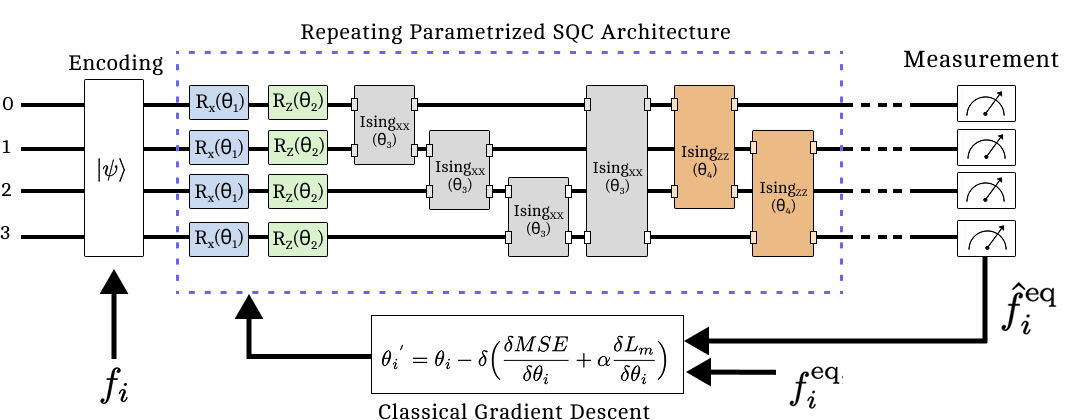}
    \caption{End-to-end training loop for the SQC. The pre-collision populations $f_i$ are first encoded into a quantum state. The SQC then acts on this state to generate the post-collision state, from which the post-collision predictions $\hat f_i^{\text{eq}}$ are obtained after measurement in the computational basis. These predictions are evaluated using the combined loss $L = \mathrm{MSE} + \alpha L_m$, and the circuit parameters $\theta_i$ are updated via classical gradient descent.}
    \label{fig:SQCloop}
\end{figure}

\noindent
The final loss function combines these terms: 

\begin{equation}
\label{eq:loss_cons}
L
=
\mathrm{MSE}
\;+\;
\alpha\,L_{m}
\end{equation}

\noindent
where $\alpha$ is a regularization weight that increases the relative strength of the momentum penalty over the training iterations. A schematic of the complete training loop is given in Figure \ref{fig:SQCloop}.

\section{Training Results}
\label{chap:trainingResults}

In this section, we perform several experiments to determine the optimal circuit architecture for the SQC and its best training configuration. Section \ref{subsec:optimalarc} compares several architecture designs to determine which arrangement of single-qubit rotation layers and two-qubit Ising entangling layers results in the most accurate SQC for the BGK operator. Building on these findings, Section \ref{subsec:nrlayers} presents a circuit depth analysis that quantifies the balance between collision accuracy and total gate count. In Section \ref{subsec:momentumcon}, we investigate whether including a momentum penalty in the loss function guides the training towards learning an SQC that better conserves momentum. A complete list of the fixed training parameters used in the above sections is provided in Appendix \cref{App:CircAcc,App:depth,App:cons}. Finally, Section \ref{subsec:optimaltraining} specifies the SQC architecture and training parameters that we use to benchmark the SQC on the LBM test cases in Section \ref{sec:SQCevaluation}.

\subsection{Optimal Circuit Architecture}
\label{subsec:optimalarc}

The number of possible SQC architectures is too large to explore exhaustively, therefore we focus on designs built from identical blocks that consist of two successive single-qubit rotation layers ($X$ and $Z$), and either one or two successive Ising entangling layers drawn from $ \{\, XX^{A},XX^{D},ZZ^{A},ZZ^{D}\}$. Since \(X\) and \(Z\) do not commute, the order of the rotation layers matters. We therefore evaluate each design in both orders, $X\to Z$ and $Z\to X$. When two Ising entangling layers are included, only the pairs \((XX^A,ZZ^D)\) and \((XX^D,ZZ^A)\) fail to commute. These commutation properties result in exactly 24 distinct circuit blocks for constructing the SQC. In total, combining the two possible rotation orders with the four available Ising entangling layers and accounting for the two non-commuting entangler pairs, each of which must be tested in both possible sequences, results in 24 distinct circuit blocks for constructing the SQC.

Each architecture is configured to have the same total number of trainable parameters. If one circuit has more free parameters than another, any observed performance gain might simply result from its greater number of degrees of freedom rather than from its specific gate arrangement. To enforce an equal number of parameters, circuits with a single Ising entangling layer use seven repeating blocks, while those with two Ising entangling layers use five blocks plus an additional single-qubit rotation layer at the end. That extra rotation layer restores the missing degree of freedom so that every design, regardless of its Ising entangling layer count, maintains an identical learning capacity.

\begin{figure}
    \centering
    \includegraphics[width=0.95\linewidth]{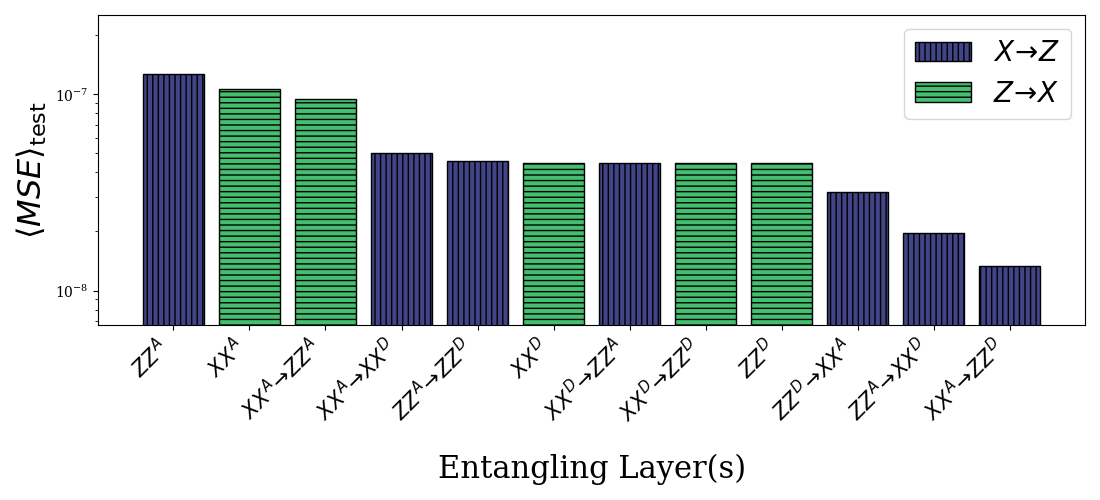}
    \caption{Average test‐set MSE loss for each Ising entangling‐layer configuration, showing only the rotation‐order variant ($X \to Z$ or $Z \to X$) that achieved the lowest loss.}
    \label{fig:exp1-MSELOSS}
\end{figure}

\begin{figure}
    \centering
    \includegraphics[width=0.95\linewidth]{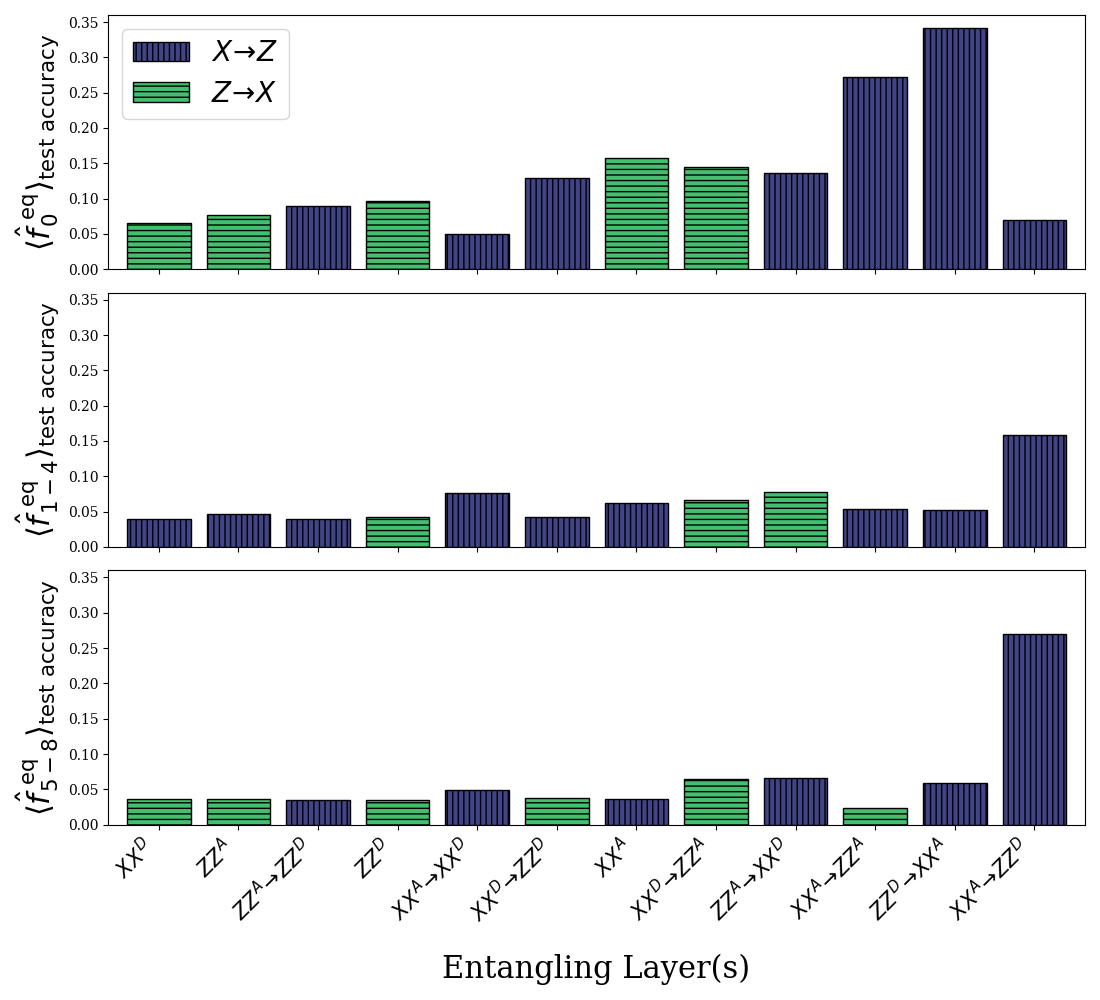}
    \caption{Average test‐set accuracy for the predicted post-collision rest population $\hat f_0^{\text{eq}}$, the averaged axial populations $\langle \hat f_{1-4}^{\text{eq}}\rangle$, and the averaged diagonal populations $\langle \hat f_{5-8}^{\text{eq}}\rangle$, obtained for each Ising entangling layer configuration. Shown is only the rotation‐order variant ($X \to Z$ or $Z \to X$) that achieved the highest accuracy.}
    \label{fig:exp1-ACC}
\end{figure}

For each architecture, we perform three independent training runs and report the average test-set MSE loss (excluding momentum penalization), along with the average test-set accuracy for the following three predicted post-collision population groups: the rest population $\hat f_0^{\text{eq}}$, the averaged axial populations $\langle \hat f_{1-4}^{\text{eq}}\rangle$, and the averaged diagonal populations $\langle \hat f_{5-8}^{\text{eq}}\rangle$. Figure \ref{fig:exp1-MSELOSS} shows, for each Ising entangling layer configuration, the lower of the two MSE losses obtained under the $X \to Z$ or $Z \to X$ rotation orders, arranged from highest loss on the left to lowest loss on the right. Figure \ref{fig:exp1-ACC} displays, for the same configurations, the higher of the two test-set accuracies obtained for each of the three population groups under both rotation orders. The configurations are arranged from left to right in order of increasing combined accuracy. The best performing architecture is the one in which the repeating blocks follow the sequence $\{ X, Z, XX^{A}, ZZ^{D} \}$. It achieves both the lowest average test-set MSE and the highest combined accuracy across the three population groups. We therefore select this SQC architecture for all subsequent training experiments. For reference, this architecture is exactly the one shown in the training loop in Figure \ref{fig:SQCloop}.

Across all analyzed architectures, a clear trade-off exists between the accuracy attainable for the post-collision population $\hat f_0^{\text{eq}}$ and that for the axial and diagonal populations. This behavior can be understood from the analytical form of the equilibrium distribution for $f_0^{\text{eq}}$. Since its associated discrete lattice velocity is $\mathbf{e}_0=(0,0)$, its equilibrium distribution is given by:

\begin{equation}
    \label{eq:f0equilibrium}
    f_{0}^{\text{eq}} = w_{0}\,\rho
    \left[1 -\frac{(\mathbf{u}\cdot \mathbf{u})}{2\,c_{s}^{2}} \right].
\end{equation}

\noindent 
This functional form differs fundamentally from those of the axial and diagonal equilibria, implying that the circuit must learn two distinct functional relationships. First, $f_{0}^{\text{eq}}$ lacks both the linear term in the velocity $(\mathbf{e}_i\cdot \mathbf{u})$ and the directional quadratic term $(\mathbf{e}_i\cdot \mathbf{u})^2$, that appear in the equilibria of the other populations (see Eq. \ref{eq:equilibrium}). Its sole dependence on the scalar quadratic term $(\mathbf{u} \cdot \mathbf{u})$, makes $f_0^{\text{eq}}$ purely nonlinear, requiring the circuit to capture a stronger quadratic mapping. Second, $f_{0}^{\text{eq}}$ depends only on the magnitude of the velocity and not on its direction, whereas the other equilibria vary with directional projections through the inner products $(\mathbf{e}_i\cdot \mathbf{u})$. Therefore, $f_{0}^{\text{eq}}$ is isotropic since its value remains the same for all flow directions, while the axial and diagonal equilibria are anisotropic, showing direction-dependent behavior. The circuit must therefore represent two qualitatively different behaviors: an isotropic quadratic dependence for $f_{0}^{\text{eq}}$, and directionally varying responses for the axial and diagonal equilibria. Because the latter constitute the majority of the lattice directions, achieving high accuracy for these anisotropic equilibria provides a more representative measure of the circuit’s overall predictive capability than optimizing solely for $f_0^{\text{eq}}$. The chosen SQC architecture ($\{ X, Z, XX^{A}, ZZ^{D} \}$) attains the best overall accuracy when this weighting toward the axial and diagonal equilibria is taken into account.

Architectures with two Ising entangling layers generally outperform those with a single layer, particularly when the layers differ in orientation (one axial and one diagonal) and in type (one $XX$ and  one $ZZ$). Combining different Ising entangling layers in this way, allows these circuits to correlate amplitudes across basis states more effectively and to explore a larger portion of Hilbert space via rotations about multiple axes. Several architectures that achieve high population accuracies, such as those relaying on the entangling layer configurations $XX^A$, $ZZ^D \to XX^A$ and $XX^A \to ZZ^A$, do so at the cost of larger MSE losses. Recall that the MSE loss is computed over all 16 basis states (see Eq. \ref{eq:MSE}), including those with zero target amplitudes. An architecture can lead to misleading results by transferring amplitude on the seven basis states which should remain unoccupied. This inflates the accuracies of the other nine basis states while penalizing the MSE loss. Our chosen SQC architecture avoids this trade-off entirely, resulting in both a low MSE loss and high population accuracies.

\subsection{Circuit Depth Analysis}
\label{subsec:nrlayers}

\begin{figure}[htp]
  \centering
  \begin{subfigure}[t]{0.46\textwidth}  
    \includegraphics[width=\textwidth]{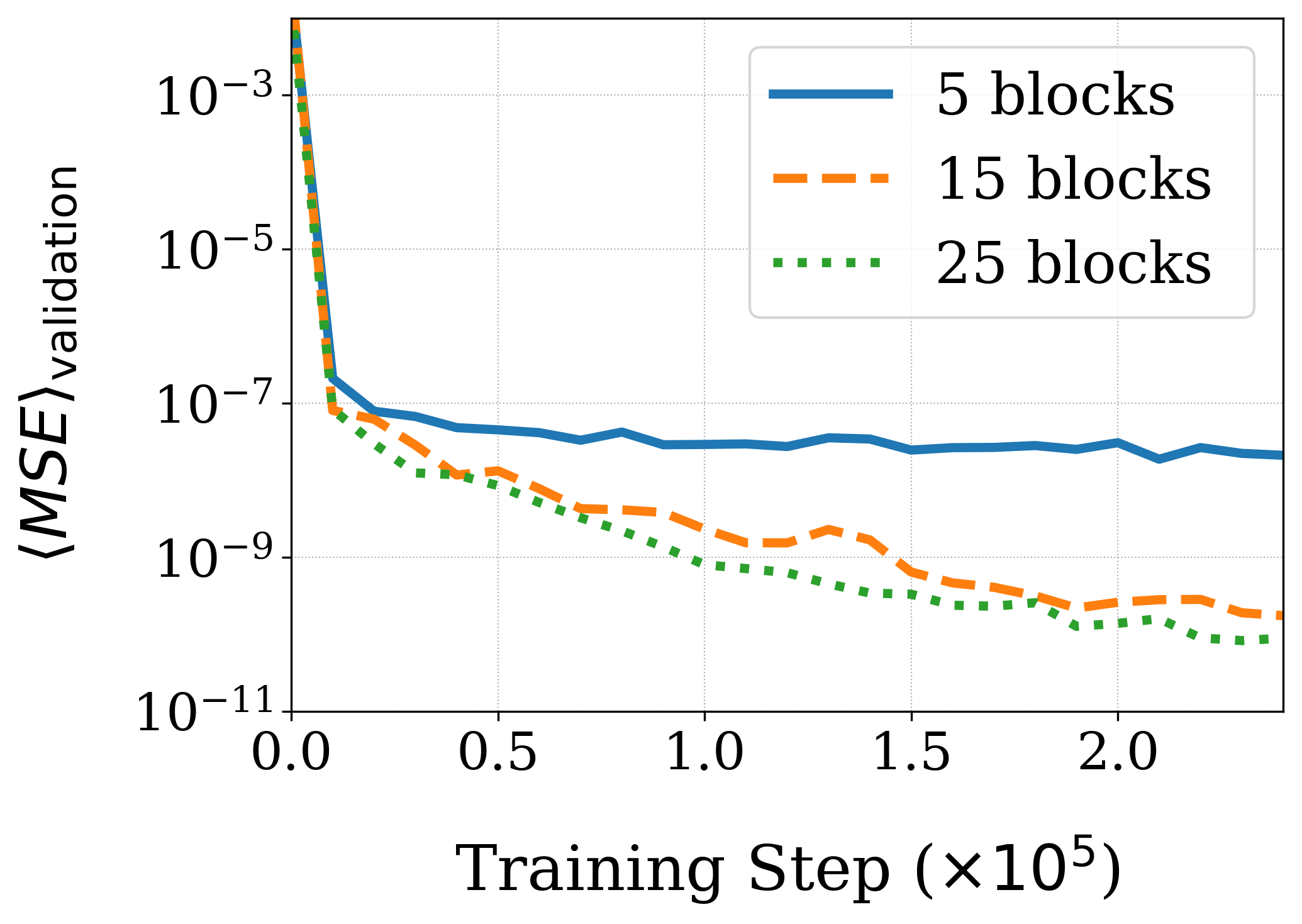}
    \caption{Average validation-set MSE loss vs.\ training step for the different circuit depths.}
    \label{fig:mse_curve}
  \end{subfigure}
  \hfill
  \begin{subfigure}[t]{0.46\textwidth}  
    \includegraphics[width=\textwidth]{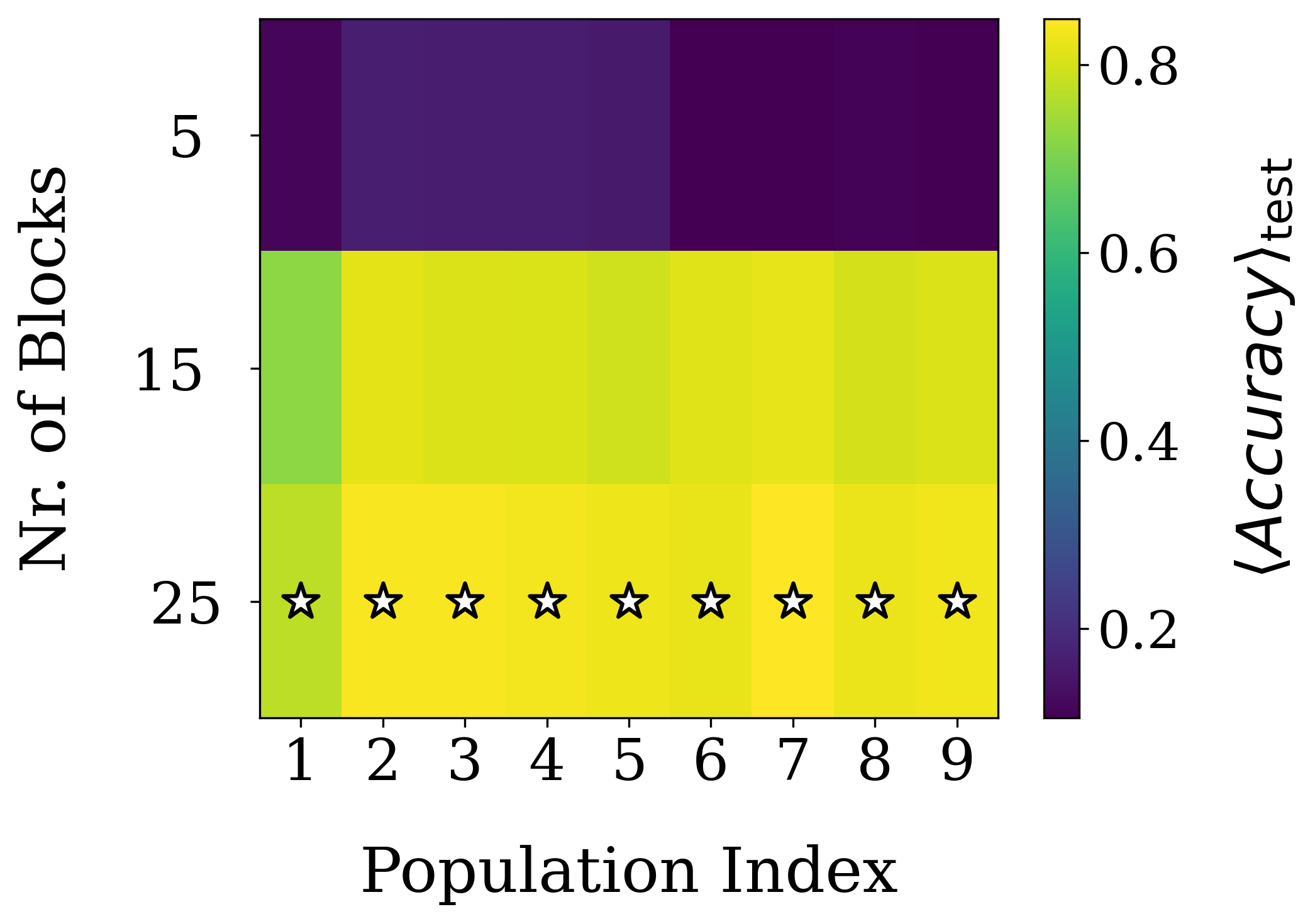}
    \caption{Average test-set population accuracy for different circuit depths. A star marks the highest accuracy attained across all depths.}
    \label{fig:acc_curve}
  \end{subfigure}
  \caption{Training results for SQC circuit depth experiments.}
  \label{fig:exp-depth}
\end{figure}

To evaluate how the SQC performance varies with circuit depth, we build SQC circuits with 5, 15, and 25 repeating blocks based on the best architecture design ($\{X,Z,XX^{A},ZZ^{D}\}$) identified in the previous section. For each circuit depth we conduct three independent training runs. Figure \ref{fig:mse_curve} shows the validation-set MSE loss as a function of the training step, with values averaged over all runs. Figure \ref{fig:acc_curve} presents the corresponding test-set accuracy for all nine populations, also averaged across runs.

\captionsetup[subfigure]{%
  width=\textwidth,   
  singlelinecheck=false   
}

These results confirm that increasing circuit depth reduces the validation-set MSE loss and improves the average test-set accuracy across all nine post-collision populations. The final MSE loss decreases from $2.1 \times 10^{-8}$ for 5 blocks to $1.7 \times 10^{-10}$ for 15 blocks, while a further increase to 25 blocks results only in a slight reduction to $8.9 \times 10^{-11}$. The mean accuracy across all nine populations rises sharply from 0.13 for 5 blocks to 0.80 for 15 blocks, with a modest gain to 0.83 for 25 blocks. The improvement from 5 to 15 blocks is therefore much larger than that from 15 to 25, indicating that the circuit’s expressivity effectively saturates beyond this depth.

During training, the gradient norms decreased as the MSE loss approached its minimum. For 5 blocks, they typically fluctuated around $1\times 10^{-5}$ and reached minima near $1 \times 10^{-6}$, while for 15 and 25 blocks they fluctuated around $1 \times 10^{-6}$ with minima close to $1 \times 10^{-7}$. 
This moderate decrease with depth indicates that parameter updates become smaller as the circuit depth increases. However, the gradient norms remain well above the level associated with barren plateaus. The slower improvement beyond 15 blocks is therefore primarily due to the limited increase in circuit expressivity rather than the onset of barren plateaus. Beyond this depth, adding more layers does not significantly enhance the circuit’s ability to reproduce the BGK operator, as it cannot fully approximate the quadratic nonlinearity of the equilibrium distribution (further discussed in Section \ref{sec:nonlinearity}).

The SQC gate set $\{X, Z, XX^{A}, ZZ^{D}\}$ includes gates that are not all native to current quantum hardware. To obtain a realistic estimate of the actual circuit depth, we compiled the 15-block and 25-block SQC circuits into their corresponding unitary matrices and subsequently transpiled these unitaries using Qiskit’s transpiler \cite{IBMQuantumTranspiler} into the native gate set of IBM’s Heron quantum processor. This native gate set is $\{RZ, SX, CZ\}$, where $CZ$ is the only two-qubit gate \cite{ibm2025nativegates}. Transpilation was performed at Qiskit’s highest optimization level, assuming full all-to-all qubit connectivity. This setting allows the compiler to merge and cancel redundant rotations across layers, resulting in a globally optimized estimate of the native gate counts. The 25-block circuit transpiles to \textbf{733} native gates, while the 15-block circuit requires \textbf{724}, meaning that the total native gate count increases by only about 1.2\% despite adding ten additional blocks. In both cases, the number of two-qubit $CZ$ gates is \textbf{95}. The nearly identical native gate counts indicate that, once compiled into a full unitary, the transpiler no longer distinguishes between the nominal circuit depths. The decomposition cost depends primarily on the algebraic complexity of the overall unitary rather than on the number of layers used to construct it. Thus, increasing the number of blocks does not necessarily result in a more complex unitary.

Although the 25-block circuit attains slightly lower MSE losses during training, the 15-block circuit already achieves comparably low values and high population accuracies. Because training becomes increasingly expensive with circuit depth, the shorter circuit is preferable for testing different configurations and conducting detailed analyses. For these reasons, the 15-block circuit is used for all subsequent training experiments and simulations presented in Section \ref{sec:SQCevaluation}.

\subsection{Momentum Loss Penalization}
\label{subsec:momentumcon}

To evaluate whether the penalty term in the loss function (see Eq. \ref{eq:loss_cons}) helps the SQC learn to better conserve momentum during training, we perform a series of experiments with the penalty enabled. Starting from an initial regularization weight $\alpha=10^{-4}$, we increase $\alpha$ every 10,000 iterations as follows: in experiment 1 up to $\alpha=0.25$, in experiment 2 up to $\alpha=0.5$, in experiment 3 up to $\alpha=1.0$.

\begin{figure}[htp]
  \centering
  \begin{subfigure}[t]{0.46\textwidth}  
    \includegraphics[width=\textwidth]{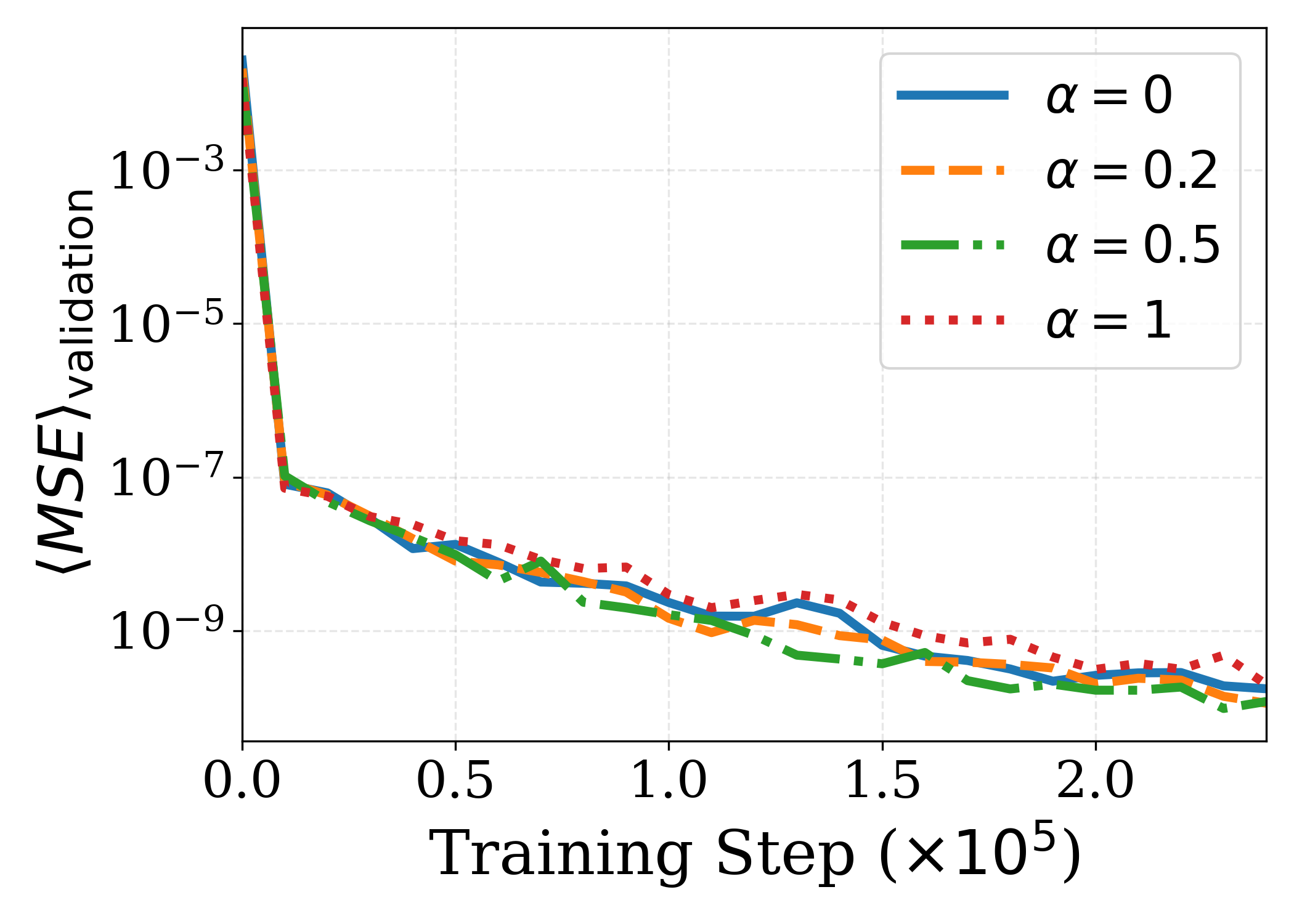}
    \caption{Average validation-set MSE loss vs. training step for different regularization weights $\alpha$.}
    \label{fig:penalty_mse}
  \end{subfigure}
  \hfill
  \begin{subfigure}[t]{0.46\textwidth}  
    \includegraphics[width=\textwidth]{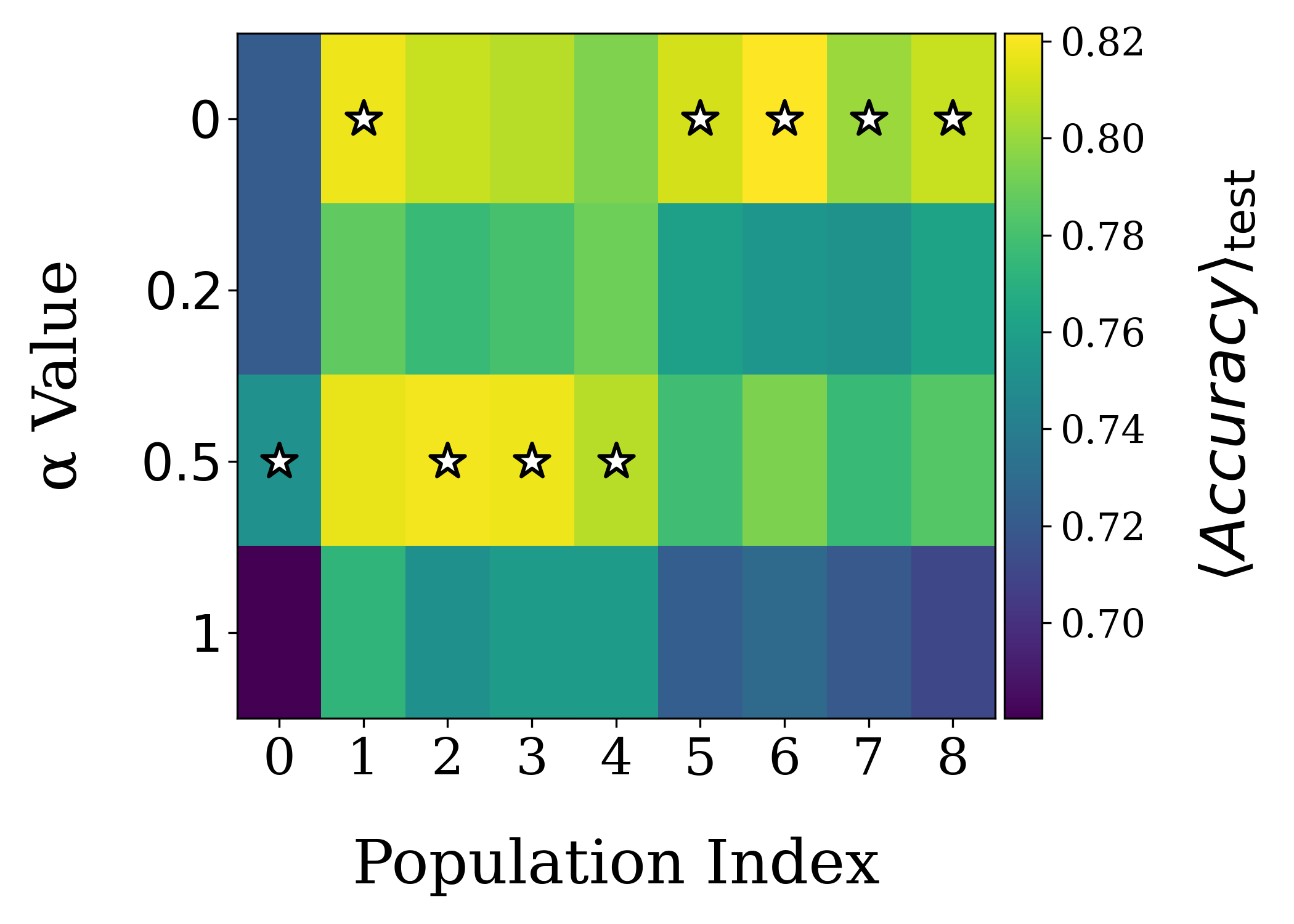}
    \caption{Average test-set population accuracy for different regularization weights $\alpha$. Star indicates the highest accuracy across all experiments.}
    \label{fig:penalty_acc}
  \end{subfigure}
  \caption{Training results for momentum penalty loss experiments.}
  \label{fig:exp-cons}
\end{figure}

We conduct three independent training runs for each experiment. Figures \ref{fig:penalty_mse} and \ref{fig:penalty_acc} show the average validation‐set MSE loss and the average test-set population accuracy, respectively, each computed over the three runs. For comparison, both plots include the 15-block SQC trained without a momentum conservation penalty ($\alpha=0$) as a baseline. As we increase the penalty weight from $\alpha = 0$ to $\alpha = 0.5$, the validation-set MSE loss shows a modest decrease during training, while the test-set accuracy remains essentially unchanged or improves slightly in populations $\hat f_0^{\text{eq}}$ and $\hat f_{2-4}^{\text{eq}}$. Over the same range, the average relative momentum loss on the test-set drops from $0.0028$ at $\alpha = 0$ to $0.0021$ at $\alpha = 0.5$, as shown in Table \ref{tab:momentum_loss}. Raising $\alpha$ further to $1.0$ continues to lower the momentum loss to $0.0019$, but at the cost of increased validation-set MSE loss and reduced accuracy across all populations. This result indicates that a penalty weight of 1.0 overconstrains the learning process. Although the SQC trained without the penalty term ($\alpha = 0$) achieves the highest overall accuracy for most populations ($\hat f_1^{\text{eq}}$ and $\hat f_{5-8}^{\text{eq}}$), the configuration with $\alpha = 0.5$ offers the best balance, improving momentum conservation with only a minimal impact on accuracy.

\begin{table}[H]
  \centering
  \caption{Average test-set relative momentum loss for different regularization weights $\alpha$.}
  \label{tab:momentum_loss}
  \begin{tabular}{cc}
    \toprule
    $\mathbf{\alpha}$ & \textbf{Average Relative Momentum Loss} \\
    \midrule
    0.0 & 0.0028 \\
    0.2 & 0.0027 \\
    0.5 & 0.0021 \\
    1.0 & 0.0019 \\
    \bottomrule
  \end{tabular}
\end{table}

\subsection{Optimal Training Conditions}
\label{subsec:optimaltraining}

Based on the experiments conducted throughout this section, we can now specify the best-performing training configuration and architecture for the SQC. These are summarized in Table \ref{tab:optimal_conditions}. 

\begin{table}[ht]
  \centering
  \caption{Best-performing SQC training configuration.}
  \label{tab:optimal_conditions}
  \sisetup{
    table-number-alignment = center,
    detect-all
  }
  \begin{tabular}{
    l           
    S[table-format=1.2]  
  }
    \toprule
    \textbf{Training Configuration} 
      & {\textbf{Setting}} \\
    \midrule
    Architecture & {$\{X,Z,XX^{A},ZZ^{D}\}$} \\  
    Number of blocks & {15} \\
    Optimizer       
      & {JAX Gradient Descent} \\
    Learning Rate   
      & {0.05} \\
    Training Iterations 
      & {750000} \\
    Batch Size ($B$)      
      & {5} \\
    Momentum Conservation Penalty 
      & {ON} \\
    Regularization weight ($\alpha$) & {0.5} \\
    Parameter Initialization ($\theta$) 
      & {Uniform on $[-\pi,\pi]$} \\
    \bottomrule
  \end{tabular}
\end{table}

\section{Emergence of Nonlinearity and Dissipation in the SQC Framework}
\label{sec:nonlinearity}

In this section, we show that the SQC framework, which combines the rooted-density encoding, the SQC itself, and measurement in the computational basis, can approximate more than a purely linearized relaxation toward equilibrium, even though the circuit’s evolution is strictly unitary and therefore linear in Hilbert space. More specifically, we demonstrate that this framework can partially capture the nonlinear and dissipative behavior of the BGK operator. The nonlinearity arises from the nonlinear mapping between the classical populations and the quantum amplitudes that encode them, together with the measurement step that converts those amplitudes back into populations in physical space. The apparent dissipation originates from the same measurement process, which collapses the quantum state and thereby induces an irreversible relaxation toward equilibrium in physical space. In the remainder of this section, we explicitly derive these effects and discuss their implications for extending the framework to avoid measurement and re-initialization at every time step when applied within a QLBM algorithm.

\subsection{Nonlinear Relaxation Towards Equilibrium}

\begin{figure}[H]
    \centering
    \includegraphics[width=\linewidth]{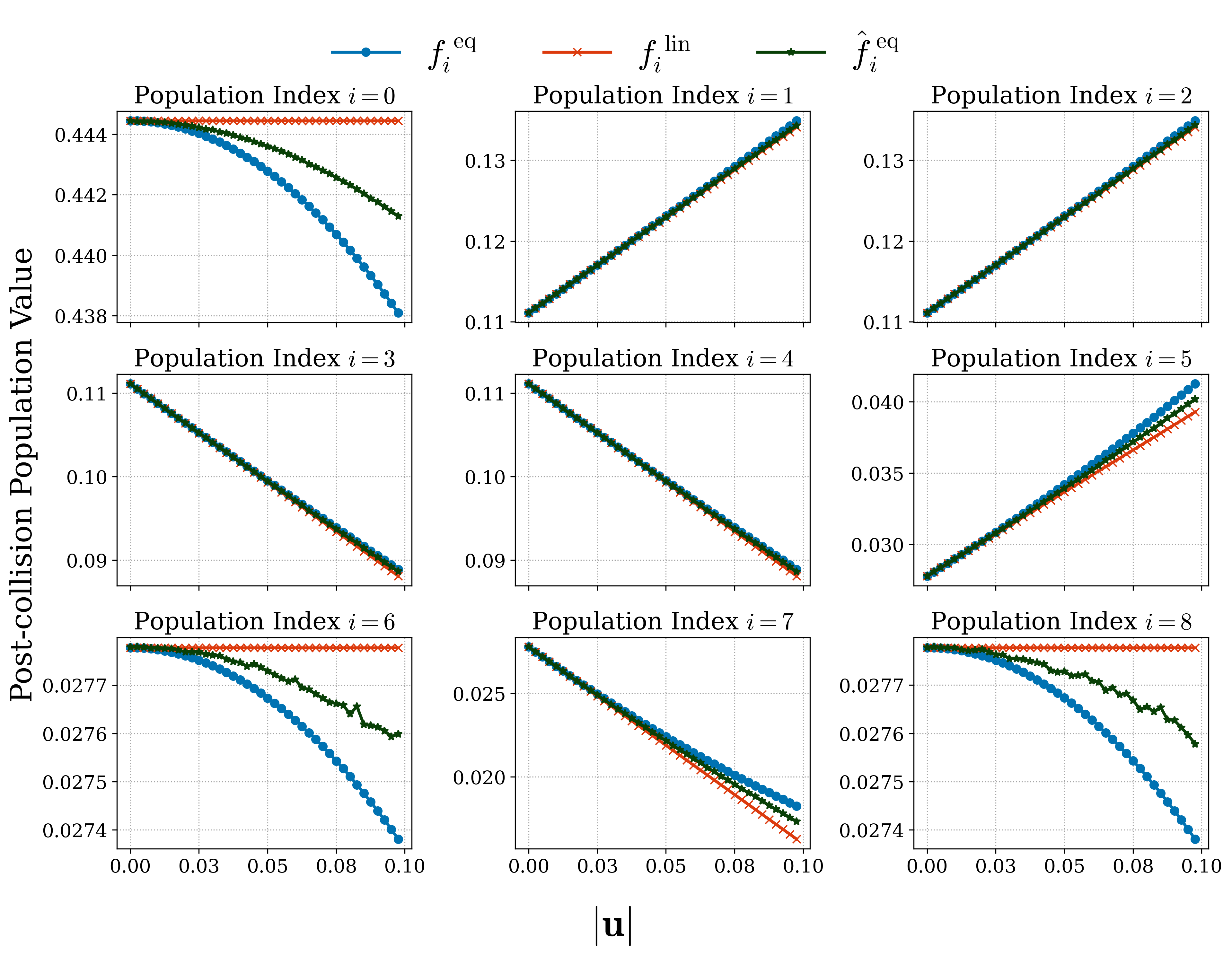}
    \caption{Comparison of post-collision population profiles obtained from the nonlinear LBM equilibrium $f_i^{\text{eq}}$, the SQC predictions $\hat f_i^{\text{eq}}$, and the linearized equilibrium $f_i^{\text{lin}}$ for velocity magnitudes in the range $|\textbf{u}|\in[0,0.1]$.}
    \label{fig:nonlinearity}
\end{figure}

To evaluate the SQC framework’s ability to reproduce the nonlinear relaxation behavior of the BGK operator, we test the trained SQC on a new synthetic dataset generated according to the procedure described in Section \ref{subsec:training}, using the optimal training configuration summarized in Section \ref{subsec:optimaltraining}. The evaluation spans the full velocity magnitude range $|\mathbf{u}|\in[0,0.1]$ usually encountered in LBM simulations. The velocity components\footnote{Varying $\theta$ changes the individual population profiles in Figure \ref{fig:nonlinearity}, yet the overall ability to capture the underlying nonlinearity remains unchanged.} are defined as $u_x = |\mathbf{u}| \cos(\theta)$ and $u_y = |\mathbf{u}| \sin(\theta)$, with $\theta = \pi/4$.

Each pre-collision population $f_i$ in the synthetic dataset is generated with a fixed density $\rho = 1$ and includes a small non-equilibrium perturbation drawn from a zero-mean Gaussian distribution with standard deviation $\sigma_{\text{neq}} = 5\times10^{-4}$. For each velocity magnitude in the range $|\mathbf{u}|\in[0,0.1]$, we compute three sets of post-collision populations: the nonlinear equilibrium populations $f_i^{\text{eq}}$ defined by Eq. \ref{eq:equilibrium}, the predicted populations $\hat f_i^{\text{eq}}$ obtained from the SQC, and a linearized reference $f_i^{\text{lin}}$ obtained by truncating the LBM equilibrium to first order in the velocity $\mathbf{u}$, thereby removing all quadratic nonlinear terms:

\begin{equation}
    f_i^{\text{lin}} = 
  w_{i}\,\rho
  \left[
  1
  + \frac{\mathbf{e}_{i}\cdot\mathbf{u}}{c_{s}^{2}}
  \right],  \quad \ i = 0, \ldots, 8.
  \label{eq:lin-equilibrium}
\end{equation}

Figure \ref{fig:nonlinearity} compares the three sets of post-collision populations over the velocity magnitude range $|\mathbf{u}|\in[0,0.1]$. Unlike the linearized equilibrium, both the nonlinear equilibrium and the SQC results show clear curvature, indicating a nonlinear dependence on $\mathbf{u}$. The SQC predictions follow the nonlinear equilibrium closely across the entire velocity range, accurately reproducing both the curvature and the relative scaling between populations. The different shapes of the population profiles come from how the discrete velocities in the D$_2$Q$_9$ lattice are oriented relative to the diagonal flow direction at $\theta = \pi/4$. 

Population $f_6$ and $f_8$ are associated with the discrete velocity vectors $\mathbf{e}_6=(-1,1)$ and $\mathbf{e}_8=(1,-1)$ which are perpendicular to the flow direction at $\theta = \pi/4$. This means that the linear and quadratic terms $(\mathbf{e}_i \cdot \mathbf{u})$ and $(\mathbf{e}_i \cdot \mathbf{u})^2 $ are both equal to zero in their respective equilibrium distributions. Consequently, their equilibrium distributions depend only on the purely quadratic term $(\mathbf{u}\cdot \mathbf{u})$. This explains why the curvature of their corresponding post-collision values appears more pronounced in the figure. A similar argument holds for the $f_0$ population, whose equilibrium distribution also depends only on the quadratic term $(\mathbf{u}\cdot \mathbf{u})$, since the other two terms vanish due to its discrete lattice velocity vector being $\mathbf{e}_0=(0,0)$. Populations $f_5$ and $f_7$ have discrete velocity vectors $\mathbf{e}_5=(1,1)$ and $\mathbf{e}_7=(-1,-1)$, which are aligned and anti-aligned, respectively, with the flow direction at $\theta = \pi/4$. Therefore, their equilibria include contributions from both the linear and quadratic terms, resulting in less curved profiles in the figure. Finally, for a flow direction at $\theta=\pi/4$ the axial populations have relatively large and symmetric projections on $\mathbf{u}$, such that their equilibria are dominated by the linear term $(\mathbf{e}_i \cdot \mathbf{u})$ and their post-collision populations have nearly linear profiles. 

Altogether, these results confirm that although the SQC itself implements a unitary and therefore linear evolution, the overall SQC framework can partially reproduce the nonlinear relaxation behavior of the BGK operator.

\subsection{Origin of Nonlinearity and Dissipation in the SQC Framework}
\label{subsec:origin-nonlinearity}

In this section we analyze in detail how the SQC framework can partially reproduce the nonlinear and dissipative relaxation behavior of the BGK operator. At first glance this behavior appears counterintuitive. The SQC performs a unitary and therefore linear transformation in Hilbert space, yet the resulting mapping between pre- and post-collision populations shows a partially nonlinear dependence, as demonstrated in the previous section. This apparent contradiction can be understood by noting that the nonlinearity does not arise from the unitary evolution itself, but from how the classical populations are connected to the quantum state through the encoding and measurement steps. The rooted-density encoding and the measurement in the computational basis are both nonlinear operations that link the classical and quantum descriptions. The encoding introduces a nonlinear relationship between the classical populations and the amplitudes of the quantum state, while the measurement converts the evolved amplitudes back into populations in physical space. This measurement step also breaks unitarity by collapsing the quantum state, introducing an effective relaxation process that mimics the dissipation of the BGK operator. Together, these elements preserve the linearity of the circuit evolution while giving rise to the effective nonlinearity and apparent dissipation observed in the measured outputs.

To make this precise, we start by examining the rooted-density encoding, which maps the classical populations in physical space into the amplitudes of the quantum state:

\begin{equation}
\label{eq:statevector}
\ket{\psi} = \frac{1}{\sqrt{\rho}}\sum_{i=0}^{15} \sqrt{f_i(\mathbf{x},t) }\ket{\textbf{e}_i}, \quad \rho = \sum_{i=0}^{15} f_i(\mathbf{x},t).
\end{equation}

\noindent
Here, only the first nine populations are nonzero, so the remaining seven amplitudes ($i=9, \ldots,15$) are initialized to zero. The amplitude associated with each velocity basis state is therefore $a_i = \sqrt{f_i(\mathbf{x},t)/\rho}$. These amplitudes depend nonlinearly on the classical populations through the square-root mapping $f_i(\mathbf{x},t) \mapsto \sqrt{f_i(\mathbf{x},t)}$. This nonlinearity is confined to the encoding stage, after which the quantum state evolves linearly under the SQC and produces the post-collision amplitudes:

\begin{equation}
\label{eq:aj-finalstate-evolution2}
a_j
= \sum_{i=0}^{15} U^{\text{SQC}}_{ji} \sqrt{\frac{f_i(\mathbf{x},t)}{\rho}},
 \quad \ j = 0, \ldots, 15.
\end{equation}

\noindent
As shown in Section \ref{subsec:measurement}, we can also express these amplitudes in the following form:

\begin{equation}
\label{eq:aj-finalstate2}
a_j = \sqrt{\frac{\hat f_j^{\text{eq}}(\mathbf{x},t)}{\rho}} e^{i\varphi_j},
\quad \ j = 0, \ldots, 15.
\end{equation}

Finally, the measurement in the computational basis defines a nonlinear mapping between the populations $\hat f_j^{\text{eq}}(\mathbf{x},t)$ present in the quantum state and their classical counterparts in physical space. According to the measurement procedure described in Section \ref{subsec:measurement}, Eq. \ref{eq:probability} and Eq. \ref{eq:probability2} together define:

\begin{equation}
\label{eq:fj}
\hat f_j^{\text{eq}}(\mathbf{x},t) = \rho |a_j|^2,  \quad \ j = 0, \ldots, 15,
\end{equation}

\noindent
which makes the nonlinearity explicit, since the measured populations are obtained from the squared magnitudes of the amplitudes. Substituting Eq. \ref{eq:aj-finalstate-evolution2} into Eq. \ref{eq:fj}, we obtain:

\begin{equation}
\begin{split}
\hat f_j^{\text{eq}}(\mathbf{x},t) 
&= \rho \left| 
\sum_{i=0}^{15} U^{\text{SQC}}_{ji}\sqrt{\frac{f_i(\mathbf{x},t)}{\rho}} 
\right|^2 \\
&= \rho 
\left( \sum_{i=0}^{15} U^{\text{SQC}}_{ji}\sqrt{\frac{f_i(\mathbf{x},t)}{\rho}} \right)
\left( \sum_{k=0}^{15} (U^{\text{SQC}}_{jk})^* \sqrt{\frac{f_k(\mathbf{x},t)}{\rho}} \right),  \quad \ j = 0, \ldots, 15.
\end{split}
\end{equation}

\noindent
Here, $(U^{\text{SQC}}_{jk})^*$ denotes the complex conjugate of the matrix element $U^{\text{SQC}}_{jk}$. We now separate the diagonal terms ($i=k$) from the off-diagonal terms ($i \neq k$):

\begin{equation}
\label{eq:fj-final-alt}
\begin{split}
\hat f_j^{\text{eq}}(\mathbf{x},t)
&= \sum_{i=0}^{15} |U^{\text{SQC}}_{ji}|^2 f_i(\mathbf{x},t) \\
&+ \sum_{i=0}^{15}\sum_{\substack{ k = 0\\ k\ne i}}^{15} 
U^{\text{SQC}}_{ji} (U^{\text{SQC}}_{jk})^*
\sqrt{f_i(\mathbf{x},t) f_k(\mathbf{x},t)}, 
 \quad \ j = 0, \ldots, 15.
\end{split}
\end{equation}

\noindent
Since $\hat f_j^{\text{eq}}(\mathbf{x},t)$ represent post-collision populations which are by definition real quantities, it is convenient to symmetrize the off-diagonal sum:

\begin{equation}
\label{eq:fj-final}
\begin{split}
\hat f_j^{\text{eq}}(\mathbf{x},t) 
&= \underbrace{\sum_{i=0}^{15} |U^{\text{SQC}}_{ji}|^2 f_i(\mathbf{x},t)}_{\text{linear}} \\
&+ \underbrace{2\sum_{k=1}^{15}\sum_{i=0}^{k-1}  
\Re\!\big(U^{\text{SQC}}_{ji} (U^{\text{SQC}}_{jk})^*\big)\sqrt{f_i(\mathbf{x},t) f_k(\mathbf{x},t)}}_{\text{nonlinear}},
  \quad \ j = 0, \ldots, 15.
\end{split}
\end{equation}

\noindent
Here, $\Re(\cdot)$ denotes the real part of a complex number. The final expression in Eq. \ref{eq:fj-final} contains two distinct contributions. The first is a \emph{linear} redistribution term, in which each pre-collision population $f_i(\mathbf{x},t)$ contributes linearly to the post-collision population $\hat f_j^{\text{eq}}(\mathbf{x},t)$, with weights given by the squared magnitudes of the unitary matrix elements. The second, \emph{nonlinear} term represents interference between populations $f_i(\mathbf{x},t)$ and $f_k(\mathbf{x},t)$, as it depends on the product $\sqrt{f_i(\mathbf{x},t) f_k(\mathbf{x},t)}$. This interference is controlled by the relative phase between the matrix elements $U^{\text{SQC}}_{ji}$ and $(U^{\text{SQC}}_{jk})^*$, which determines whether the contribution to $\hat f_j^{\text{eq}}(\mathbf{x},t)$ is constructive or destructive.

The structure of Eq. \ref{eq:fj-final} closely mirrors that of the nonlinear LBM equilibrium in Eq. \ref{eq:equilibrium}, which also combines linear and nonlinear terms. The difference lies in the mathematical form of the nonlinear coupling. In the LBM equilibrium, the quadratic dependence on velocity generates terms proportional to $f_i(\mathbf{x},t) f_k(\mathbf{x},t)$, whereas in the SQC-derived expression the coupling enters through $\sqrt{f_i(\mathbf{x},t) f_k(\mathbf{x},t)}$. This square-root dependence arises directly from the rooted-density encoding. This is why the SQC predictions shown in Figure \ref{fig:nonlinearity} reproduce the same qualitative nonlinear behavior as the LBM equilibrium, but not its exact functional form.

Furthermore, the measurement step acts as an effective dissipative process. Projecting the post-collision quantum amplitudes onto the classical populations in physical space is inherently non-unitary, as it collapses the superposed amplitudes into definite outcomes. This collapse leads to an apparent relaxation of the measured populations toward an equilibrium-like distribution as shown in Eq. \ref{eq:fj-final}. The effect is analogous to the dissipative action of the BGK collision operator, which irreversibly drives the system toward local equilibrium.

The SQC framework therefore shows that a strictly linear quantum evolution can give rise to both effective nonlinearity and dissipation in the measured classical populations. The nonlinear encoding and measurement mappings, together with phase-dependent interference between amplitudes, generate the observed nonlinear dependence in physical space, while the projection inherent to measurement introduces the effective dissipation associated with relaxation toward equilibrium. As a result, the SQC framework reproduces the essential features of the BGK collision process, even though the underlying circuit evolution remains strictly unitary.

\subsection{Multi-time Step Extension}

As shown in the previous section, the rooted-density encoding, together with the unitary evolution implemented by the SQC and the measurement in the computational basis, produces an effective nonlinearity and dissipation in physical space that partially reproduces the relaxation toward the LBM equilibrium. It is important to clarify what this means for a multi-time step QLBM simulation, in particular when we want to avoid measurement and re-initialization after every time step. If Eq. \ref{eq:aj-finalstate2} is substituted into Eq. \ref{eq:fj}, the following relation is obtained:

\begin{equation}
\label{eq:equivalence}
\hat f_j^{\text{eq}}(\mathbf{x},t) = \rho \left| \sqrt{\frac{\hat f_j^{\text{eq}}(\mathbf{x},t)}{\rho}} e^{i\varphi_j} \right|^2,  \quad \ j = 0, \ldots, 15,
\end{equation}

\noindent
where the left-hand side represents the measured post-collision populations in physical space, while the right-hand side represents the same quantities encoded in the quantum state. Eq. \ref{eq:equivalence} shows that these two quantities are identical. The post-collision populations in physical space are precisely those already contained in the quantum state before measurement. Measurement therefore simply retrieves the post-collision populations from the amplitudes of the quantum state. If we were to measure and immediately re-encode, we would recover the same quantum state as before measurement, up to the relative phases.

This equivalence explains why measurement may not be required after every time step. The nonlinear and dissipative relaxation that appears in physical space is not a separate process applied after each collision. It is the classical description of a unitary transformation that has already taken place in Hilbert space. The SQC evolves the encoded amplitudes coherently, and the apparent relaxation arises only when these amplitudes are interpreted as populations through the nonlinear mapping defined by the encoding and measurement. From this perspective, the same evolution can be viewed in two complementary ways. In Hilbert space, it is a strictly linear and unitary transformation of amplitudes. In physical space, it appears as a nonlinear and dissipative relaxation of populations toward equilibrium. The effective dissipation and nonlinearity are therefore properties of the mapping between the two descriptions and not of the underlying quantum evolution.

The subsequent streaming operation simply permutes the basis states and therefore leaves the amplitudes unchanged. It does not alter their magnitudes or relative phases, but only reassigns them across lattice sites according to their discrete velocities \cite{SchalkersMoller2024}. As a result, the overall unitary evolution of the circuit remains intact, and it is reasonable to hypothesize that multiple time steps could be carried out coherently, with measurement performed only at the end of the simulation. However, it must still be shown explicitly that the combined application of the SQC and streaming over several coherent time steps can reproduce the same form of nonlinear relaxation that the BGK operator produces when coupled with streaming in the classical setting. An additional consideration concerns the introduction of relative phases. As discussed in the previous section, these phases play an important role because they expand the range of interference patterns the SQC can represent and thereby tune the effective nonlinearity it can model.

If, at the next time step, the input state to the SQC already contains relative phases, the circuit’s predictions will change significantly. The SQC was trained only on real amplitudes, so additional phases in the inputs modify the interference patterns it has learned to reproduce. Extending the framework to a measurement-free, fully coherent evolution over multiple time steps would therefore require either a circuit capable of handling complex amplitudes or one that remains entirely real and prevents the introduction of phases. A purely real circuit, however, would lose the interference capability of the current SQC and would need to be designed to recover the missing degrees of expressivity. Designing such circuits lies outside the scope of the present study. What is demonstrated here is that by constructing a framework that reproduces nonlinearity and apparent dissipation in physical space while maintaining strictly unitary evolution in Hilbert space, it is possible to realize a low-depth circuit that provides an accurate approximation of the BGK operator.

\section{SQC Evaluation}
\label{sec:SQCevaluation}

In this section, we compare the collision predictions obtained using the SQC with those from the BGK operator for the Taylor-Green vortex decay case in Section \ref{sec:taylor_green} and the lid-driven cavity flow case in Section \ref{sec:lid_driven_cavity}. For simulations that use the SQC to predict collisions, the streaming and boundary conditions are handled classically, while the collision step at each grid point is performed by the SQC. Both flow cases are simulated at Reynolds numbers Re = 10 and Re = 50, using a characteristic velocity of $U=0.05$. This velocity is significantly higher than those encountered during training, allowing us to evaluate the SQC’s ability to extrapolate and accurately predict collisions when nonlinear terms in the LBM equilibrium distribution become dominant. All flow variables in this section are expressed in lattice units.

\subsection{Taylor-Green Vortex Decay}
\label{sec:taylor_green}

The Taylor-Green vortex decay case models the decay of perfectly symmetric counter-rotating vortices in a fully periodic domain. In LBM the streaming step simply propagates populations between neighboring lattice nodes, introducing no physical dissipation. Vortex decay is therefore driven entirely by the collision process, which relaxes populations toward the local equilibrium and, in so doing, injects exactly the viscous dissipation that gives rise to the Navier-Stokes viscous-stress tensor in the continuum limit. Without a properly formulated collision step, the flow remains effectively inviscid and the vortices never decay. For each Reynolds number (Re = 10 and 50), we run two simulations on a two-dimensional periodic square domain with sides of length $L$. Specifically, we use $L=34$ for Re = 10 and $L=168$ for Re = 50 and set $\Delta x= \Delta t = 1$. In one simulation, the collision step is carried out using the classical BGK operator, while in the other the SQC is used. Both simulations are initialized with a uniform density $\rho = 1$ and the velocity field: 

\begin{equation}
u_x(x,y) \;=\; u_0 \,\sin(k_x x)\,\cos(k_y y), 
\end{equation}

\begin{equation}
u_y(x,y) \;=\; -u_0 \,\cos(k_x x)\,\sin(k_y y),
\end{equation}

\noindent
where the initial velocity amplitude is $u_0 = 0.05$ and wave‐numbers are $k_x = 2 \pi / L,\; k_y = 2\pi/L$. Each simulation is run up to a non-dimensional time $t^* = 0.1$, defined as $t^* = tu_0/L$.

\begin{figure}[htp]
  \centering

  \begin{subfigure}[b]{0.48\textwidth}
    \centering
    \includegraphics[width=\textwidth]{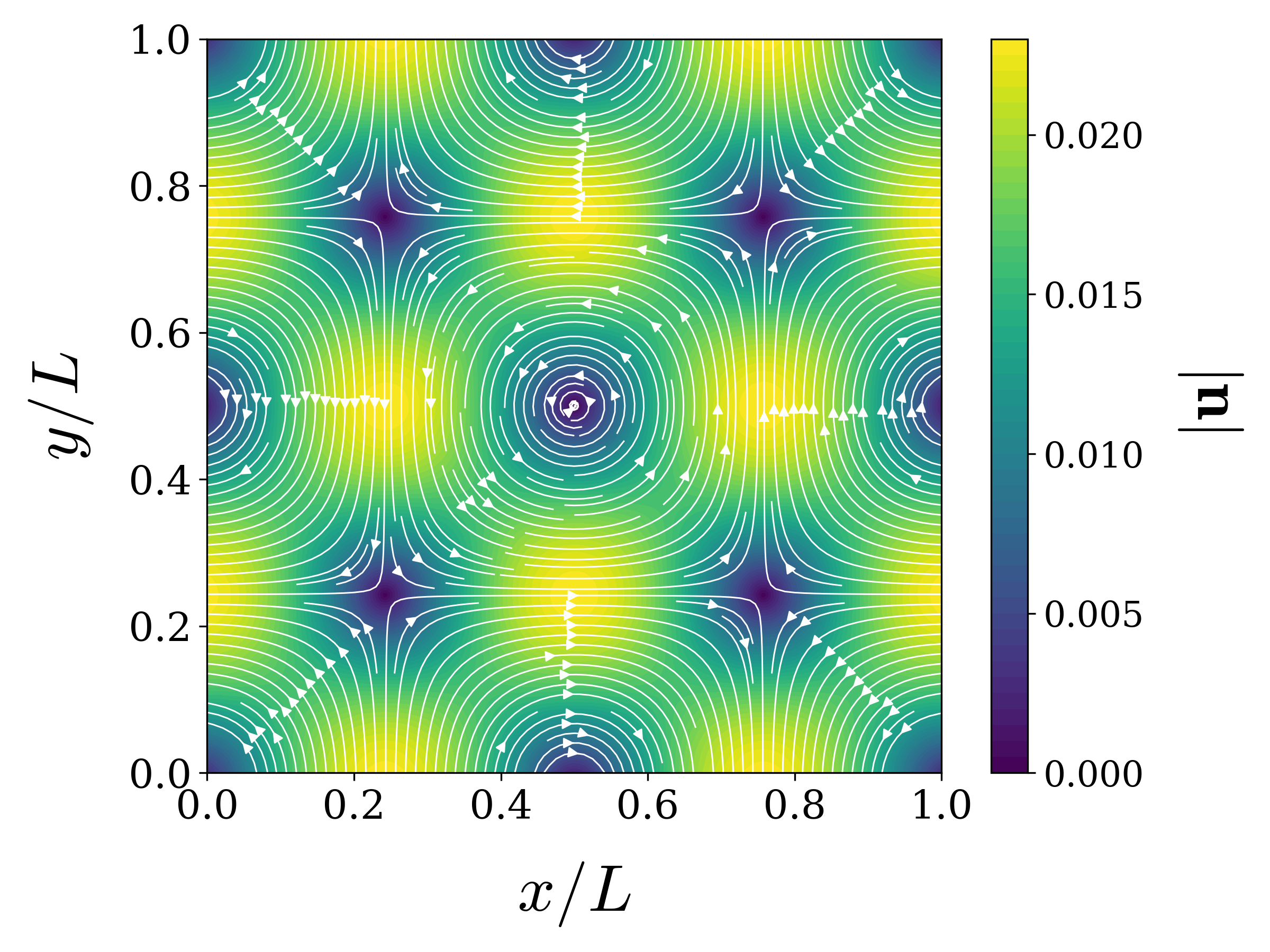}
    \caption{Velocity magnitude field $|\mathbf{u}|$ and streamlines for the SQC simulation at Re = 10.}
    \label{fig:Taylor-u-SQC-Re10}
  \end{subfigure}
  \hfill
    \begin{subfigure}[b]{0.48\textwidth}
    \centering
    \includegraphics[width=\textwidth]{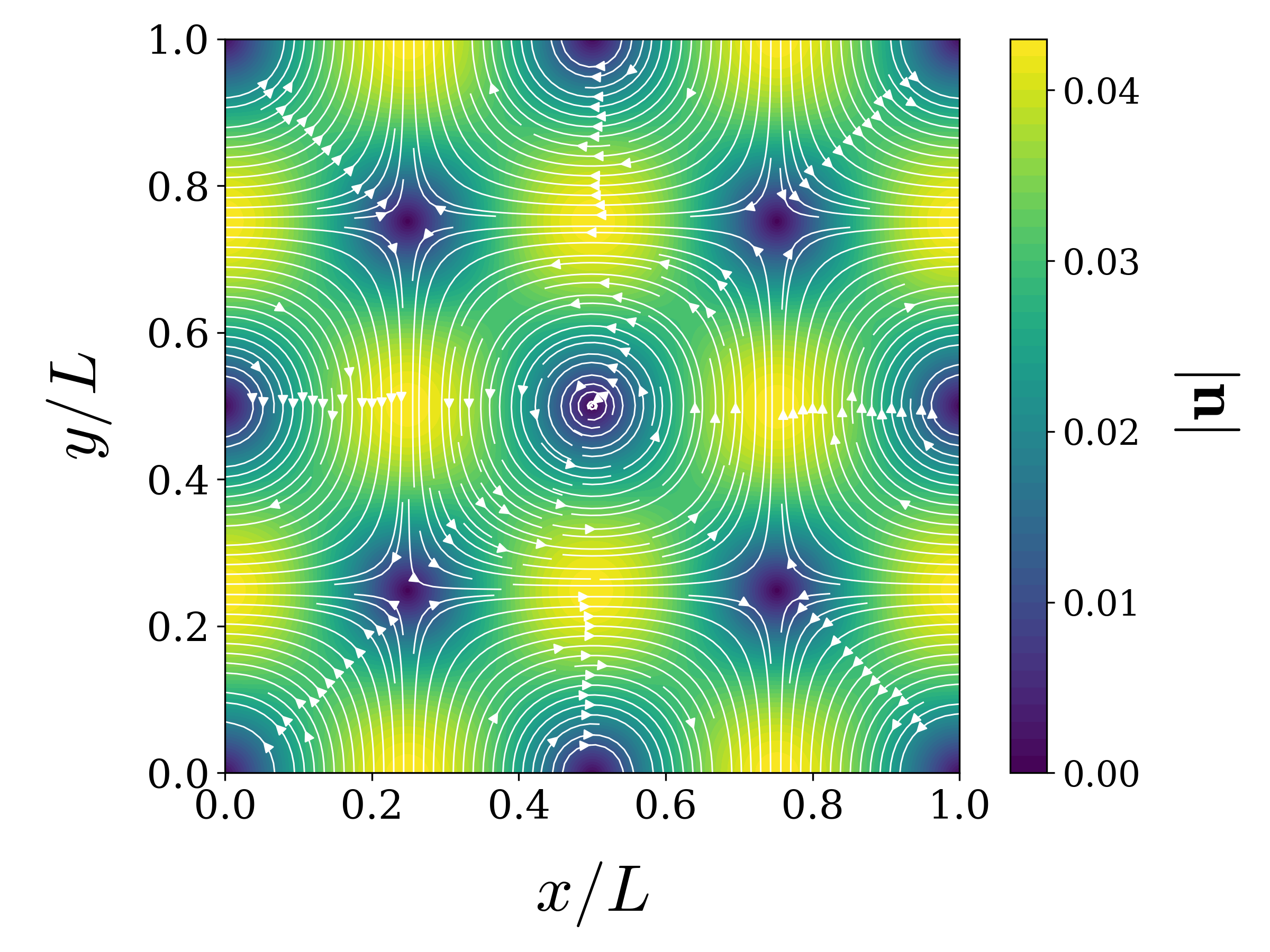}
    \caption{Velocity magnitude field $|\mathbf{u}|$ and streamlines for the SQC simulation at Re = 50.}
    \label{fig:Taylor-u-SQC-Re50}
    \end{subfigure}

  \vspace{1em} 

  \begin{subfigure}[b]{0.48\textwidth}
    \centering
    \includegraphics[width=\textwidth]{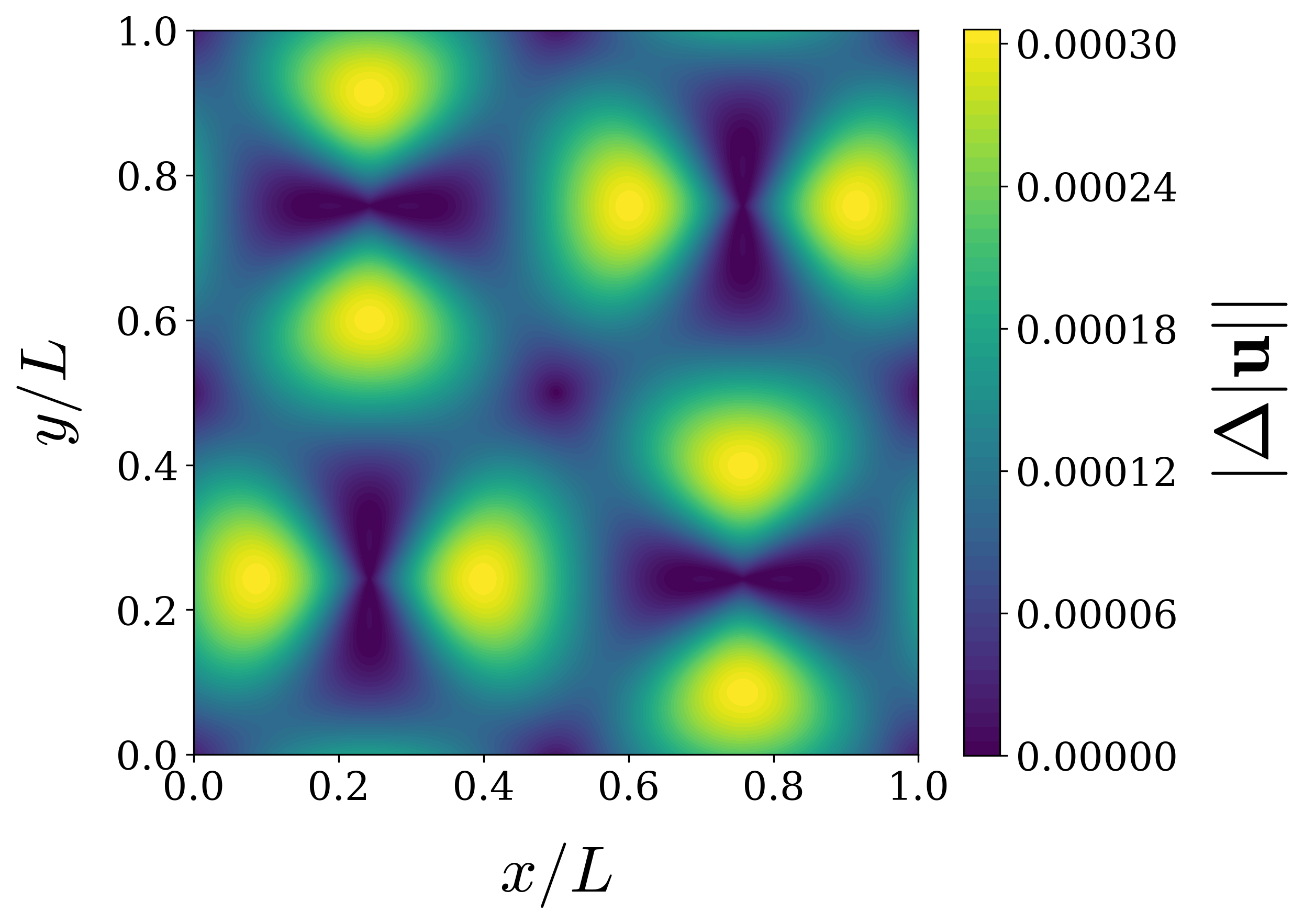}
    \caption{Absolute error in velocity magnitude between SQC and BGK simulations, $|\Delta|\mathbf{u}||$, at Re = 10.}
    \label{fig:Taylor-u-error-Re10}
  \end{subfigure}
    \hfill
  \begin{subfigure}[b]{0.48\textwidth}
    \centering
    \includegraphics[width=\textwidth]{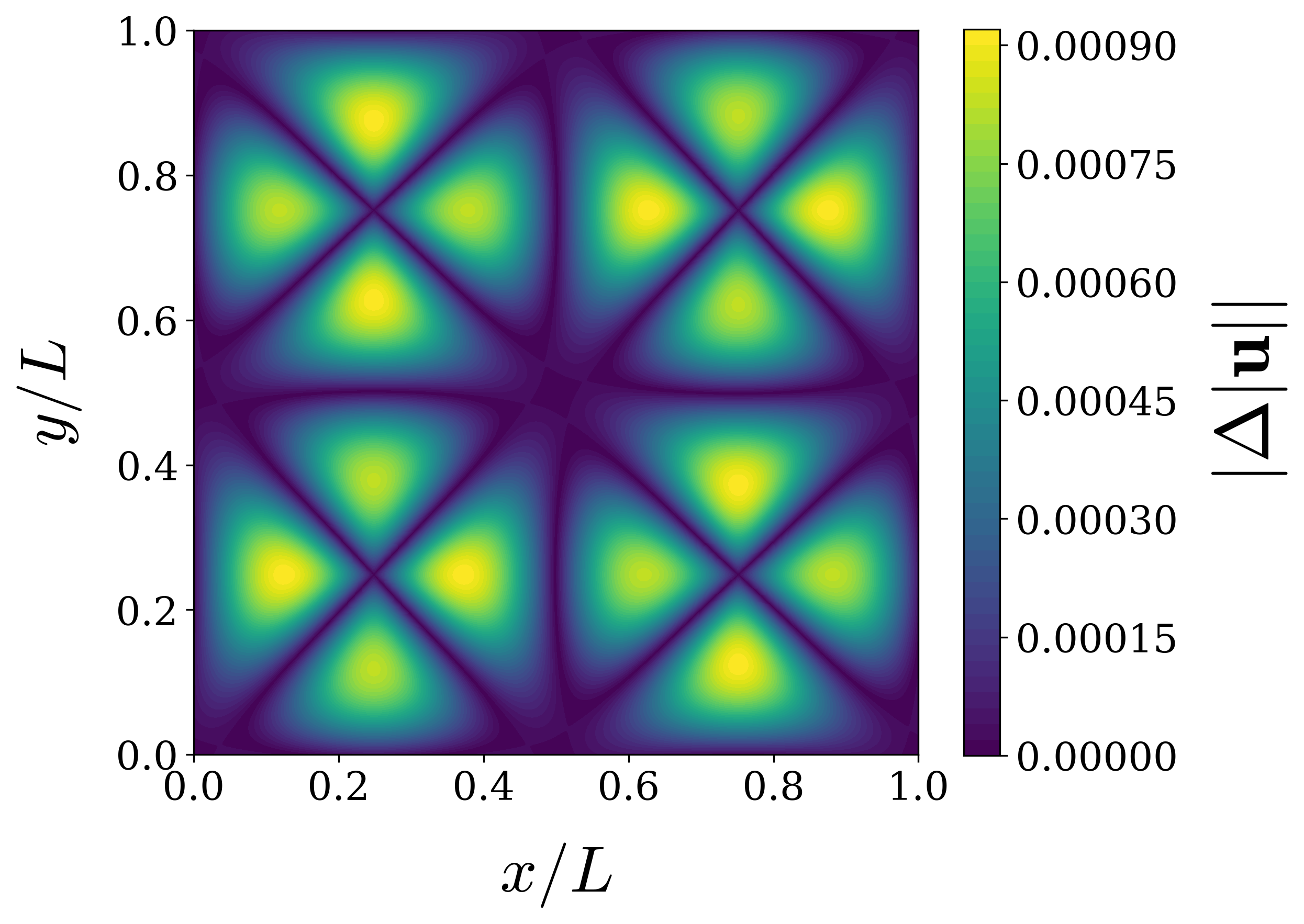}
    \caption{Absolute error in velocity magnitude between SQC and BGK simulations, $|\Delta|\mathbf{u}||$, at Re = 50.}
    \label{fig:Taylor-u-error-Re50}
  \end{subfigure}

  \caption{Comparison of velocity magnitude fields and streamlines obtained from the SQC simulation with the corresponding absolute error fields in velocity magnitude between the SQC and BGK simulations for the Taylor-Green vortex decay case at $t^* = 0.1$, for Re = 10 and Re = 50.}
  \label{fig:Taylor-u}
\end{figure}

\begin{figure}[htbp]
  \centering
  \begin{subfigure}[b]{0.83\textwidth}
      \centering
       \includegraphics[width=\textwidth]{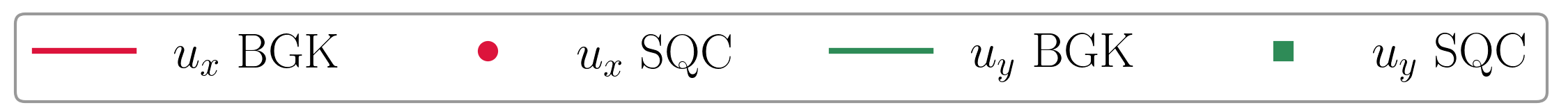}
  \end{subfigure}
  
  \begin{subfigure}[b]{0.48\textwidth}
    \centering
    \includegraphics[width=\textwidth]{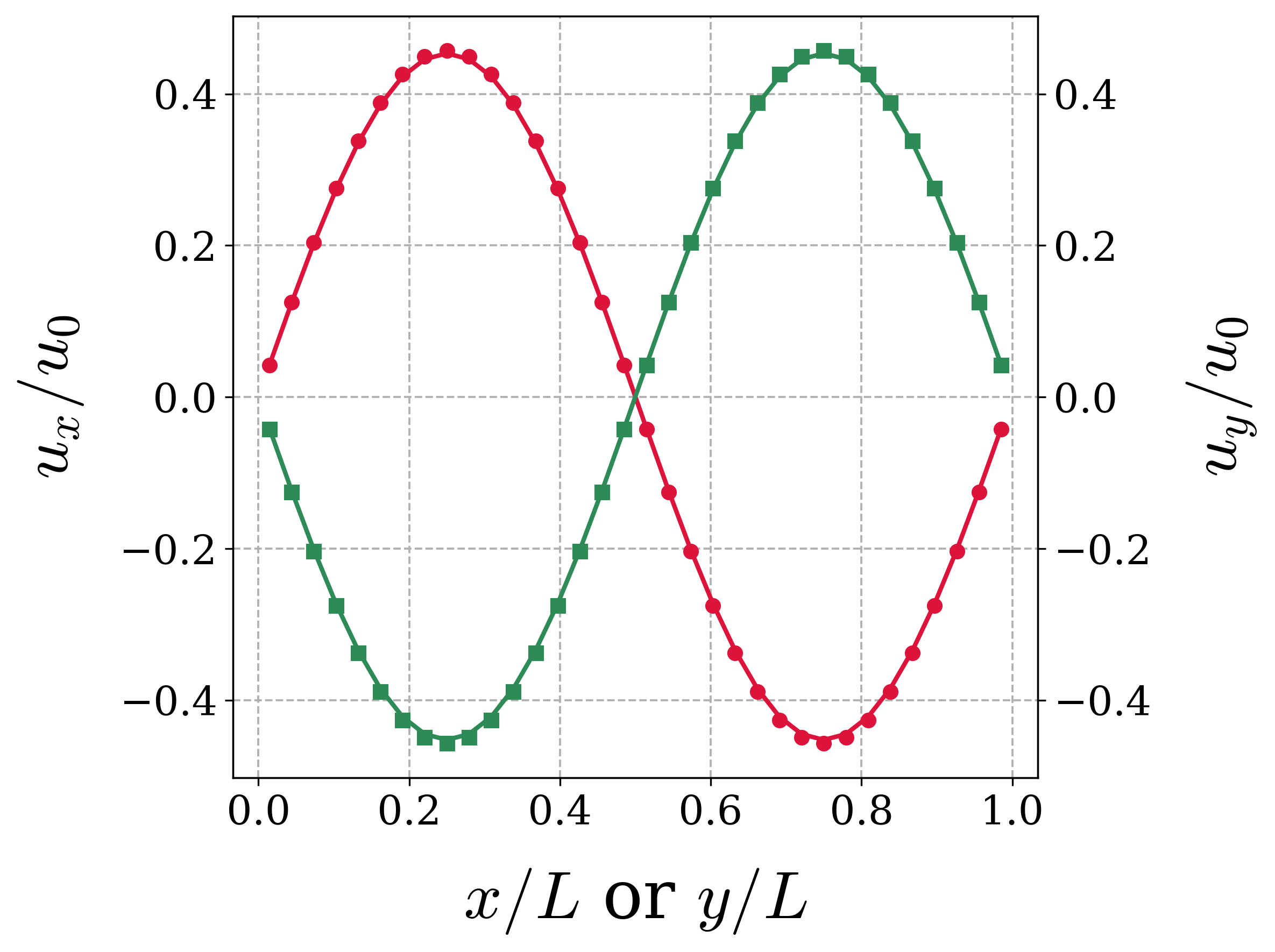}
    \caption{Profiles of the normalized horizontal velocity $u_x/u_0$ and normalized vertical velocity $u_y/u_0$ along the domain mid‐line at Re = 10.}
    \label{fig:Taylor-centerline-Re10}
  \end{subfigure}
  \hfill
  \begin{subfigure}[b]{0.48\textwidth}
    \centering
    \includegraphics[width=\textwidth]{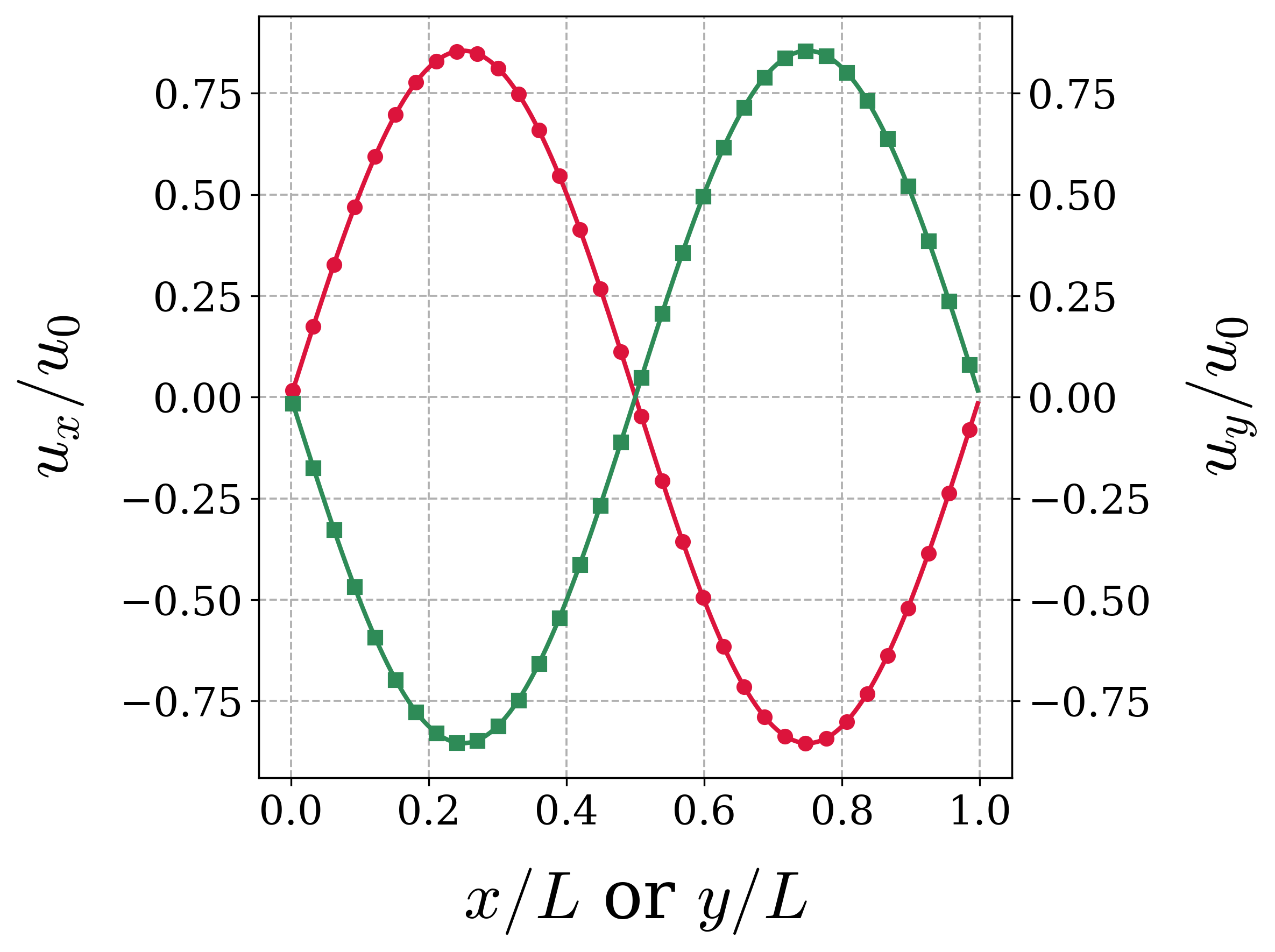}
    \caption{Profiles of the normalized horizontal velocity $u_x/u_0$ and normalized vertical velocity $u_y/u_0$ along the domain mid‐line at Re = 50.}
    \label{fig:Taylor-centerline-Re50}
  \end{subfigure}

  \caption{Comparison of centerline velocity profiles obtained at $t^*=0.1$ for the SQC and BGK simulations of the Taylor-Green vortex decay case at Re = 10 and Re = 50.}
  \label{fig:Taylor-centerline}
\end{figure}

The velocity magnitude fields obtained at $t^* = 0.1$ for the two SQC simulations at Re = 10 and Re = 50 are shown in Figures \ref{fig:Taylor-u-SQC-Re10} and \ref{fig:Taylor-u-SQC-Re50}, respectively. The SQC preserves the symmetry of the vortices throughout their decay process due to its D$_8$ equivariant design and adapts well to the different dissipation levels associated with the two Reynolds numbers. For the simulation at Re = 10, the maximum velocity amplitude decays to $u_{\text{max}}$ = 0.023, while at Re = 50 it remains higher at $u_{\text{max}}$ = 0.043. The slower decay at Re = 50 results from the reduced effective viscosity associated with higher Reynolds numbers. This shows that the SQC can adjust its relaxation strength across different Reynolds number regimes, which explains the excellent agreement between the centerline horizontal and vertical velocity profiles shown in Figure \ref{fig:Taylor-centerline}.

Figures \ref{fig:Taylor-u-error-Re10} and \ref{fig:Taylor-u-error-Re50} show the corresponding absolute error fields in velocity magnitude between the SQC and BGK simulations. The largest errors are concentrated near regions of strong velocity gradients, mainly along the separation lines between adjacent vortices (approximately at $x/L=0.25,0.75$ and $y/L=0.25,0.75$). Since the SQC can only partially reproduce the nonlinear terms of the LBM equilibrium, it cannot fully represent the higher-order moments that are needed to recover the viscous stresses in the Navier-Stokes equations. As a result, its predictions are less accurate in these regions of strong velocity gradients. Nevertheless, the absolute errors in the velocity magnitude fields remain small in both cases, with maximum values of $|\Delta|\mathbf{u}||_{\text{max}}=3.0 \times 10^{-4}$ for Re = 10 and $|\Delta|\mathbf{u}||_{\text{max}} =9.2 \times 10^{-4}$ for Re = 50. This is noteworthy given that the SQC is extrapolating in large parts of the domain due to the relatively high velocities at which the simulations are run. At Re = 10, although all vortices decay at the same rate, the absolute error field shows an alternating pattern between the four quadrants of the domain. This pattern arises because the vortices are counter-rotating, such that the direction of the local velocity gradients changes from one quadrant to the next, producing alternating regions of higher and lower error. At Re = 50, this alternating pattern is still present but less pronounced, as the overall error increases more uniformly across the domain. Although the SQC can adapt its damping strength across the two Reynolds numbers, the higher errors at Re = 50 indicate that this adjustment is not perfect.

\begin{figure}[htp]
  \centering
  
  \begin{subfigure}[b]{\textwidth}
    \centering
    \includegraphics[width=\textwidth]{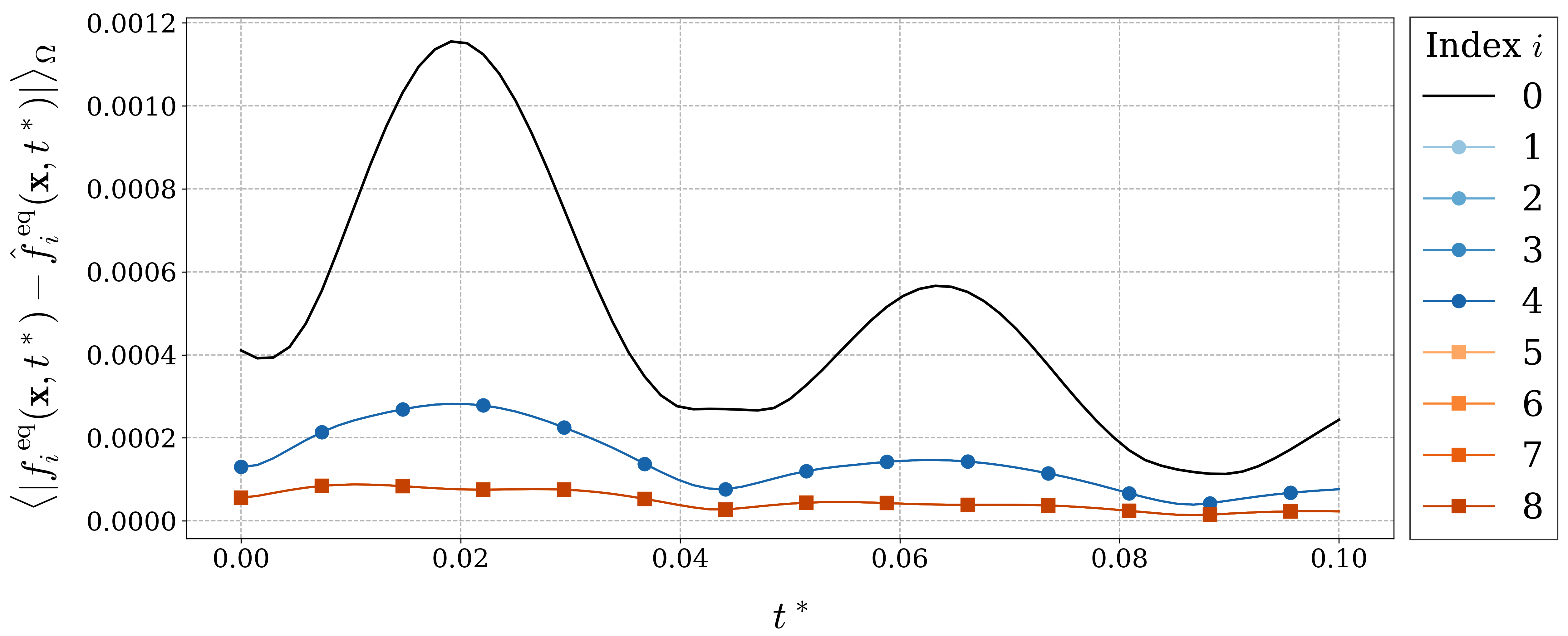}
    \caption{Time evolution of the absolute errors between the post-collision populations predicted by the SQC, $\hat f_i^{\text{eq}}(\mathbf{x},t^*)$, and those computed using the BGK operator, $f_i^{\text{eq}}(\mathbf{x},t^*)$, averaged over the entire domain $\Omega$, for the Re = 10 simulation.}
    \label{fig:Taylor-pop-Re-10}
  \end{subfigure}
  \hfill

  \vspace{1em}

  \begin{subfigure}[b]{\textwidth}
    \centering
    \includegraphics[width=\textwidth]{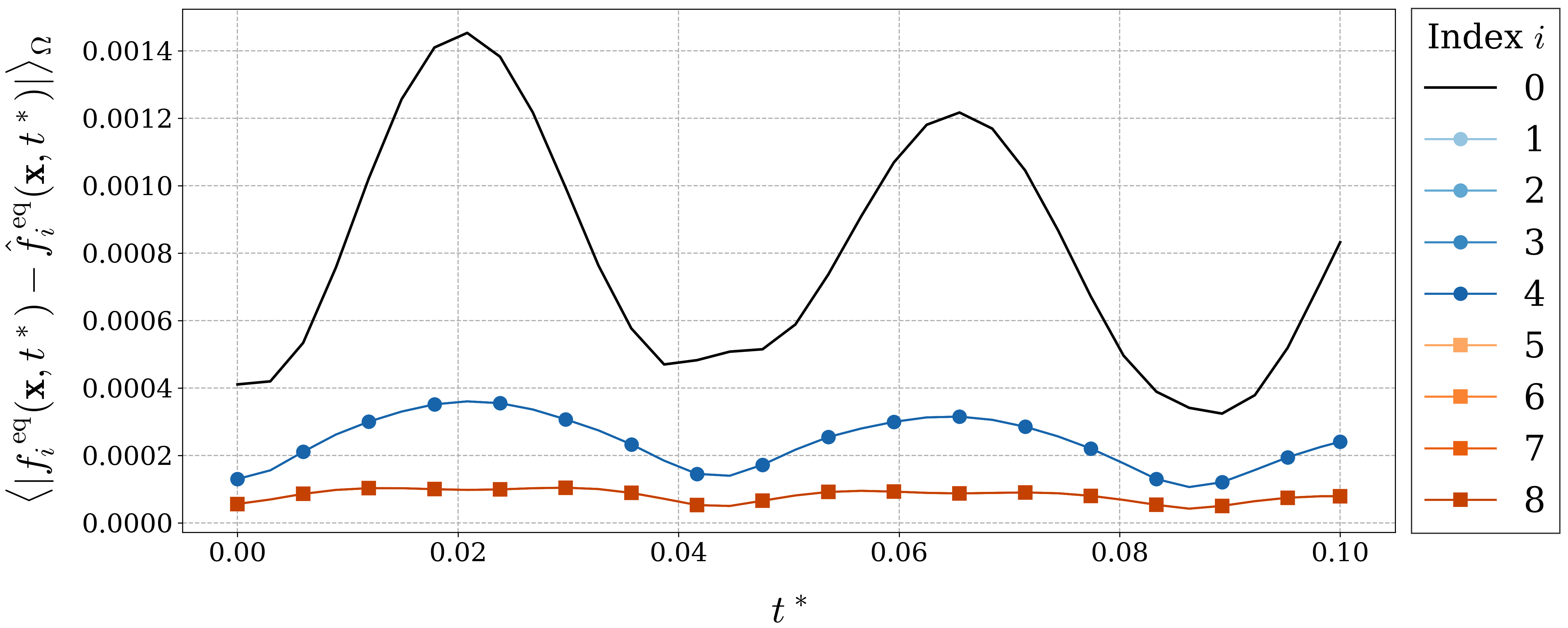}
    \caption{Time evolution of the absolute errors between the post-collision populations predicted by the SQC, $\hat f_i^{\text{eq}}(\mathbf{x},t^*)$, and those computed using the BGK operator, $f_i^{\text{eq}}(\mathbf{x},t^*)$, averaged over the entire domain $\Omega$, for the Re = 50 simulation.}
    \label{fig:Taylor-pop-Re-50}
  \end{subfigure}
  \hfill

  \caption{Time evolution of the domain-averaged absolute errors between the post-collision populations obtained during the SQC and BGK simulations of the Taylor-Green vortex decay case at Re = 10 and Re = 50.}
  \label{fig:Taylor-populations}
\end{figure}

To better understand the errors in the prediction of each post-collision population during the course of the simulation, Figure \ref{fig:Taylor-populations} shows the domain-averaged time evolution of the absolute error between the SQC-predicted post-collision populations, $\hat f_i^{\text{eq}}(\mathbf{x},t^*)$, and those computed using the BGK operator, $f_i^{\text{eq}}(\mathbf{x},t^*)$, for the simulations at Re = 10 and Re = 50. The magnitudes of the error curves are consistently higher for the Re = 50 case than for Re = 10, in agreement with the trends previously observed in the absolute error fields of the velocity magnitude. The magnitudes of the error curves follow a clear three-level hierarchy: the rest population shows the largest error, followed by the axial populations, and finally the diagonal populations. This ordering results from several factors, one of which is the relative magnitude of the equilibrium distribution that is used to compute each post-collision population. As given in Eq. \ref{eq:latticeweight}, in the D$_2$Q$_9$ lattice, the rest population equilibrium $f_0^{\text{eq}}$ has the largest lattice weight ($w_0 = 4/9$), which is four times that of the axial population equilibria ($w_{1-4} = 1/9$) and sixteen times that of the diagonal population equilibria ($w_{5-8} = 1/36$). As a result, similar fractional deviations between the SQC and BGK predictions appear as progressively smaller absolute errors when moving from the rest to the diagonal populations. Another contributing factor is the SQC circuit architecture itself, which is inherently less accurate for the rest population, as discussed in Section \ref{subsec:optimalarc}. 

Furthermore, the equilibrium distribution for the rest population depends only quadratically on the velocity through the term $(\mathbf{u}\cdot \mathbf{u})$ (see Eq. \ref{eq:f0equilibrium}), unlike the other equilibria, which also contain the term ($\mathbf{e}_i\cdot \mathbf{u}$) which is linear in the velocity. Because the SQC cannot fully reproduce this nonlinear dependence, the resulting discrepancy is strongest for this population. The slightly higher errors observed in the axial populations compared with the diagonals can be explained by the structure of the velocity field. Along the $x$ and $y$ directions, the velocity components $u_x$ and $u_y$ vary significantly, which produces the strongest velocity gradients along the lines separating adjacent vortices. The axial populations have discrete velocity vectors oriented along these same directions, so they are more sensitive to small mismatches in how accurately the SQC captures the local velocity gradients. Along the diagonal directions, the flow varies more smoothly, and the velocity components $u_x$ and $u_y$ often act in opposite directions due to the counter-rotating vortices. These opposing contributions partially cancel, leading to smaller overall errors in the diagonal populations.

Figure \ref{fig:Taylor-populations} also shows that, for both Re = 10 and Re = 50, the absolute error magnitudes for all populations decrease over time. This happens because, as the flow decays, the overall velocity in the domain decreases. Consequently, the nonlinear terms in the LBM equilibrium distribution become less significant compared with the linear term. Since the SQC cannot fully reproduce these nonlinear terms, the reduction in nonlinearity during the decay naturally leads to smaller prediction errors. The gradual rise and fall of the absolute error magnitudes reflects the evolution of the strain and vorticity fields during the vortex decay. As the flow evolves, kinetic energy is transiently redistributed between the velocity components, while the total energy decreases monotonically due to viscous dissipation. This redistribution produces small fluctuations in the local error amplitudes during the early stages, even though the overall decay of kinetic energy remains smooth and continuous. As a result, the apparent oscillations in the error curves are progressively damped, with the decay occurring more rapidly at Re = 10 than at Re = 50, as seen most clearly in the curves corresponding to the rest population. Furthermore, the errors for the axial and diagonal populations are identical within each group. This is a direct result of the D$_8$ equivariant design of the SQC and the symmetric structure of the vortices, which together ensure that populations related by rotations and reflections within the D$_8$ group experience identical local flow dynamics and therefore show identical errors.

\begin{figure}[htp]
  \centering
  
  \begin{subfigure}[b]{\textwidth}
    \centering
    \includegraphics[width=\textwidth]{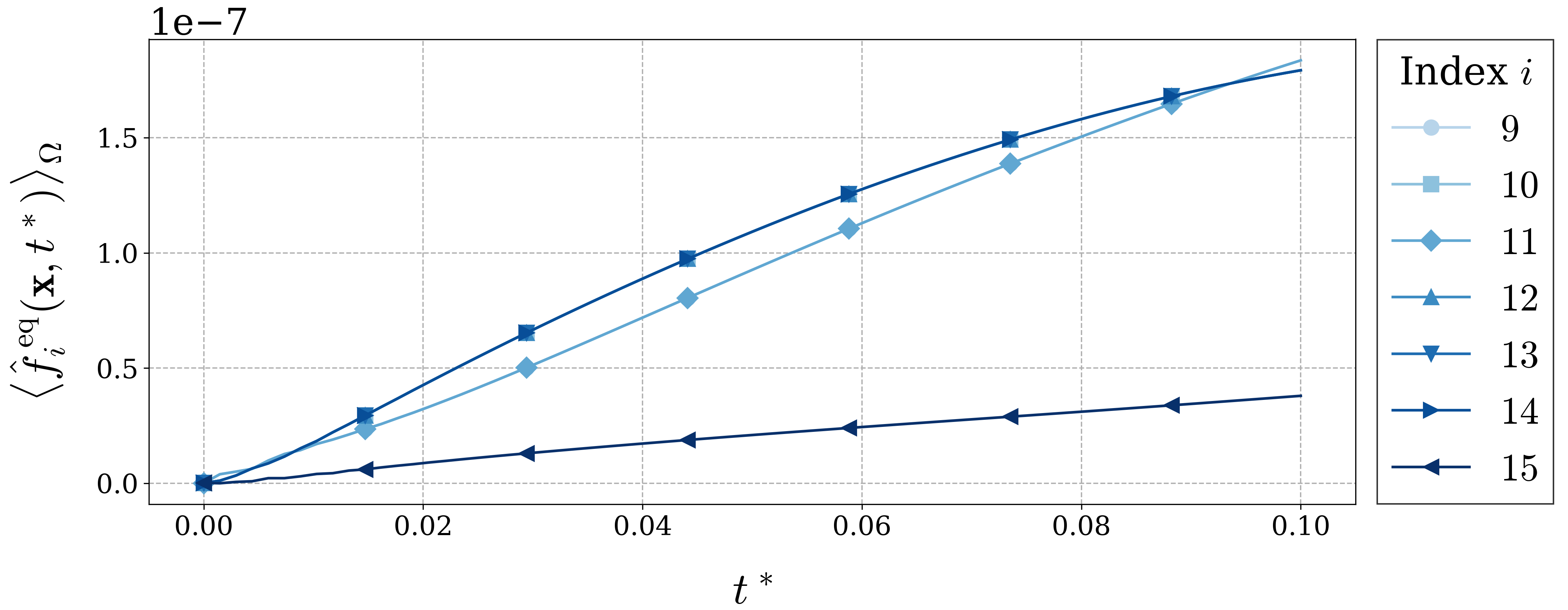}
    \caption{Time evolution of the additional post-collision populations $\hat f_{9-15}^{\text{eq}}(\mathbf{x},t^*)$, averaged over the entire domain $\Omega$, at Re = 10.}
    \label{fig:Taylor-unused-Re-10}
  \end{subfigure}
  \hfill

  \vspace{1em}

  \begin{subfigure}[b]{\textwidth}
    \centering
    \includegraphics[width=\textwidth]{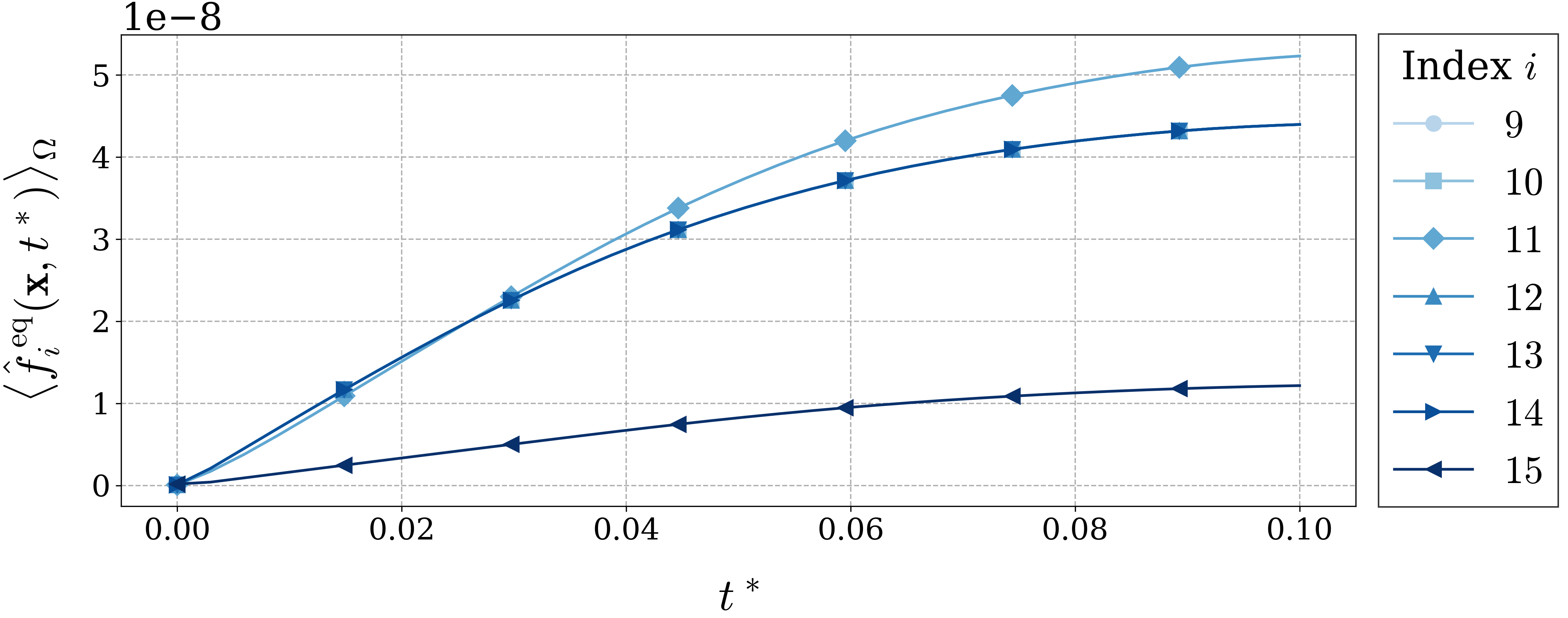}
    \caption{Time evolution of the additional post-collision populations $\hat f_{9-15}^{\text{eq}}(\mathbf{x},t^*)$, averaged over the entire domain $\Omega$, at Re = 50.}
    \label{fig:Taylor-unused-Re-50}
  \end{subfigure}
  \hfill

  \caption{Time evolution of the domain-averaged additional post-collision populations during the Taylor-Green vortex decay simulations at Re = 10 and Re = 50.}
  \label{fig:Taylor-unused}
\end{figure}

To quantify the mass transfer into the amplitudes associated with the seven additional basis states, Figure \ref{fig:Taylor-unused} shows the domain-averaged time evolution of the corresponding fictitious post-collision populations, $\hat f_{9-15}^{\text{eq}}(\mathbf{x},t^*)$, for the Re = 10 and Re = 50 simulations. As discussed in Section \ref{subsec:encoding}, these fictitious populations contribute to the evaluation of macroscopic quantities such as density and momentum in the same way as the $f_0$ population. The results show that the mass transferred into the amplitudes of these additional basis states is extremely small-on the order of $1 \times 10^{-7}$ for the Re = 10 case and $1 \times 10^{-8}$ for Re = 50. The reason for the smaller mass transfer observed at Re = 50 is not entirely clear. Several factors may contribute to this behavior, such as differences in how the SQC adjusts its predictions to conserve momentum, how it adapts to the effective dissipation level, or how it tries to reproduce the nonlinear terms in the equilibrium distribution.

Although the flow is transient the profiles level off over time, indicating that the mass transfer into the amplitudes of the additional basis states tends to plateau after an initial period. This indicates that the exchange of mass between the amplitudes of the basis states corresponding to the D$_2$Q$_9$ populations and those corresponding to the the additional basis states reaches a steady balance. At the start of the simulation, the additional basis states are initialized with zero amplitude. As the simulation progresses, mass is transferred both into and out of these amplitudes. This exchange continues until an approximate equilibrium is reached, where the inflow and outflow of mass balance each other, leading to the plateaus in mass accumulation.

\subsection{Lid-Driven Cavity}
\label{sec:lid_driven_cavity}

The lid‐driven cavity case simulates a fluid enclosed in a square cavity with its top lid moving at a constant speed, driving recirculation within the cavity. For each Reynolds number (Re = 10 and 50), we run two simulations on a two-dimensional square domain with sides of length $L$. Specifically, we use $L=34$ for Re = 10 and $L=168$ for Re = 50 and set $\Delta x= \Delta t = 1$. In one simulation, the collision step is carried out using the classical BGK operator, while in the other the SQC is used. Bounce‐back boundary conditions are applied to the stationary walls, while the moving lid is modeled with a modified bounce‐back condition imposing $u_{\mathrm{lid}}=0.05$. The flow is initialized with a uniform density $\rho = 1$ and advanced in time until convergence. Convergence is monitored using the relative $L_2$ norm of the change in the velocity field between successive iterations. The solution is declared converged once this relative $L_2$ difference remains below a prescribed tolerance of $1 \times 10^{-8}$ for several consecutive checks.

\begin{figure}[htp]
  \centering
  
  \begin{subfigure}[b]{0.48\textwidth}
    \centering
    \includegraphics[width=\textwidth]{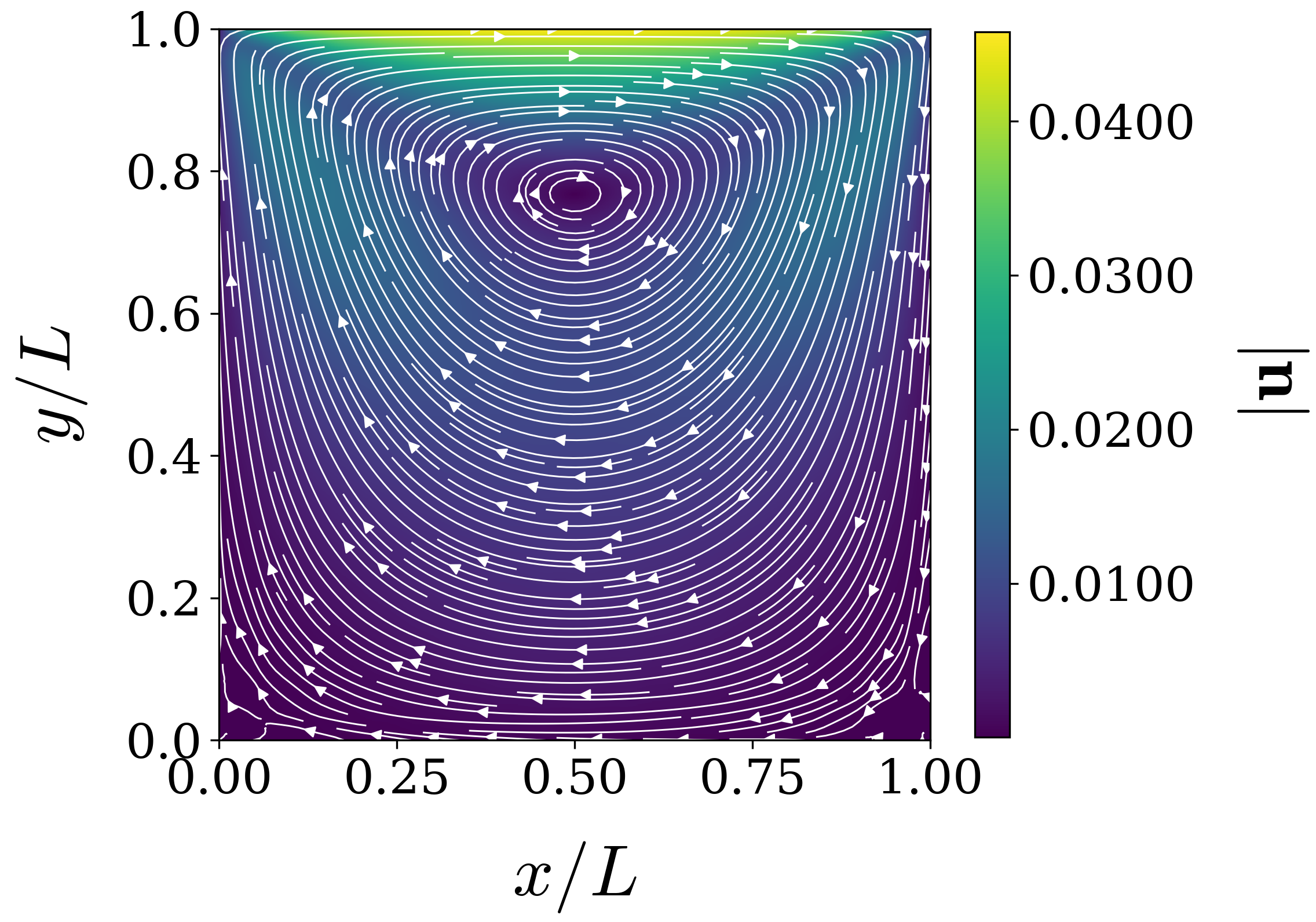}
    \caption{Velocity magnitude field $|\mathbf{u}|$ and streamlines for the SQC simulation at Re = 10.}
    \label{fig:Lid-u-SQC-Re10}
  \end{subfigure}
  \hfill
    \begin{subfigure}[b]{0.48\textwidth}
    \centering
    \includegraphics[width=\textwidth]{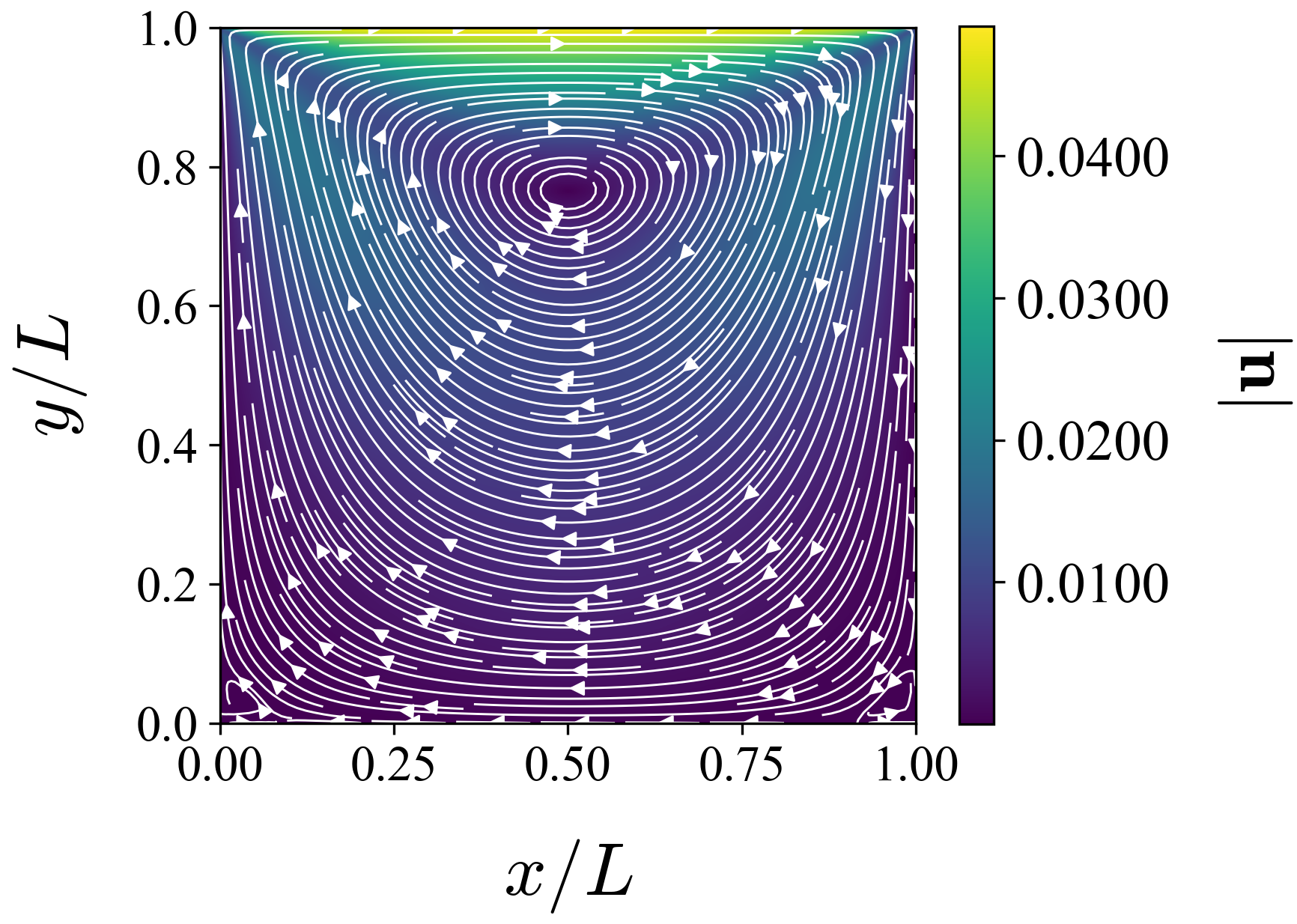}
    \caption{Velocity magnitude field $|\mathbf{u}|$ and streamlines for the SQC simulation at Re = 50.}
    \label{fig:Lid-u-SQC-Re50}
  \end{subfigure}

  \vspace{1em} 

  \begin{subfigure}[b]{0.48\textwidth}
    \centering
    \includegraphics[width=\textwidth]{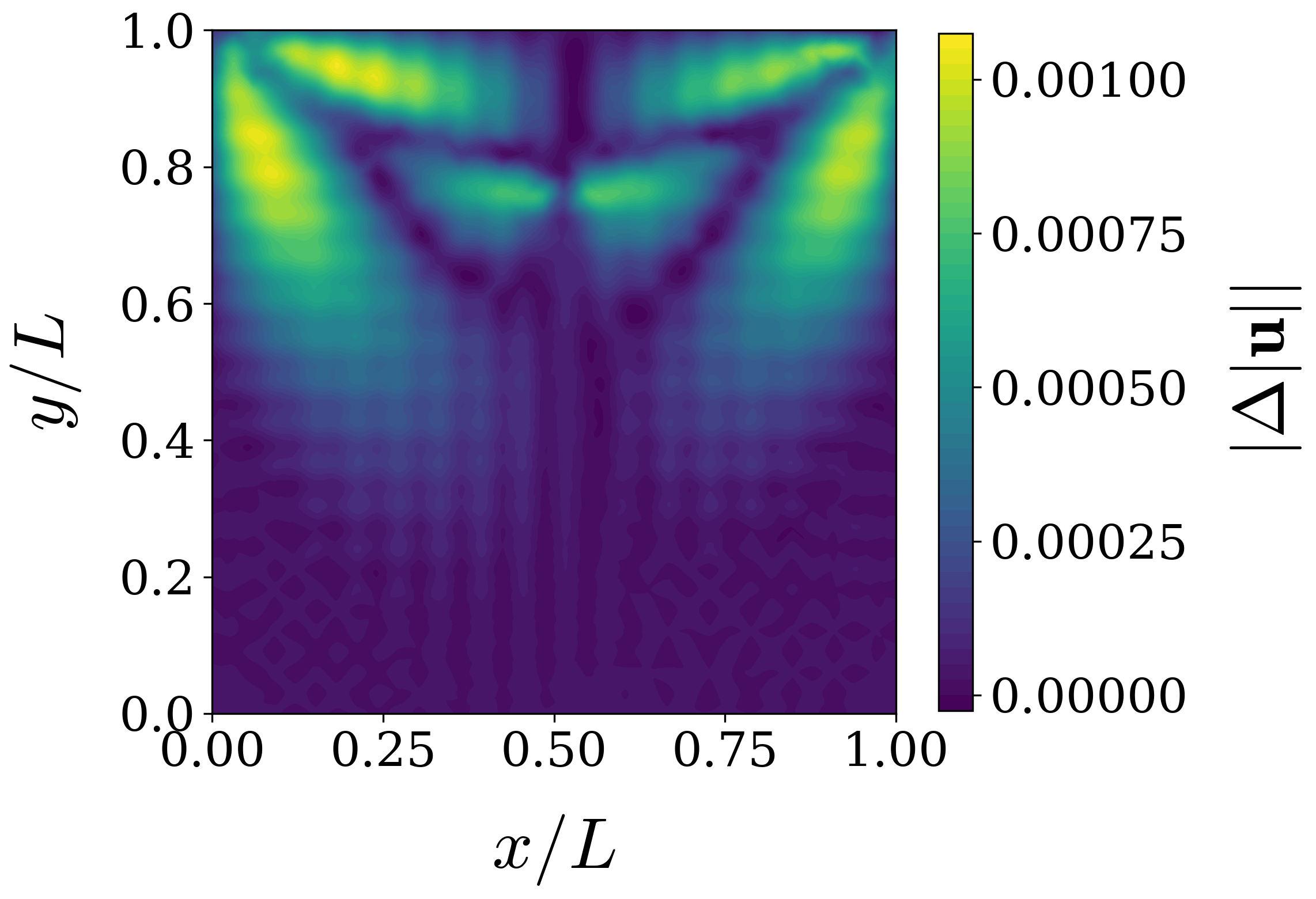}
    \caption{Absolute error in velocity magnitude between SQC and BGK simulations, $|\Delta|\mathbf{u}||$, at Re = 10.}
    \label{fig:Lid-u-error-Re10}
  \end{subfigure}
  \hfill
  \begin{subfigure}[b]{0.48\textwidth}
    \centering
    \includegraphics[width=\textwidth]{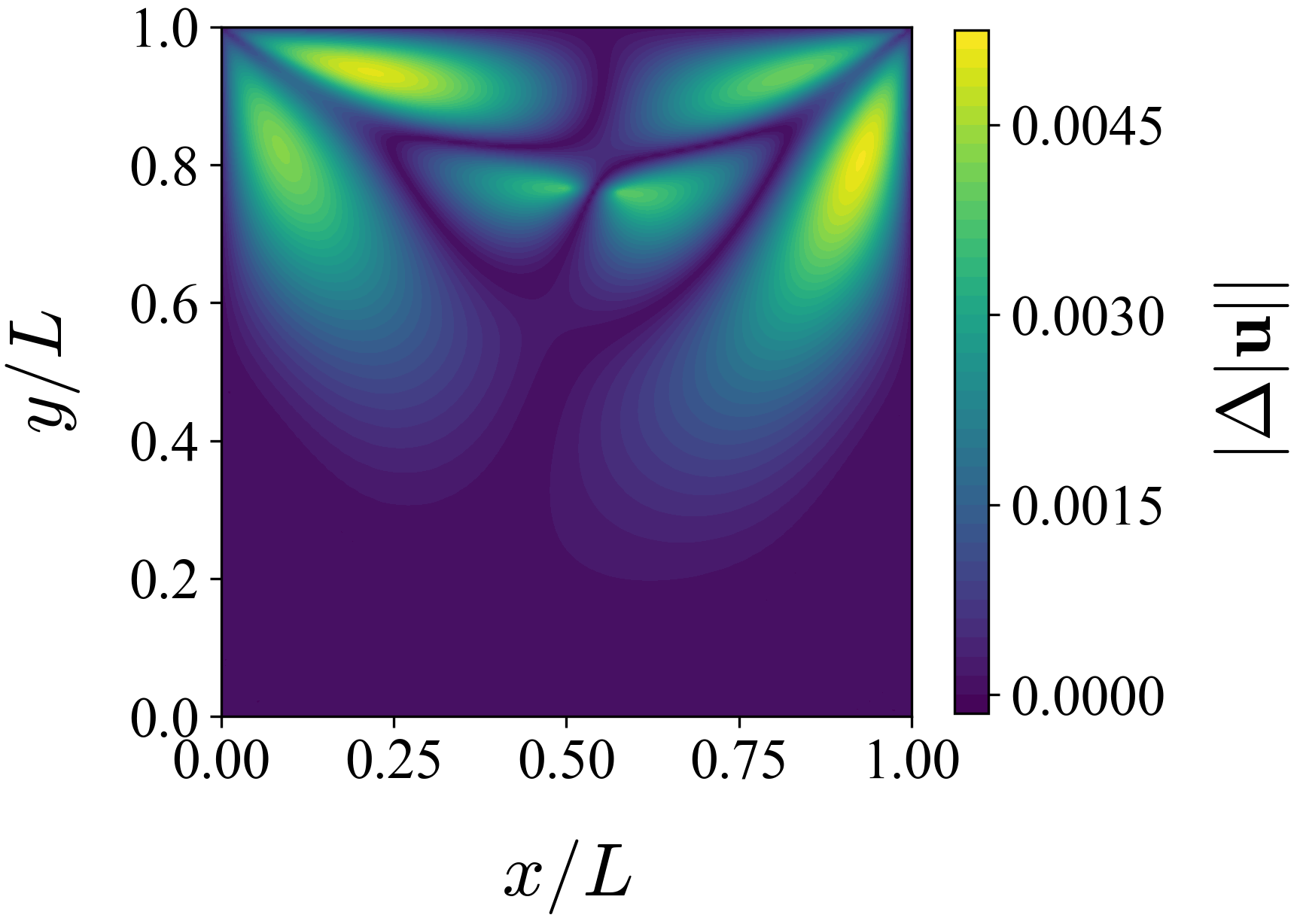}
    \caption{Absolute error in velocity magnitude between SQC and BGK simulations, $|\Delta|\mathbf{u}||$, at Re = 50.}
    \label{fig:Lid-u-error-Re50}
  \end{subfigure}

  \caption{Comparison of velocity magnitude fields and streamlines obtained from the SQC simulation with the corresponding absolute error fields in velocity magnitude between the SQC and BGK simulations for the lid-driven cavity case, obtained after convergence for Re = 10 and Re = 50.}
  \label{fig:Lid-u}
\end{figure}

The velocity magnitude fields obtained after convergence of the simulations for Re = 10 and Re = 50 are shown in Figures \ref{fig:Lid-u-SQC-Re10} and \ref{fig:Lid-u-SQC-Re50}, respectively. In both cases, the SQC successfully captures the primary recirculating flow within the cavity and the formation of secondary vortices near the bottom corners of the domain. A comparison of the absolute error in the velocity magnitude fields between the SQC and BGK simulations, shown in Figures \ref{fig:Lid-u-error-Re10} and \ref{fig:Lid-u-error-Re50}, reveals that the overall errors increase significantly at Re = 50. The maximum absolute errors in the domain are $|\Delta|\mathbf{u}||_{\text{max}} = 1.1 \times 10^{-3}$ for Re = 10  and  $|\Delta|\mathbf{u}||_{\text{max}} = 5.1 \times 10^{-3}$ for Re = 50. The absolute error distributions indicate that the largest discrepancies occur directly below the moving lid, where the fluid velocity decreases sharply from $u_{\text{lid}}$ to nearly zero due to the no-slip boundary condition. Elevated errors are also observed near the top corners, where the moving lid meets the stationary sidewalls and the flow turns abruptly, generating strong shear and steep velocity gradients. Along the vertical walls, additional errors arise within the boundary layers, which introduce further variations in the velocity field. In contrast, smaller errors are found near the bottom corners, where weaker recirculating vortices produce milder velocity gradients.

\begin{figure}[htbp]
  \centering
  \begin{subfigure}[b]{0.85\textwidth}
      \centering
       \includegraphics[width=\textwidth]{Taylor-Legend.png}
  \end{subfigure}
  
  \begin{subfigure}[b]{0.48\textwidth}
    \centering
    \includegraphics[width=\textwidth]{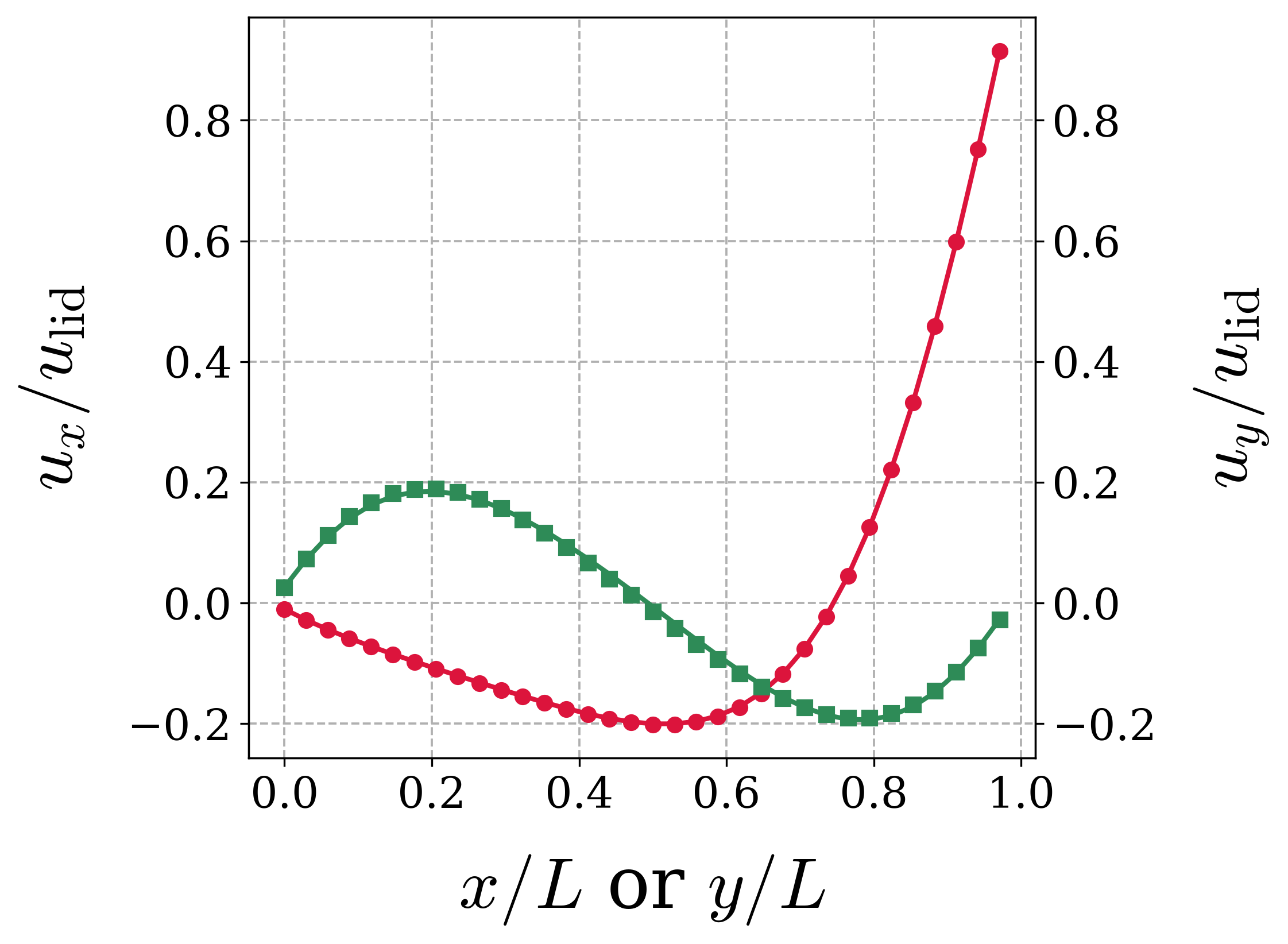}
    \caption{Profiles of the normalized horizontal velocity $u_x/u_{\text{lid}}$ and normalized vertical velocity $u_y/u_{\text{lid}}$ along the domain mid‐line at Re = 10.}
    \label{fig:Lid-centerline-Re10}
  \end{subfigure}
  \hfill
  \begin{subfigure}[b]{0.48\textwidth}
    \centering
    \includegraphics[width=\textwidth]{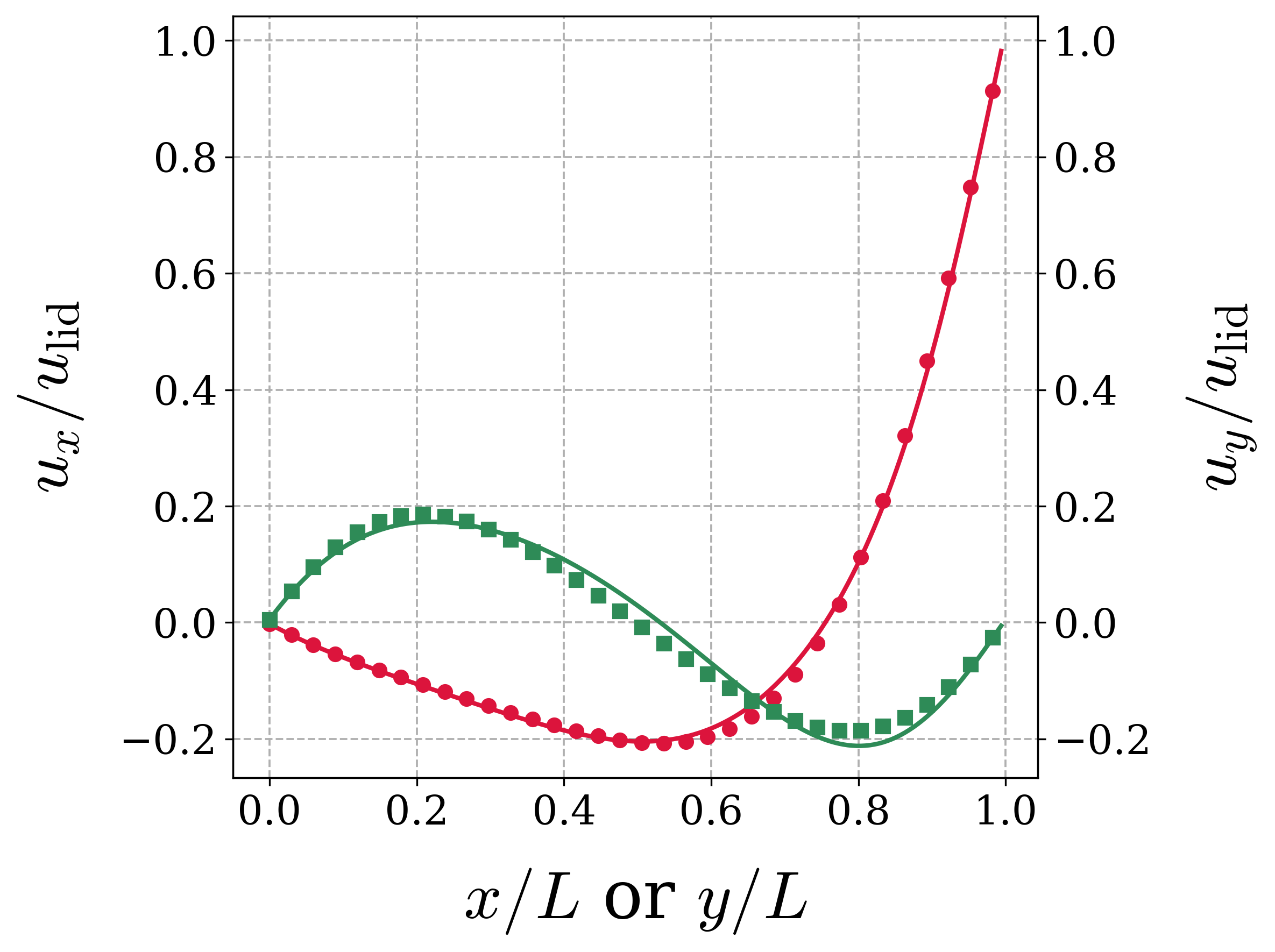}
    \caption{Profiles of the normalized horizontal velocity $u_x/u_{\text{lid}}$ and normalized vertical velocity $u_y/u_{\text{lid}}$ along the domain mid‐line at Re = 50.}
    \label{fig:Lid-centerline-Re50}
  \end{subfigure}

  \caption{Comparison of centerline velocity profiles obtained after convergence for the SQC and BGK simulations of the lid-driven cavity case at Re = 10 and Re = 50.}
  \label{fig:Lid-centerline}
\end{figure}

When the Reynolds number increases, the velocity gradients near the moving lid and top corners steepen, and the boundary layers that develop along the walls become thinner. Flow separation at the corners intensifies, resulting in sharper shear layers and stronger vorticity. Therefore, the errors for Re = 50 are highest along the right sidewall and in the upper region of the cavity near the moving lid. As discussed previously for the Taylor-Green case, the SQC cannot fully capture the nonlinear terms in the LBM equilibrium and therefore cannot reproduce the complete Navier-Stokes stress tensor. This limitation leads to inaccuracies in resolving the velocity gradients that develop in the flow. As the Reynolds number increases and the influence of these gradients becomes more significant, the discrepancy between the SQC and BGK results correspondingly grows, as seen in Figure \ref{fig:Lid-u-error-Re50}. This behavior also explains why the mismatch between the BGK and SQC centerline velocity profiles in Figure \ref{fig:Lid-centerline} is larger for the Re = 50 case than for Re = 10.

\begin{figure}[htp]
  \centering
  
  \begin{subfigure}[b]{\textwidth}
    \centering
    \includegraphics[width=\textwidth]{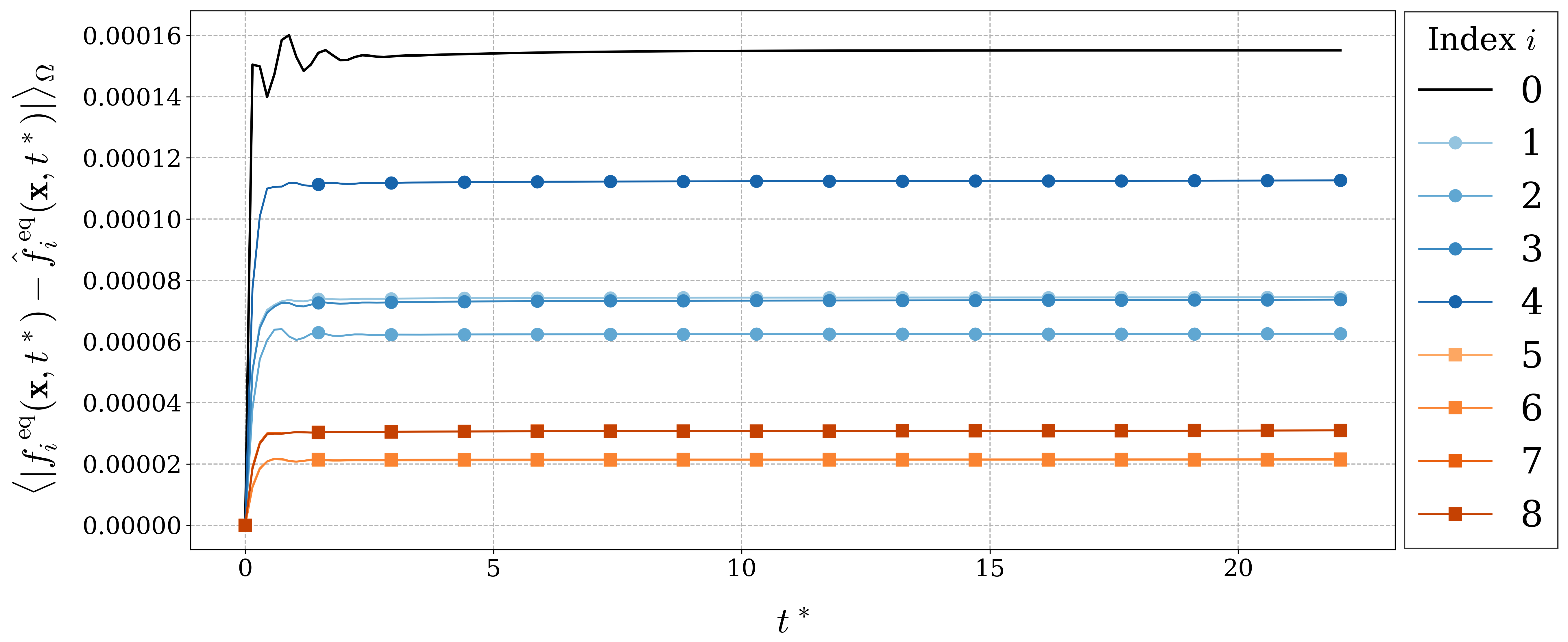}
    \caption{Time evolution of the absolute errors between the post-collision populations predicted by the SQC, $\hat f_i^{\text{eq}}(\mathbf{x},t^*)$, and those computed using the BGK operator, $f_i^{\text{eq}}(\mathbf{x},t^*)$, averaged over the entire domain $\Omega$, for Re = 10.}
    \label{fig:Lid-pop-Re-10}
  \end{subfigure}
  \hfill

  \vspace{1em}

  \begin{subfigure}[b]{\textwidth}
    \centering
    \includegraphics[width=\textwidth]{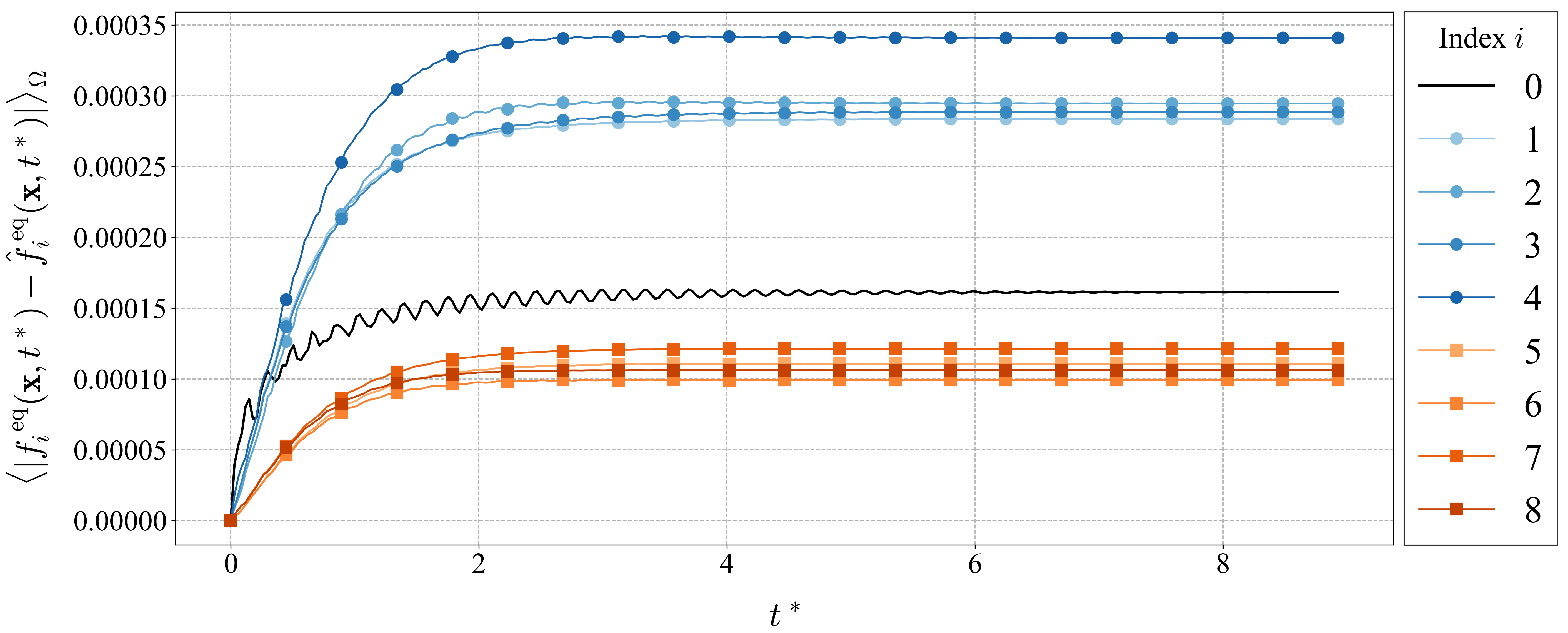}
    \caption{Time evolution of the absolute errors between the post-collision populations predicted by the SQC, $\hat f_i^{\text{eq}}(\mathbf{x},t^*)$, and those computed using the BGK operator, $f_i^{\text{eq}}(\mathbf{x},t^*)$, averaged over the entire domain $\Omega$, for Re = 50.}
    \label{fig:Lid-pop-Re-50}
  \end{subfigure}
  \hfill

  \caption{Time evolution of the domain-averaged absolute errors between the post-collision populations obtained during the SQC and BGK simulations of the lid-driven cavity case at Re = 10 and Re = 50.}
  \label{fig:Lid-populations}
\end{figure}

Similar to the Taylor-Green case, we track the time evolution of the prediction errors for each post-collision population over the dimensionless time $t^*=t u_{\text{lid}}/L$. Figure \ref{fig:Lid-populations} presents the domain-averaged absolute error between the SQC-predicted post-collision populations, $\hat f_i^{\text{eq}}(\mathbf{x},t^*)$ and those computed using the BGK operator, $f_i^{\text{eq}}(\mathbf{x},t^*)$, for the simulations at Re = 10 and Re = 50. In both cases, the errors increase sharply during the initial stage of the simulation as the lid begins to drive the flow and strong velocity gradients develop, before leveling off once the recirculating flow becomes fully developed. This behavior is evident in the rapid rise of the error curves during $t^*<1$, followed by a clear plateau once the flow reaches steady state.

\begin{figure}[htp]
  \centering
  
  \begin{subfigure}[b]{\textwidth}
    \centering
    \includegraphics[width=\textwidth]{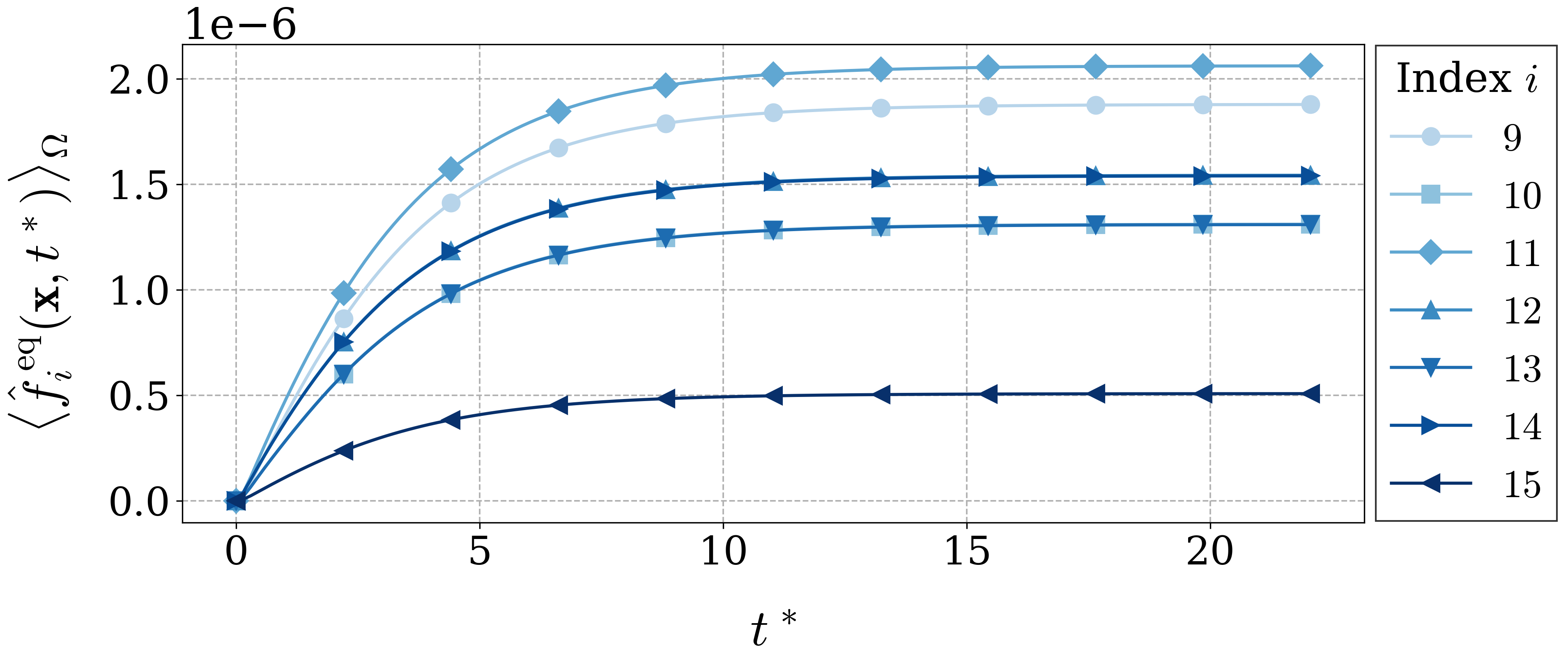}
    \caption{Time evolution of the additional post-collision populations $\hat f_{9-15}^{\text{eq}}(\mathbf{x},t^*)$, averaged over the entire domain $\Omega$, at Re = 10.}
    \label{fig:Lid-unused-Re-10}
  \end{subfigure}
  \hfill

  \vspace{1em}

  \begin{subfigure}[b]{\textwidth}
    \centering
    \includegraphics[width=\textwidth]{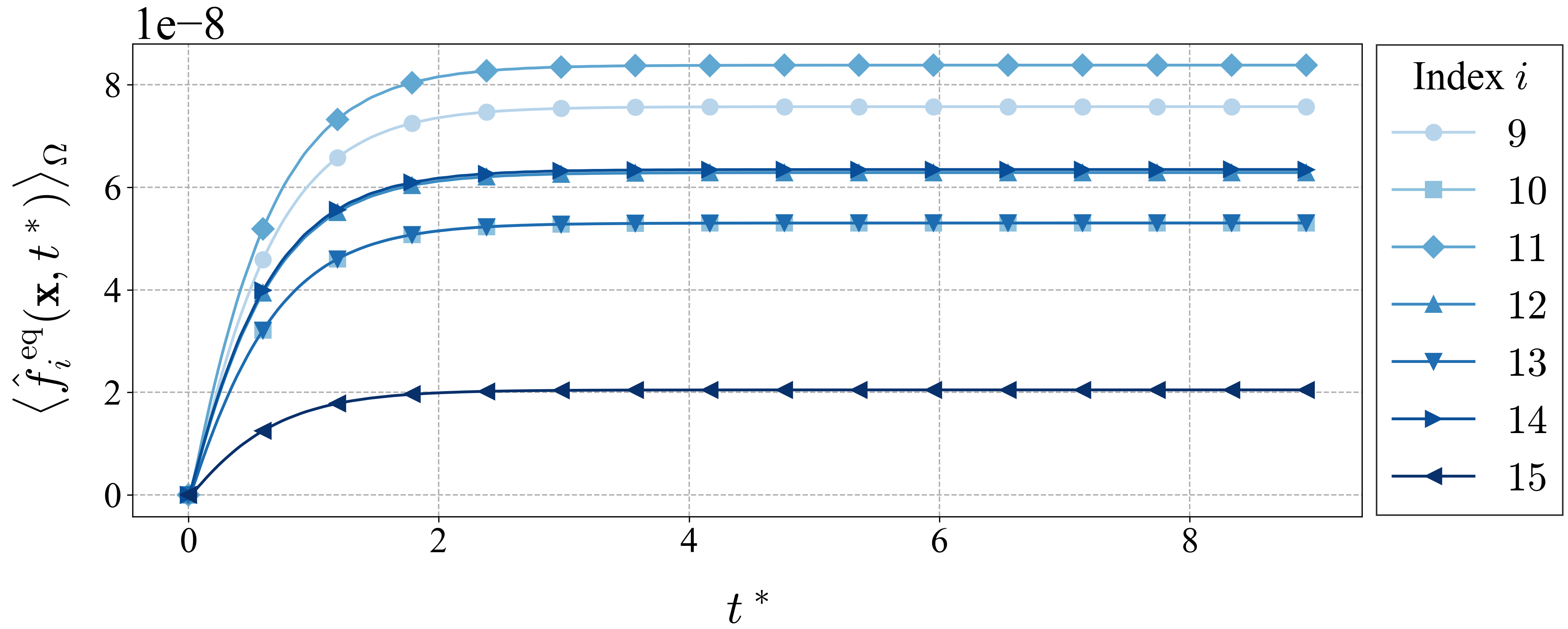}
    \caption{Time evolution of the additional post-collision populations $\hat f_{9-15}^{\text{eq}}(\mathbf{x},t^*)$, averaged over the entire domain $\Omega$, at Re = 50.}
    \label{fig:Lid-unused-Re-50}
  \end{subfigure}
  \hfill

  \caption{Time evolution of the domain-averaged additional post-collision populations during the lid-driven cavity simulations at Re = 10 and Re = 50.}
  \label{fig:Lid-unused}
\end{figure}

In the Re = 10 case, the highest errors are obtained for the prediction of the rest post-collision population, followed by the axial and then the diagonal populations, consistent with the behavior observed in the Taylor-Green case. At Re = 50, however, the largest errors appear in the axial populations. Figure \ref{fig:Lid-populations} also shows that the largest differences among the axial populations occur for the $\hat f_2^{\text{eq}}(\mathbf{x},t^*)$ and $\hat f_4^{\text{eq}}(\mathbf{x},t^*)$ post-collision populations, which are associated with the lattice velocity directions $\mathbf{e}_2 = (0,1)$ and $\mathbf{e}_4 = (0,-1)$, oriented vertically along the coordinate axis. These populations play a dominant role in determining the $u_y$ velocity component. Because the SQC scheme struggles to capture the steep velocity gradients and strong flow turning near the top-right corner and along the right wall, the predicted post-collision populations for these vertical directions show the largest errors. The other two post-collision axial populations, $\hat f_1^{\text{eq}}(\mathbf{x},t^*)$ and $\hat f_3^{\text{eq}}(\mathbf{x},t^*)$, corresponding to $\mathbf{e}_1 = (1,0)$ and $\mathbf{e}_3 = (-1,0)$ and contributing primarily to the $u_x$ component, are less affected but still show increased errors due to the enhanced horizontal shear near the moving lid and in the recirculating region. At Re = 50, where the velocity gradients are even steeper, the errors in the axial populations become significantly larger and exceed those of the rest population. This also explains the larger deviations in the $u_y$ centerline velocity for the Re = 50 case shown in Figure \ref{fig:Lid-centerline-Re50}. The diagonal populations consistently show the smallest errors because they depend on both $u_x$ and $u_y$ velocity components. In regions where the two velocity components vary in opposite directions, these opposing contributions partially cancel, leading to smaller overall errors. Consequently, the diagonal populations are less sensitive to the steep local gradients that dominate the axial populations.

To quantify the mass transfer into the amplitudes associated with the seven additional basis states, Figure \ref{fig:Lid-unused} shows the domain-averaged time evolution of the fictitious post-collision populations, $\hat f_{9-15}^{\text{eq}}(\mathbf{x},t^*)$, for the Re = 10 and Re = 50 simulations. The results show that the mass transferred into the amplitudes of these additional basis states is small-on the order of $1 \times 10^{-6}$ for the Re = 10 case and $1 \times 10^{-8}$ for Re = 50. The reason for the smaller mass transfer observed at Re = 50, which was also observed in the Taylor-Green case, is not entirely clear. As previously explained, several factors may contribute to this behavior such as differences in how the SQC adjusts its predictions to conserve momentum, how it adapts to the effective dissipation level, or how it tries reproduce the nonlinear terms in the equilibrium distribution. Similar to the Taylor-Green case, the profiles level off over time, indicating that the mass transfer into and out of the amplitudes associated with these basis states reaches a balance over the course of the simulations.

Overall, the results obtained for both the Taylor-Green and lid-driven cavity cases provide clear insight into the SQC’s performance. First, the SQC provides accurate predictions even when extrapolating to velocity ranges beyond those seen during training, where nonlinear terms in the LBM equilibrium distribution become significant. These findings support the analysis in Section \ref{subsec:origin-nonlinearity}, which shows that the SQC can partially capture the nonlinear behavior of the BGK operator. Second, the SQC successfully reproduces the effective dissipative behavior of the BGK operator and adapts its level of dissipation across different Reynolds numbers, as demonstrated in the Taylor-Green case. Finally, the lid-driven cavity results show that the SQC is sensitive to the strong velocity gradients that develop in the flow. This limitation arises because the SQC only partially captures the nonlinear terms of the equilibrium distribution and therefore cannot fully reproduce the Navier-Stokes stress tensor, leading to larger errors in regions with strong velocity gradients.

\section{Conclusion}
\label{conclusion}

This study presented the development and validation of a low-depth SQC for the BGK collision operator on the D$_2$Q$_9$ lattice. The proposed framework combines a rooted-density encoding of the particle populations into the quantum state, the evolution of that state under the SQC, and a final measurement in the computational basis to recover the post-collision populations in physical space. Through analytical derivations and numerical validation, we showed that this framework can partially reproduce both the nonlinear and dissipative behavior of the BGK operator without relying on probabilistic algorithms such as LCU or on multiple copies of the quantum state. Instead, these effects emerge naturally through the mapping between the physical and quantum representations.

By enforcing mass conservation, D$_8$ equivariance and scale equivariance in the circuit design, and encouraging momentum conservation during training, the SQC achieves an accurate and physically consistent approximation of the BGK operator. The final 15-block architecture, built from alternating single-qubit rotation and two-qubit entangling layers, achieves a substantially lower circuit depth than previous quantum collision circuits while still capturing part of the essential nonlinear and dissipative effects. When compiled to the native gate set of IBM’s Heron quantum processor (assuming all-to-all connectivity), the circuit requires only 724 native gates. Moreover, since the collision operation is local, the same circuit can be applied independently at each lattice site, making the total gate count independent of the lattice size.

The SQC was validated on the Taylor-Green vortex and lid-driven cavity flows at Re = 10 and Re = 50. The results demonstrate that the SQC generalizes well beyond its training range, accurately predicting flow behavior in regimes where nonlinear effects in the equilibrium distribution become significant. The SQC also reproduces the dissipative character of the BGK operator, adapting its effective dissipation with Reynolds number. At the same time, the results reveal the SQC’s sensitivity to strong velocity gradients. Since the SQC can only partially capture the nonlinear dependence of the equilibrium, it cannot fully reproduce the Navier-Stokes stress tensor, leading to larger errors in regions where the velocity gradients are strong.

To overcome the current need for re-initialization and measurement at every time step, future work should verify that coherent application of the SQC over multiple time steps, in combination with streaming, can reproduce the nonlinear relaxation behavior characteristic of the BGK operator in the classical setting. Furthermore, it will be necessary to address the accumulation of relative phases that occurs when the SQC is applied repeatedly. This may require training a circuit that can handle complex amplitudes or designing one that avoids phase buildup altogether while maintaining sufficient expressivity. Finally, one could explore whether the seven additional basis states, currently used to effectively contribute to the rest population when computing macroscopic quantities, could be assigned a more physically meaningful role. Since these states expand the accessible Hilbert space, they may offer a way to capture a larger portion of the dissipation represented by the BGK operator if used more effectively.

\section{CRediT authorship contribution statement}

\textbf{Monica L\u{a}c\u{a}tu\c{s}:} Conceptualization, Methodology, Software, Data curation, Investigation, Validation, Visualization, Writing - original draft, Writing - review \& editing. \textbf{Matthias M\"{o}ller:} Supervision, Funding acquisition, Writing - review \& editing.

\section{Acknowledgment}

We gratefully acknowledge support from the joint research project Quantum Computational Fluid Dynamics by Fujitsu Limited and Delft University of Technology, co-funded by the Netherlands Enterprise Agency under project number PPS23-3-03596728.

\bibliographystyle{elsarticle-num}

\section*{References}

\bibliography{paper}

@article{Bedrunka2024,
  author       = {Mario Christopher Bedrunka and Tobias Horstmann and Ben Picard and Dirk Reith and Holger Foysi},
  title        = {Machine Learning Enhanced Collision Operator for the Lattice Boltzmann Method Based on Invariant Networks},
  journal      = {arXiv preprint},
  volume       = {arXiv:2412.08229},
  year         = {2024},
  note         = {physics.comp-ph},
  doi          = {10.48550/arXiv.2412.08229},
}

@article{SucciSanavioLove2025,
  author  = {Sauro Succi and Claudio Sanavio and Peter Love},
  title   = {The foundational value of quantum computing for classical fluids},
  journal = {arXiv preprint arXiv:2510.09178},
  year    = {2025},
  doi     = {10.48550/arXiv.2510.09178},
  url     = {https://arxiv.org/abs/2510.09178}
}

@article{Ortali2024LENNs,
  author       = {Giulio Ortali and Alessandro Gabbana and Imre Atmodimedjo and Alessandro Corbetta},
  title        = {Enhancing lattice kinetic schemes for fluid dynamics with Lattice-Equivariant Neural Networks},
  journal      = {arXiv preprint},
  volume       = {arXiv:2405.13850},
  year         = {2024},
  note         = {physics.comp-ph},
  doi          = {10.48550/arXiv.2405.13850},
}

@article{Horstmann2024,
  author       = {Jan Tobias Horstmann and Mario Christopher Bedrunka and Holger Foysi},
  title        = {Lattice Boltzmann method with artificial bulk viscosity using a neural collision operator},
  journal      = {Computers \& Fluids},
  volume       = {272},
  pages        = {106191},
  year         = {2024},
  doi          = {10.1016/j.compfluid.2024.106191},
  url          = {https://www.sciencedirect.com/science/article/pii/S0045793024000239}
}

@article{Meyer2023Symmetry,
  author = {Meyer, Johannes Jakob and Mularski, Marian and Gil-Fuster, Elies and Mele, Antonio Anna and Arzani, Francesco and Wilms, Alissa and Eisert, Jens},
  title = {{E}xploiting {S}ymmetry in {V}ariational {Q}uantum {M}achine {L}earning},
  journal = {PRX Quantum},
  volume = {4},
  number = {1},
  pages = {010328},
  year = {2023},
  doi = {10.1103/PRXQuantum.4.010328},
}

@article{Mitarai2018Quantum,
  title        = {Quantum circuit learning},
  author       = {Mitarai, Kosuke and Negoro, Makoto and Kitagawa, Masahiro and Fujii, Keisuke},
  journal      = {Phys. Rev. A},
  volume       = {98},
  number       = {3},
  pages        = {032309},
  year         = {2018},
  month        = sep,
  doi          = {10.1103/PhysRevA.98.032309},
  note         = {Published 10 September 2018}
}

@article{McClean2018Barren,
  title        = {Barren plateaus in quantum neural network training landscapes},
  author       = {McClean, Jarrod R. and Boixo, Sergio and Smelyanskiy, Vadim N. and Babbush, Ryan and Neven, Hartmut},
  journal      = {Nature Communications},
  volume       = {9},
  pages        = {4812},
  year         = {2018},
  month        = nov,
  doi          = {10.1038/s41467-018-07090-4},
  note         = {Published 16 November 2018}
}

@article{Nguyen2024Theory,
  title        = {Theory for {E}quivariant {Q}uantum {N}eural {N}etworks},
  author       = {Nguyen, Quynh T. and Schatzki, Louis and Braccia, Paolo and Ragone, Michael and Coles, Patrick J. and Sauvage, Frédéric and Larocca, Martín and Cerezo, M.},
  journal      = {PRX Quantum},
  volume       = {5},
  pages        = {020328},
  year         = {2024},
  month        = may,
  doi          = {10.1103/PRXQuantum.5.020328},
  note         = {Published 6 May 2024}
}

@book{krugerLBM,
  author       = {Krüger, Timm and Kusumaatmaja, Halim and Kuzmin, Alexandr
                  and Shardt, Orest and Silva, Goncalo and Viggen, Erlend Magnus},
  title        = {The {L}attice {B}oltzmann {M}ethod: {P}rinciples and {P}ractice},
  series       = {Graduate Texts in Physics},
  edition      = {1st},
  publisher    = {Springer},
  year         = {2017},
  doi          = {10.1007/978-3-319-44649-3},
  isbn         = {978-3-319-83103-9},
}

@book{Chuang,
  author       = {Nielsen, Michael A. and Chuang, Isaac L.},
  title        = {Quantum {C}omputation and {Q}uantum {I}nformation},
  edition      = {1st},
  series       = {Cambridge Series on Information and the Natural Sciences},
  publisher    = {Cambridge University Press},
  address      = {Cambridge, UK},
  year         = {2000},
  isbn         = {0-521-63503-9},
  doi          = {10.1017/CBO9780511976667},
}

@article{corbetta2023learning,
  author  = {Corbetta, Alessandro and Gabbana, Alessandro and Gyrya, Vitaliy and Livescu, Daniel and Prins, Joost and Toschi, Federico},
  title   = {Toward learning Lattice {B}oltzmann collision operators},
  journal = {European Physical Journal E},
  series  = {Soft Matter},
  volume  = {46},
  number  = {3},
  pages   = {10},
  month   = mar,
  year    = {2023},
  doi     = {10.1140/epje/s10189-023-00267-w},
  issn    = {1292-895X},
}

@article{Theis2017End,
  author       = {Theis, Thomas N. and Wong, H.-S. Philip},
  title        = {{The {E}nd of {M}oore's {L}aw: {A} {N}ew {B}eginning for {I}nformation {T}echnology}},
  journal      = {Computing in Science \& Engineering},
  volume       = {19},
  number       = {2},
  pages        = {41--50},
  month        = mar,
  year         = {2017},
  doi          = {10.1109/MCSE.2017.29},
  issn         = {1521-9615}
}

@article{Love2019QuantumExtensions,
  author  = {Love, Peter},
  title   = {{On Quantum Extensions of Hydrodynamic Lattice Gas Automata}},
  journal = {Condensed Matter},
  volume  = {4},
  number  = {2},
  pages   = {48},
  year    = {2019},
  doi     = {10.3390/condmat4020048},
  eissn   = {2410-3896},
}

@article{Fonio2025Quantum,
  author  = {Fonio, Niccol\`o and Sagaut, Pierre and Di Molfetta, Giuseppe},
  title   = {{Quantum collision circuit, quantum invariants and quantum phase estimation procedure for fluid dynamic lattice gas automata}},
  journal = {Computers \& Fluids},
  volume  = {299},
  pages   = {106688},
  month   = jun,
  year    = {2025},
  doi     = {10.1016/j.compfluid.2025.106688},
  issn    = {0045-7930}
}

@article{TodorovaSteijl2020,
  author  = {Todorova, B.\ N. and Steijl, R.},
  title   = {Quantum algorithm for the collisionless {B}oltzmann equation},
  journal = {Journal of Computational Physics},
  volume  = {409},
  pages   = {109347},
  year    = {2020},
  doi     = {10.1016/j.jcp.2020.109347},
}

@article{SchalkersMoller2024,
  author       = {Schalkers, Merel A. and Möller, Matthias},
  title        = {Efficient and fail-safe quantum algorithm for the transport equation},
  journal      = {Journal of Computational Physics},
  volume       = {502},
  pages        = {Article 112816},
  year         = {2024},
  doi          = {10.1016/j.jcp.2024.112816},
  issn         = {0021-9991},
}

@article{SchalkersMoller2024b,
  author       = {Schalkers, Merel A. and M{\"o}ller, Matthias},
  title        = {Momentum exchange method for quantum {B}oltzmann methods},
  journal      = {Computers \& Fluids},
  volume       = {285},
  pages        = {Article 106453},
  year         = {2024},
  doi          = {10.1016/j.compfluid.2024.106453},
  issn         = {0045-7930},
}

@article{ItaniSucci2024,
  author  = {Itani, Wael and Sreenivasan, Katepalli R. and Succi, Sauro},
  title   = {Quantum {A}lgorithm for {L}attice {B}oltzmann {(QALB)} {S}imulation of {I}ncompressible {F}luids with a {N}onlinear {C}ollision {T}erm},
  journal = {Physics of Fluids},
  volume  = {36},
  number  = {1},
  pages   = {017112},
  year    = {2024},
  doi     = {10.1063/5.0176569},
}

@article{SanavioSucci2024,
  author  = {Sanavio, Claudio and Succi, Sauro},
  title   = {Lattice {B}oltzmann–{C}arleman quantum algorithm and circuit for fluid flows at moderate {R}eynolds number},
  journal = {AVS Quantum Science},
  volume  = {6},
  number  = {2},
  pages   = {023802},
  year    = {2024},
  doi     = {10.1116/5.0195549},
}

@misc{WangMengZhaoYang2025,
  title        = {Quantum lattice {B}oltzmann method for simulating nonlinear fluid dynamics},
  author       = {Wang, Boyuan and Meng, Zhaoyuan and Zhao, Yaomin and Yang, Yue},
  year         = {2025},
  eprint       = {arXiv:2502.16568},
  archivePrefix= {arXiv},
  primaryClass = {physics.flu-dyn},
  doi          = {10.48550/arXiv.2502.16568},
}

@article{SanavioSucci2025,
  author       = {Sanavio, Claudio and Simon, William A. and Ralli, Alexis and Love, Peter and Succi, Sauro},
  title        = {Carleman--lattice--{B}oltzmann quantum circuit with matrix access oracles},
  journal      = {Physics of Fluids},
  volume       = {37},
  number       = {3},
  pages        = {037123},
  year         = {2025},
  doi          = {10.1063/5.0254588},
}

@article{KumarFrankel2025,
  author  = {Kumar, E. Dinesh and Frankel, Steven H.},
  title   = {Quantum unitary matrix representation of the lattice {B}oltzmann model for low {R}eynolds fluid flow simulation},
  journal = {AVS Quantum Science},
  volume  = {7},
  number  = {1},
  pages   = {013802},
  year    = {2025},
  doi     = {10.1116/5.0245082},
}

@article{KocherBryngelson2024,
  title   = {A multiple-circuit approach to quantum resource reduction with application to the quantum lattice {B}oltzmann method},
  author  = {Lee, Melody and Song, Zhixin and Kocherla, Sriharsha and Adams, Austin and Alexeev, Alexander and Bryngelson, Spencer H.},
  journal = {arXiv preprint arXiv:2401.12248},
  year    = {2024},
  doi     = {10.48550/arXiv.2401.12248},
}

@article{WawrzyniakEtAl2024,
  author  = {Wawrzyniak, David and Winter, Josef and Schmidt, Steffen and Indinger, Thomas and Janßen, Christian F. and Schramm, Uwe and Adams, Nikolaus A.},
  title   = {A quantum algorithm for the lattice--{B}oltzmann method advection--diffusion equation},
  journal = {Computer Physics Communications},
  volume  = {306},
  pages   = {109373},
  year    = {2024},
  doi     = {10.1016/j.cpc.2024.109373},
}

@article{Li2009,
  title   = {{DNS} of a spatially developing turbulent boundary layer with passive scalar transport},
  author  = {Li, Qiang and Schlatter, Philipp and Brandt, Luca and Henningson, Dan S.},
  journal = {International Journal of Heat and Fluid Flow},
  volume  = {30},
  number  = {5},
  pages   = {916--929},
  year    = {2009},
  doi     = {10.1016/j.ijheatfluidflow.2009.06.007},
}

@article{Yao2023,
  title   = {Direct numerical simulations of turbulent pipe flow up to {$Re_\tau \approx 5200$}},
  author  = {Yao, Jie and Rezaeiravesh, Saleh and Schlatter, Philipp and Hussain, Fazle},
  journal = {Journal of Fluid Mechanics},
  volume  = {956},
  pages   = {A18},
  year    = {2023},
  doi     = {10.1017/jfm.2022.1013},
}

@inproceedings{Grover1996,
  author    = {Grover, Lov K.},
  title     = {A {F}ast {Q}uantum {M}echanical {A}lgorithm for {D}atabase {S}earch},
  booktitle = {Proceedings of the Twenty-Eighth Annual ACM Symposium on Theory of Computing (STOC ’96)},
  year      = {1996},
  pages     = {212--219},
  doi       = {10.1145/237814.237866},
}

@inproceedings{Shor1994,
  author    = {Shor, Peter W.},
  title     = {Algorithms for {Q}uantum {C}omputation: {D}iscrete {L}ogarithms and {F}actoring},
  booktitle = {Proceedings of the 35th Annual Symposium on Foundations of Computer Science (FOCS)},
  year      = {1994},
  month     = {Nov},
  pages     = {124--134},
  address   = {Santa Fe, NM, USA},
  publisher = {IEEE},
  doi       = {10.1109/SFCS.1994.365700},
  isbn      = {0-8186-6580-7},
}

@article{Kandala2019,
  title   = {Error mitigation extends the computational reach of a noisy quantum processor},
  author  = {Kandala, Abhinav and Temme, Kristan and C{\'{o}}rcoles, Antonio D. and Mezzacapo, Antonio and Chow, Jerry M. and Gambetta, Jay M.},
  journal = {Nature},
  volume  = {567},
  pages   = {491--495},
  year    = {2019},
  doi     = {10.1038/s41586-019-1040-7},
}

@article{Yu2014,
  title   = {{GPU} accelerated lattice {B}oltzmann simulation for rotational turbulence},
  author  = {Yu, H. and Chen, R. and Wang, H. and Yuan, Z. and Zhao, Y. and An, Y. and Xu, Y. and Zhu, L.},
  journal = {Computers \& Mathematics with Applications},
  volume  = {67},
  number  = {2},
  pages   = {437--451},
  year    = {2014},
  doi     = {10.1016/j.camwa.2013.09.017},
}

@article{Budinski2021,
  author  = {Budinski, Ljubomir},
  title   = {Quantum algorithm for the advection--diffusion equation simulated with the lattice {B}oltzmann method},
  journal = {Quantum Information Processing},
  volume  = {20},
  pages   = {57},
  year    = {2021},
  month   = {feb},
  doi     = {10.1007/s11128-021-02996-3},
}

@article{Xu2025,
  author  = {Xu, Li and Li, Ming and Zhang, Lei and Sun, Hai and Yao, Jun},
  title   = {Improved quantum lattice {B}oltzmann method for advection-diffusion equations with a linear collision model},
  journal = {Physical Review E},
  volume  = {111},
  number  = {4},
  pages   = {045305},
  year    = {2025},
  month   = {apr},
  doi     = {10.1103/PhysRevE.111.045305},
}

@misc{IBMQuantumTranspiler,
  author    = {{IBM Quantum Documentation}},
  title     = {Transpiler (`qiskit.transpiler`)},
  url       = {https://quantum.cloud.ibm.com/docs/en/api/qiskit/transpiler},
  urldate   = {2025-07-29},
  note      = {Accessed: 29 July 2025}
}

@article{SchalkersMoller2024DataEncoding,
  author  = {Schalkers, Merel A. and M{\"o}ller, Matthias},
  title   = {On the importance of data encoding in quantum {B}oltzmann methods},
  journal = {Quantum Information Processing},
  volume  = {23},
  number  = {1},
  pages   = {20},
  year    = {2024},
  doi     = {10.1007/s11128-023-04216-6},
}

@article{BastidaZamora2024EfficientQLGA,
  author  = {Bastida Zamora, Antonio David and Budinski, Ljubomir and Niemim{\"a}ki, Ossi and Lahtinen, Valtteri},
  title   = {Efficient quantum lattice gas automata},
  journal = {Computers \& Fluids,},
  year    = {2024},
  doi     = {10.1016/j.compfluid.2024.106476},
}

@article{Yepez2002Burgers,
  author  = {Yepez, Jeffrey},
  title   = {Quantum lattice-gas model for the {B}urgers equation},
  journal = {Journal of Statistical Physics},
  volume  = {107},
  number  = {1-2},
  pages   = {203--224},
  year    = {2002},
  doi     = {10.1023/A:1014514805610},
}

@article{Gaitan2020NSQuantum,
  author  = {Gaitan, Frank},
  title   = {Finding flows of a {N}avier–{S}tokes fluid through quantum computing},
  journal = {npj Quantum Information},
  volume  = {6},
  article-number = {61},
  year    = {2020},
  doi     = {10.1038/s41534-020-00291-0},
}

@article{Budinski2022,
  author       = {Budinski, Ljubomir},
  title        = {Quantum algorithm for the {N}avier--{S}tokes equations by using the streamfunction-vorticity formulation and the lattice {B}oltzmann method},
  journal      = {International Journal of Quantum Information},
  volume       = {20},
  number       = {02},
  year         = {2022},
  doi          = {10.1142/S0219749921500398},
}

@article{Liu2023VQLSStokes,
  author  = {Liu, Yangyang and Chen, Zhen and Shu, Chang and Rebentrost, Patrick and Liu, Yaguang and Chew, S. C. and Khoo, B. C. and Cui, Y. D.},
  title   = {A variational quantum algorithm–based numerical method for solving potential and {S}tokes flows},
  journal = {arXiv preprint arXiv:2303.01805},
  year    = {2023},
  doi     = {10.48550/arXiv.2303.01805},
}

@article{Song2024HybridNS,
  author  = {Song, Zhixin and Deaton, Robert and Gard, Bryan and Bryngelson, Spencer H.},
  title   = {Incompressible {N}avier–{S}tokes solve on noisy quantum hardware via a hybrid quantum–classical scheme},
  journal = {arXiv preprint arXiv:2406.00280},
  year    = {2024},
  doi     = {10.48550/arXiv.2406.00280}
}

@article{Kyriienko2020DQCsNS,
  author  = {Kyriienko, Oleksandr and Paine, Annie E. and Elfving, Vincent E.},
  title   = {Solving nonlinear differential equations with differentiable quantum circuits},
  journal = {arXiv preprint arXiv:2011.10395},
  year    = {2020},
  doi     = {10.48550/arXiv.2011.10395},
}

@misc{ibm2025nativegates,
  author    = {{IBM Quantum Documentation}},
  title     = {Native gates and operations},
  url       = {https://quantum.cloud.ibm.com/docs/guides/native-gates},
  urldate   = {2025-07-29},
  note      = {Accessed: 29 July 2025}
}

@article{childs2012hamiltonian,
  author       = {Childs, Andrew M. and Wiebe, Nathan},
  title        = {Hamiltonian {S}imulation {U}sing {L}inear {C}ombinations of {U}nitary {O}perations},
  journal      = {Quantum Inf. Comput.},
  volume       = {12},
  number       = {11–12},
  pages        = {901--924},
  year         = {2012},
  publisher    = {Rinton Press},
  doi          = {10.5555/2481569.2481570},
}

@phdthesis{Itanithesis,
  author       = {Itani, Wael},
  title        = {Towards a {Q}uantum {A}lgorithm for {L}attice {B}oltzmann ({QALB})
{S}imulation with a {N}onlinear {C}ollision {T}erm},
  school       = {New {Y}ork {U}niversity
{T}andon {S}chool of {E}ngineering},
  year         = {2025},
  address      = {New York, New York},
  type         = {Ph.D. dissertation},
  url          = {https://www.proquest.com/openview/a52e013673992dcace5c320c9d4dcb09},
}

\newpage

\begin{appendices}
\section{Data Generations}
\label{App:DataGen}

\begin{figure}[http]
    \centering

    \begin{subfigure}[b]{0.48\textwidth}
        \centering
        \includegraphics[width=\textwidth]{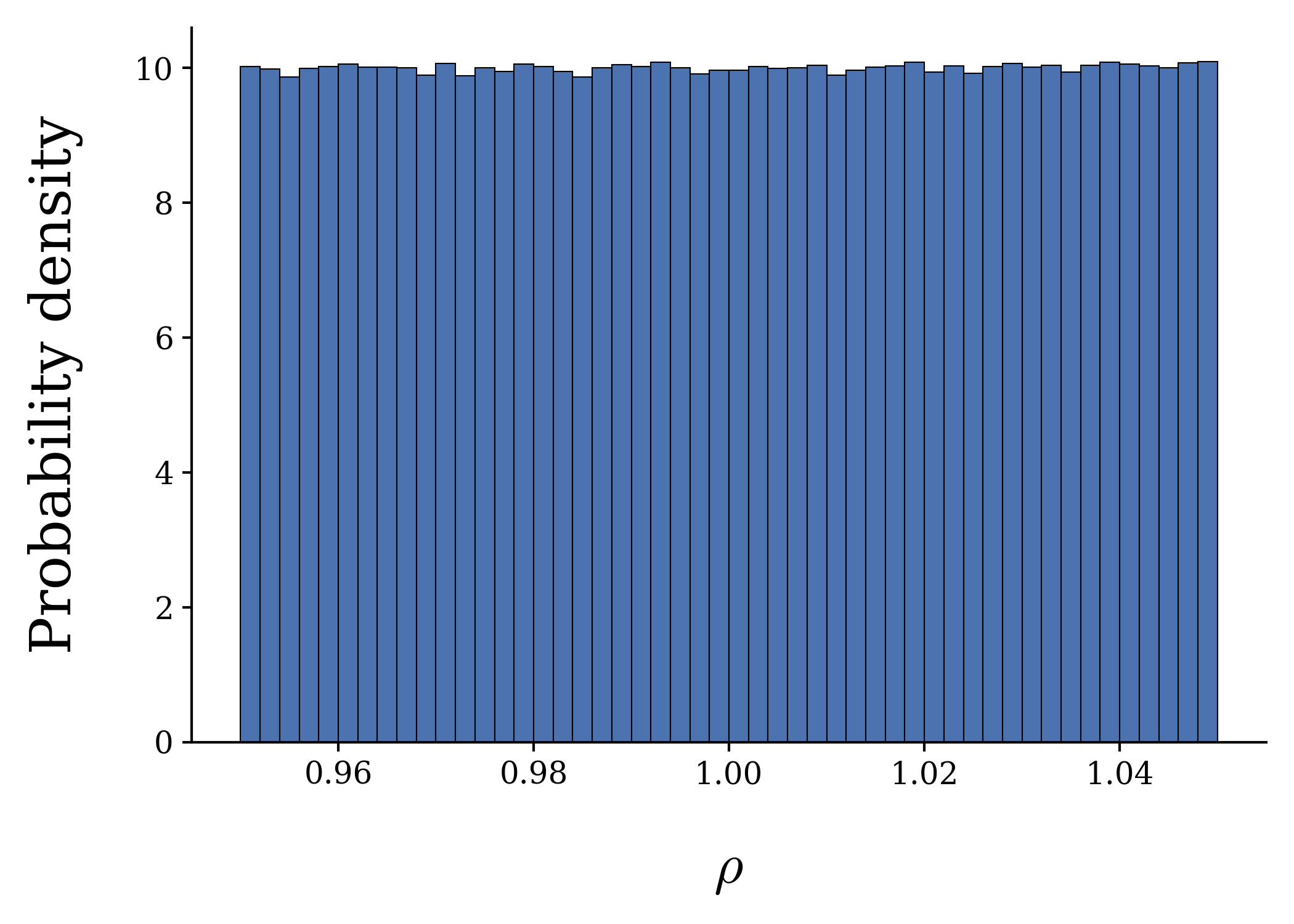}
        \caption{Density distribution $\rho$.}
        \label{fig:rho_dist}
    \end{subfigure}
    \hfill
    \begin{subfigure}[b]{0.48\textwidth}
        \centering
        \includegraphics[width=\textwidth]{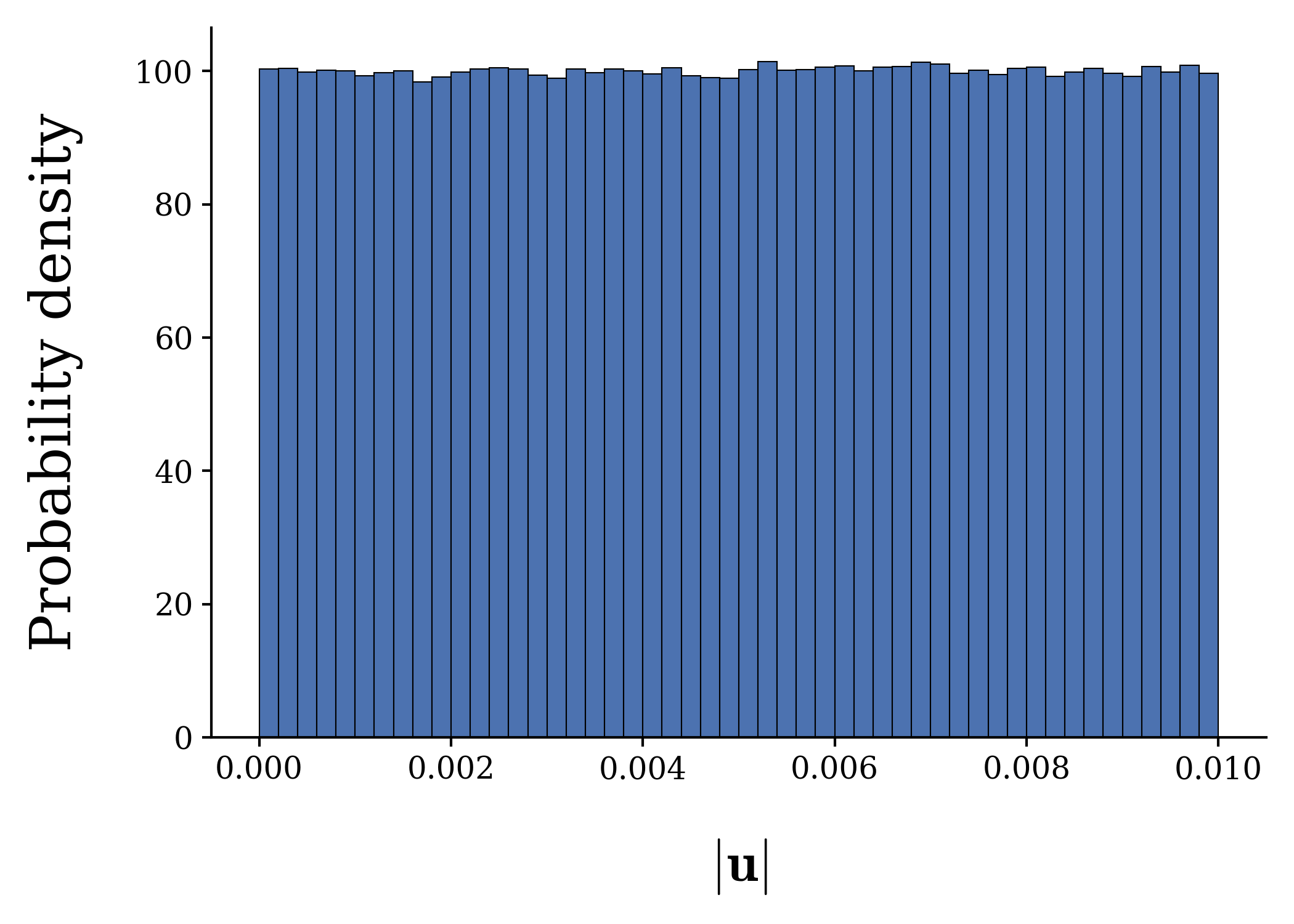}
        \caption{Velocity magnitude distribution $\lVert \mathbf{u} \rVert$.}
        \label{fig:u_mag_dist}
    \end{subfigure}

    \vspace{0.5cm}

    \begin{subfigure}[b]{0.48\textwidth}
        \centering
        \includegraphics[width=\textwidth]{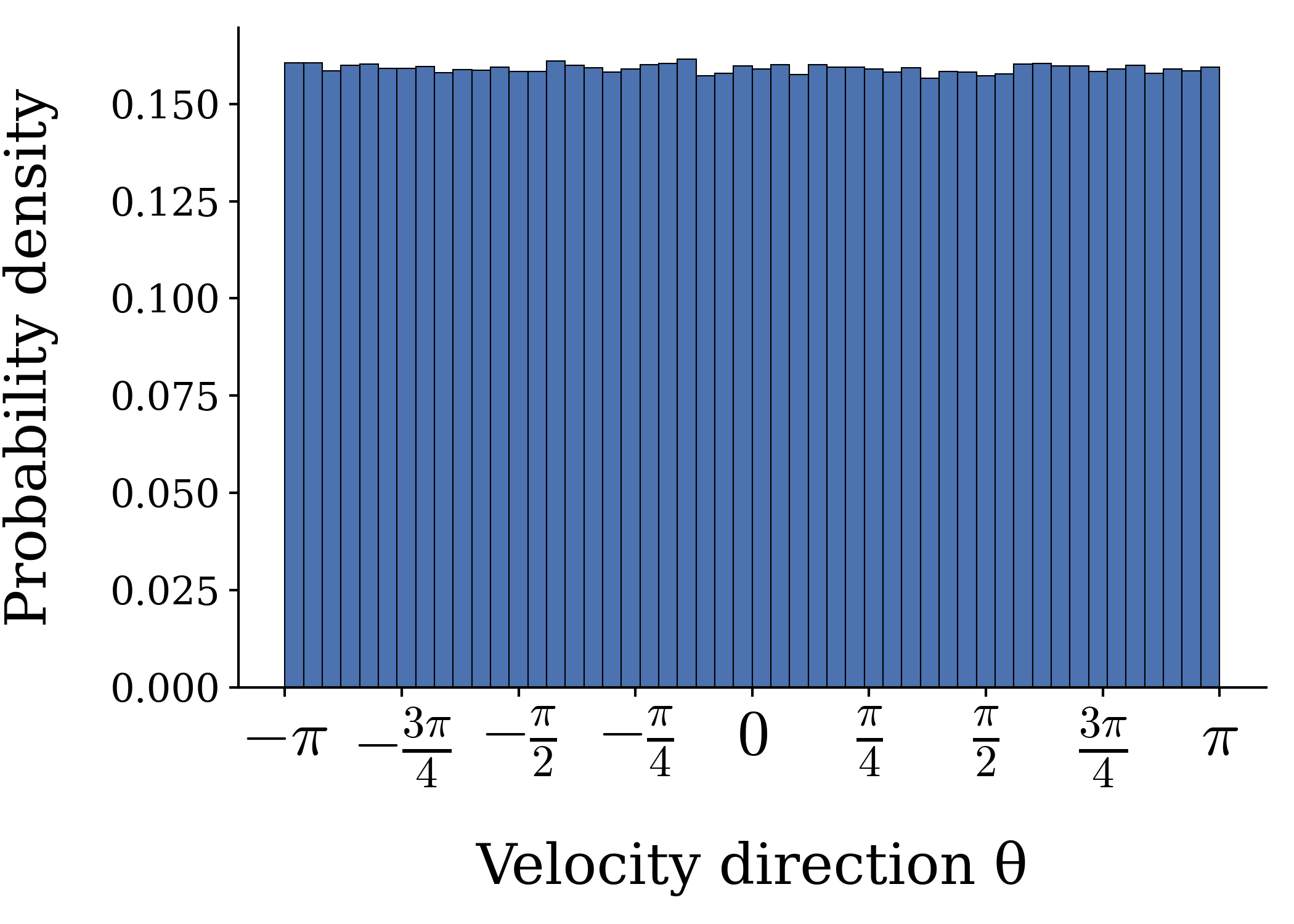}
        \caption{Velocity direction distribution \\ $\theta = \arctan2(u_y,\,u_x)$.}
        \label{fig:u_angle_dist}
    \end{subfigure}
    \hfill
    \begin{subfigure}[b]{0.48\textwidth}
        \centering
        \includegraphics[width=\textwidth]{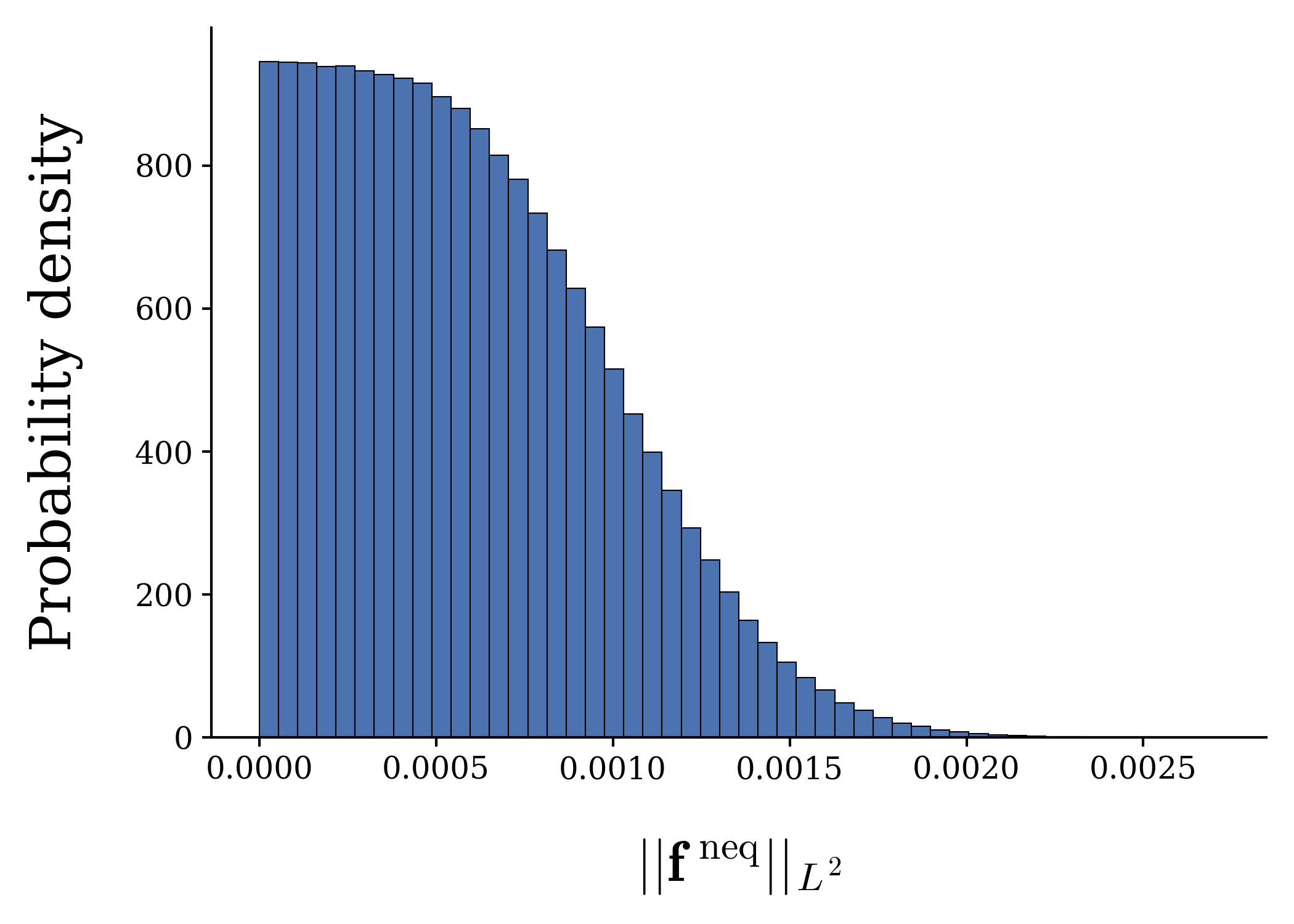}
        \caption{$L^{2}$-norm of the non-equilibrium vector $\lVert \mathbf{f}^{\mathrm{neq}} \rVert_{2}$.}
        \label{fig:fneq_dist}
    \end{subfigure}

    \caption{
        Distributions sampled for the training dataset. The plots show (a) the density distribution, (b) the magnitude of the velocity field, (c) the corresponding velocity-direction distribution, and (d) the $L^{2}$-norm of the non-equilibrium population vector. 
    }
    \label{fig:training_dists}
\end{figure}

The parameters used for generating the training dataset are summarized in Table \ref{tab:exp1}. A visualization of the sampled density distribution, the velocity-magnitude distribution together with the associated velocity-direction distribution, as well as the $L^2$-norm distribution of the non-equilibrium components, is provided in Figure \ref{fig:training_dists}.

\begin{table}[ht]
  \centering
  \caption{Parameters used in data generation.}
  \label{tab:exp1}
  \begin{tabular}{@{} l c l @{}}
    \toprule
    \textbf{Parameter}
      & \textbf{Symbol}
      & \textbf{Value} \\
    \midrule
    Number of training samples
      & $N$
      & 1,000,000 \\
    Density
      & $\rho$
      & Uniform on $[0.95,\,1.05]$ \\
    Velocity magnitude
      & $|\mathbf{u}|$
      & Uniform on $[0.00,\,0.01]$ \\
    Std.\ deviation of non-eq.\ noise
      & $\sigma_{\mathrm{neq}}$
      & Uniform on $[0.00,\,5\times10^{-4}]$ \\
    Relaxation time
      & $\tau$
      & 1 \\
    Test split fraction
      & $r_{\mathrm{split}}$
      & 0.001 \\
    \bottomrule
  \end{tabular}
\end{table}

\newpage
\section{Optimal Circuit Architecture Experiments}
\label{App:CircAcc}

\begin{table}[ht]
  \centering
  \caption{Training parameters used in the circuit architecture experiments.}
  \label{tab:exp2}
  \sisetup{
    table-number-alignment = center,
    detect-all
  }
  \begin{tabular}{
    l           
    S[table-format=1.2]  
  }
    \toprule
    \textbf{Training Parameter} 
      & {\textbf{Setting}} \\
    \midrule
    Optimizer       
      & {JAX Gradient Descent} \\
    Learning Rate   
      & {0.05} \\
    Training Iterations 
      & {250,000} \\
    Batch Size      
      & {5} \\
    Momentum Conservation Penalty 
      & {OFF} \\
    Accuracy Tolerance 
      & {$1\times10^{-5}$} \\
    Parameter Initialization  
      & {Uniform on $[-\pi,\pi]$} \\
    \bottomrule
  \end{tabular}
\end{table}

\section{SQC Depth Analysis Experiments}
\label{App:depth}

\begin{table}[ht]
  \centering
  \caption{Training parameters used in the circuit depth analysis experiments.}
  \label{tab:exp3}
  \sisetup{
    table-number-alignment = center,
    detect-all
  }
  \begin{tabular}{
    l           
    S[table-format=1.2]  
  }
    \toprule
    \textbf{Training Parameter} 
      & {\textbf{Setting}} \\
    \midrule
    Optimizer       
      & {JAX Gradient Descent} \\
    Learning Rate   
      & {0.05} \\
    Training Iterations 
      & {250,00} \\
    Batch Size      
      & {5} \\
    Momentum Conservation Penalty 
      & {OFF} \\
    Accuracy Tolerance 
      & {$1\times10^{-5}$} \\
    Parameter Initialization  
      & {Uniform on $[-\pi,\pi]$} \\
    \bottomrule
  \end{tabular}
\end{table}

\newpage
\section{Momentum Penalty Experiments}
\label{App:cons}

\begin{table}[ht]
  \centering
  \caption{Training parameters used in the momentum penalty experiments.}
  \label{tab:exp4}
  \sisetup{
    table-number-alignment = center,
    detect-all
  }
  \begin{tabular}{
    l           
    S[table-format=1.2]  
  }
    \toprule
    \textbf{Training Parameter} 
      & {\textbf{Setting}} \\
    \midrule
    Optimizer       
      & {JAX Gradient Descent} \\
    Learning Rate   
      & {0.05} \\
    Training Iterations 
      & {250,000} \\
    Batch Size      
      & {5} \\
    Momentum Conservation Penalty 
      & {ON} \\
    Accuracy Tolerance 
      & {$1\times10^{-5}$} \\
    Parameter Initialization  
      & {Uniform on $[-\pi,\pi]$} \\
    \bottomrule
  \end{tabular}
\end{table}

\end{appendices}

\end{document}